\documentclass[a4paper,11pt]{article}
\pdfoutput=1 
\usepackage{jheppub_mod} 
\usepackage{graphicx, amsmath, amssymb, amstext, amsthm, amsfonts}

\usepackage{float}
\usepackage{multicol}
\usepackage{xcolor, enumitem}

\newcommand {\pythia}       {{\tt PYTHIA8-8.2.26}~\cite{Sjostrand:2006za,Sjostrand:2014zea}}
\newcommand {\pyth}         {\tt PYTHIA8}

\newcommand {\fastjet}      {{\tt fastjet-3}~\cite{Cacciari2012}}

\newcommand {\heptoptagger} {{\tt HEPTopTagger2}~\cite{Plehn:2011tg,Kasieczka:2015jma,Plehn:2010st,Plehn:2009rk}}
\newcommand {\toptag}       {\tt HEPTopTagger2}
\newcommand {\cajet}        {{Cambridge/Aachen} (C/A)~\cite{Dokshitzer:1997in}}
\newcommand {\akjet}        {Anti-$k_t$~\cite{Cacciari:2008gp}}
\newcommand {\ktjet}        {$k_t$~\cite{Ellis:1993tq,Catani:1993hr}}

\newcommand {\archA} {\tt Arch-1}
\newcommand {\archB} {\tt Arch-2}
\newcommand {\archC} {\tt Arch-3}

\definecolor{mred}{rgb}{0.5,0.0,0.0}
\definecolor{mgreen}{rgb}{0.0,0.5,0.0}
\definecolor{mblue}{rgb}{0.0,0.0,0.5}

\newcommand {\secref}    [1] {Sec.~\ref{#1}}
\newcommand {\subsecref} [1] {Subsec.~\ref{#1}}

\newcommand {\eqnref}    [1] {Eq.~\eqref{#1}}
\newcommand {\figref}    [1] {Figure~\ref{#1}}

\newcommand{\gev}{~\text{GeV} }
\newcommand{\tev}{~\text{TeV} }

\title{A robust anomaly finder based on autoencoders}

\author[a]{Tuhin S. Roy}
\author[b]{and Aravind H. Vijay}
\affiliation[a]{Department of Theoretical Physics, Tata Institute of 
	Fundamental Research, Mumbai 400005, India}
\affiliation[b]{Department of High Energy Physics, Tata Institute of 
	Fundamental Research, Mumbai 400005, India}
\emailAdd{tuhin@theory.tifr.res.in}
\emailAdd{aravind.vijay@tifr.res.in}
\date{\today}
\preprint{TIFR-TH/19-4}

\definecolor{c1}{rgb}{0.6,0.0,0.0}
\definecolor{c2}{rgb}{0.48,0.36,0.0}
\definecolor{c3}{rgb}{0.36,0.48,0.0}
\definecolor{c4}{rgb}{0.0,0.6,0.0}
\definecolor{c5}{rgb}{0.0,0.48,0.36}
\definecolor{c6}{rgb}{0.0,0.36,0.48}
\definecolor{c7}{rgb}{0.0,0.0,0.6}
\definecolor{c8}{rgb}{0.36,0.0,0.48}
\definecolor{c9}{rgb}{0.48,0.0,0.36}

\definecolor{b1}{rgb}{0.4,0.0,0.0}
\definecolor{b2}{rgb}{0.32,0.24,0.0}
\definecolor{b3}{rgb}{0.24,0.32,0.0}
\definecolor{b4}{rgb}{0.0,0.4,0.0}
\definecolor{b5}{rgb}{0.0,0.32,0.24}
\definecolor{b6}{rgb}{0.0,0.24,0.32}
\definecolor{b7}{rgb}{0.0,0.0,0.4}
\definecolor{b8}{rgb}{0.24,0.0,0.32}
\definecolor{b9}{rgb}{0.32,0.0,0.24}

\begin{document}
    
\abstract{
	We propose a robust method to identify anomalous jets by vetoing QCD-jets.
	The robustness of this method ensures that the distribution of the proposed discriminating variable (which allows us to veto QCD-jets) remains unaffected by the phase space of QCD-jets, even if they were different from the region on which the model was trained.
	%	{\bf{\color{c1}{(Please have a look at the above change.)}}}
	%even if QCD-jets from different phase space bins (as compared to what it was trained on) are used as control samples.
	This suggests that our method can be used to look for anomalous jets in high $m/p_{T}$ bins by simply training on jets from low $m/p_{T}$ bins, where sufficient background-enriched data is available.
	The robustness follows from combining an autoencoder with a novel way of pre-processing jets.
	We use momentum rescaling followed by a Lorentz boost to find the frame of reference where any given jet is characterized by predetermined mass and energy.
	In this frame we generate jet images by constructing a set of orthonormal basis vectors using the Gram-Schmidt method to span the plane transverse to the jet axis.
	Due to our preprocessing, the autoencoder loss function does not depend on the initial jet mass, momentum or orientation while still offering remarkable performance.
	%	{\bf{
	%		{\color{c1}{{\tt{(Do we need to remove this part?)}}}}
	We also explore the application of this loss function combined (using supervised learning techniques like boosted decision trees) with few other jet observables like the mass and Nsubjettiness for the purpose of top tagging.
	This exercise shows that our method performs almost as well as existing top taggers which use a large amount of physics information associated with top decays while also reinforcing the fact that the loss function is mostly independent of the additional jet observables.
	%	}}     
}

\maketitle

%--------------------------------------------
\section{Introduction}
\label{sec:Intro} 
%------------------------------------------------------------

With the recent discovery of the $125\gev$ Higgs by CMS~\cite{Chatrchyan:2012xdj} and ATLAS~\cite{Aad:2012tfa} collaborations of the LHC, the last missing link in the standard model (SM) of particle physics is completed. The focus has now shifted to find physics beyond the standard model (BSM), for which there are plethora of candidates all designed to solve various short comings of SM. Unfortunately, we have failed to observe any trace of physics beyond the standard model (BSM) so far in all our searches for these ``well-motivated" scenarios. At this juncture, it may well be equally justifiable to concentrate on the ``Bottom-up'' approach of looking for the ``unexpected'' in data, apart from the usual ``Top-down'' approach of first hypothesizing an extension to the SM and then predicting its phenomenology. Even though this ``Bottom-up'' approach is nothing other than  looking for deviations from the SM background and there has been remarkable progress recently in automated precision (higher order) calculations (radiative corrections due to quantum chromodynamics (QCD) and electroweak (EW) interactions as well), it is still extremely difficult (impractical) to precisely estimate all backgrounds due to SM \emph{without} using data. In this context, searching for new physics (NP) as anomalies directly from data depict a qualitatively new and highly promising direction. 

Note that the philosophy of hunting for NP as anomalies where data is used to estimate background is already a common practice in jet physics. A simple example is the search for boosted massive resonances using jet-mass $m_J$. The background is generically due to QCD-jets. The well-studied and well-understood variable $\frac{d{\sigma}}{d{m_J}}$ for QCD-jets has a continuously falling distribution at high $m_J$.  Any massive resonance decaying hadronically and boosted enough such that all its decay products is contained in a jet, shows up as a bump peaked around its mass. Looking for these bumps over a falling spectrum provides one such simplistic example where one does not need to rely on the exact nature of the massive resonance (as long as it decays hadronically).  Knowing this falling distribution due to QCD is enough, in principle, to look for unknown new particles. Additionally, jet masses for boosted massive resonances are robust under various jet grooming procedures~\cite{Butterworth:2008iy, Salam:2009jx, Krohn:2009th, Ellis:2009me}, owing to their substructures, as compared to jet masses for QCD-jets. Searching for bumps in the distribution for groomed jet masses, therefore, provides a model-independent way to look for hadronically decaying NP particles with substantial discovery potential. One can even take, massive resonances of SM like the $W^{\pm}, Z$ gauge bosons, Higgs particle $h$, and top quark $t$, as test cases to hone this search procedure. A broader generalization of this philosophy, is given in Ref~\cite{Chakraborty:2017mbz}, where the authors use a large set of jet-shape observables to construct a veto for standard objects such as QCD-jets, photons, leptons, etc. Jets that pass this veto, would be termed anomalies. Examples of such anomalies include jets containing decay products (not restricted to hadrons only) of NP resonances, long-lived particles etc. 

Machine learning can be of a big help in looking for such anomalies by constructing an anti-QCD tagger~\cite{Aguilar-Saavedra:2017rzt}. The traditional applications~\cite{Cohen:2017exh,Qu:2019gqs,Datta:2017rhs,Moore:2018lsr,Komiske:2018cqr,Butter:2017cot,Komiske:2017aww,Louppe:2017ipp,Erdmann:2018shi,Hocker:2007ht,Pearkes:2017hku,Almeida:2015jua,Macaluso:2018tck,Kasieczka:2017nvn,Cogan:2014oua,Baldi:2016fql,deOliveira:2015xxd,Collins:2019jip,Kasieczka:2019dbj,Collins:2018epr} of machine learning techniques for classification problems in collider physics are, however, centered around the idea of supervised learning where the training data is labeled into various categories and the model tries to correctly classify the training data into these categories. However, this method is inadequate when a model needs to be trained directly on data which is not labeled. This is where  autoencoders, which are unsupervised learning algorithms \cite{Cerri:2018anq,Andreassen:2018apy,Farina:2018fyg,Heimel:2018mkt,Hajer:2018kqm}, can be useful. Because of training, an autoencoder learns to reproduce QCD-jets efficiently, and it is expected to be not as effective at reproducing jets originating from other processes, resulting in larger ``loss-functions".  Therefore, the loss function of the autoencoder can be used to find anomalies~\cite{Farina:2018fyg, Heimel:2018mkt}).

An essential requirement for a high performance machine learning based anomaly finder, is that it needs to be trained with control samples of QCD-jets. Since one expects anomalies to appear within samples of jets at high jet masses, one essentially needs to train using QCD-jets of high masses. Note however, as pointed out in Ref.~\cite{Dasgupta:2013ihk}, a better variable to understand the underlying physics of QCD-jets is the dimensionless variable $\rho \equiv \frac{m_J}{p_T R}$, where $R$ reflects the  size of jets. The issue of training with large jet masses, therefore, gets translated to training using jets with a large $\rho$ (given that jets are constructed using a minimum $p_T$ and an $R$-parameter).  In any case, it is difficult to obtain sufficiently pure samples of QCD-jets with high jet-mass or a high $\rho$ from data itself. Apart from contamination effects, it is also difficult to gather sufficient training data in these high mass bins. This problem can be partially mitigated if one can use trainings on jets from low $\rho$ bins, which are relatively more pure in QCD-jets and at the same time are easily available. In order to implement this strategy, however, we require an anomaly hunter that is \emph{robust}. Technically speaking, by robustness we here refer to the criteria that the distribution of the loss-function for QCD-jets remains unaltered even if training samples from different $\rho$ bins are used.  

Unfortunately, when an autoencoder is trained in a naive way, it tends to learn features which have large correlations with the jet mass. This is expected since the jet mass is a powerful observable to characterize a jet. Therefore, generically speaking, one does not expect a simple autoencoder based anomaly hunter or anti-QCD tagger to be robust. It is not a surprise that previous attempts in this direction~\cite{Farina:2018fyg, Heimel:2018mkt} find significant dependence of autoencoder loss on jet masses. Consequently, cuts on such jet mass dependent loss functions necessarily sculpt the jet mass distribution itself, which makes techniques like bump hunts (for boosted massive resonances) and side band analyses much more challenging. Ref.~\cite{Farina:2018fyg} suggests using convolution autoencoders, since these are less affected by the above problem.
%(we have also explored this in our study and found it to be consistent). 
However, the improvement in robustness is marginal\cite{Farina:2018fyg}. On the other hand,  Ref.~\cite{Heimel:2018mkt} tries to alleviate this problem using adversary training to reduce the dependence of jet mass on autoencoder response. This technique requires an additional network to try and guess the mass from the output of the first network, which is added complexity requiring significantly larger computation to train the network and leads to decrease in transparency. Also, it is not a general purpose method in that it is not applicable for  autoencoders based on Lorentz layers~\cite{Heimel:2018mkt}.

In this work, we present a robust anti-QCD tagger based on autoencoders. The robustness comes due to a series of novel preprocessing stages, and therefore can be easily generalized for many other cases such as an autoencoder based on the Lorentz layer or even supervised learning problems. Summarizing, the method proceeds as follows:
\begin{itemize}
	\item
	Given a jet, the algorithm first rescales and then boosts to a frame of reference where the mass and energy of the jet gets fixed to predetermined values (denoted here by $m_0$ and $E_0$ respectively).    
	\item
	The Histogram, representing the jet image is filled by using the dimensionless number $E_i/E_0$, where $E_i$ represents the energy of each constituent in this new frame. Also, in order to remove residual rotational symmetries present in the plane transverse to the jet-axis, the transverse plane is spanned using an orthonormal set of basis vectors, constructed using Gram-Schmidt procedure on either two hardest subjets, or two hardest constituents.     
\end{itemize}
We find that this method has a remarkable effect that it makes QCD-jets in different $\rho$ bins to appear similar as viewed by the network.
%leading to an order of magnitude improvement in the performance of the autoencoder.
As a result, an autoencoder trained on jet data in the lower $\rho$ bins (where data is surplus and pure) can be used in searches for anomalies in a set of jets at the higher $\rho$ bins (where data is scarce and may be contaminated).  Computationally, it is a far simpler and cheaper solution for removing the dependence of autoencoder loss on the jet mass compared to the proposal in Ref.~\cite{Heimel:2018mkt}. At the same time, this method yields significantly more robust performance than Ref.~\cite{Farina:2018fyg}. These highly desirable characteristics in this proposal make it highly portable and easily usable in many important techniques like mass bump hunt and side band analyses at all energy scales. 

We also take this opportunity to study the performance of fully-connected and convolution autoencoders as anomaly finders, where the jets are preprocessed using the methods proposed here. In order to benchmark, we treat top-jets and $W$-jets as anomalies. Here top-jets and $W$-jets refer to jets containing decay products of  hadronically decaying boosted top and $W$ respectively. We find that our anomaly finder, trained on only QCD-jets, yields performance comparable to Ref.~\cite{Farina:2018fyg}.
%{\bf{\color{c1}{
On the other hand, when combined with the mass of jets, it tags top as efficiently as  {\heptoptagger}, which uses a large amount of physics information specific to top decays. Additionally, when combined with Nsubjettiness, it substantially outperforms {\heptoptagger}.
%}}}
%In case of $W$-jets we find once again that the anomaly finder significantly outperforms {\tt N-Subjettiness} {\cite{Stewart:2010tn,Thaler:2011gf}}.
Summarizing, we propose a new anomaly finder that yields remarkable robustness without sacrificing performance.
Additionally, we also benchmark the effectiveness of the anomaly finder when jets consisting of all decay products of a boosted NP particle decaying to $W^+ W^-$, each of which in turn, decay hadronically, is treated as an anomalous jet. Being remarkably robust to giving high discovery potentials for finding anomalous jets, we find our method to generate impressive performances.   

The rest of this document is organized as follows: in \secref{sec:Preliminaries}, we describe in detail our method of preprocessing, and give a brief description of the autoencoder we use for this study; in \secref{sec:Results}, we present main results of this study after specifying all simulation details; and finally we conclude in  \secref{sec:Conclusion}. 

%------------------------------------------------------
\section{Our Proposed Analysis}  
\label{sec:Preliminaries}
%------------------------------------------------------

Our method can be broadly divided into two main parts. In the first part, we take jets from the event as input and produce images corresponding to these jets as described in {\subsecref{sec:procedure}}. The second part of our method involves the autoencoder. In {\subsecref{sec:autoencoder}}, we describe the general features of the autoencoder and explain how it can be trained.

%------------------------------------------------------
\subsection{From an infrared safe jet to the jet image }
\label{sec:procedure} 
%------------------------------------------------

Our method requires jets as inputs. A jet is basically a collection of four momenta of its constituents which are clustered according to any of the infrared-safe or IR-safe~\cite{Sterman:2008kj} sequential jet clustering algorithm like {\akjet}, {\cajet} and {\ktjet} or IR-safe cone based algorithms (like SISCone)~\cite{Blazey:2000qt,Salam:2009jx}. In this work, we use the notation $P_{J}^{\mu}$ ($\vec{P}_{J}$) for the four-momentum (three-momentum) of the jet, $p_{i}^{\mu}$ ($\vec{p}_{i}$) for the four-momentum (three-momentum) of the $i^{\text{th}}$  constituent in the jet. We use ``E-scheme"~\cite{Cacciari2012}, where the four-momenta of the jet is simply the vector sum of four-momenta of its constituents.
\begin{equation}
	P_J^{\mu} \  = \  \sum_{i=1}^{N_J} p_i^{\mu} \; ,  \quad 
	{m_J}^2  \ \equiv \  \left| P_J^{\mu} \right|^2 \ = \
	\left(P^0_J \right)^2 - \left(\vec{P_J} \right)^2  \; ,   \text{ and} \quad 
	E_J  \ \equiv \ P_J^0 \; , 
        \label{eqn:jetfoundation}
\end{equation}
where $N_J$ is the total number of constituents, $m_J$ the jet mass, and $E_J$ is the jet energy.

%------------------------------------
\subsubsection{Rescale the jet} 
\label{Sec:Rescale}
%------------------------------------------------

In the first step, all the jet constituent four-momenta are rescaled such that the new jet mass is given by a preassigned $m_{0}$:
\begin{equation}
 p_{i}^{\mu} \quad  \rightarrow \quad \  p_{i}^{\prime\mu}  \ = \  \frac{m_{0}}{m_{J}} \times p_{i}^{\mu} 
\qquad \Rightarrow \qquad
P_{J}^{\mu} \quad  \rightarrow \quad \ P_{J}^{\prime\mu}  \ = \  \frac{m_{0}}{m_{J}} P_{J}^{\mu} \; .
\label{eq:rescale}
\end{equation}
Note that the energy of the ``rescaled jet'' $P^{\prime\mu}_J$, is given by $E_J^{\prime} = E_J m_{0}/ m_{J}$.

%------------------------------------
\subsubsection{Lorentz boost the jet} 
\label{Sec:LorentzBoost}
%------------------------------------------------

The next step involves performing a Lorentz boost $\left( {\Lambda}^{\mu}_{\ \nu} \right)$ to a frame such that, in the new frame, all jets have a given energy $E_0$. In other words, the rescaled jets  $\left({P^{\prime \mu}_J}\right)$ are Lorentz transformed to boosted jets $\left({\mathbf{P}_J^{\mu}}\right)$ such that:
\begin{equation}
\begin{split}
   p_{i}^{\prime\mu} \quad  \rightarrow \quad \ \mathbf{p}_{i}^{\mu} 
      \ = \  \Lambda^{\mu}_{\ \nu} \: p_{i}^{\prime\nu}
    & \qquad \Rightarrow \qquad    
   P_{J}^{\prime\mu} \quad  \rightarrow \quad
        \mathbf{P}_{J}^{\mu} \ = \ 
        \Lambda^{\mu}_{\ \nu}\: P_{J}^{\prime\nu} \\
%        {\nonumber}\\
     &   \text{such that: } 
     \mathbf{E}_{J} \ \equiv \ \mathbf{P}_{J}^{0} \ = \ E_{0} \; .
        \label{eq:lorentzboostcondition}
\end{split}        
\end{equation}
Since this is a critical step in our method,  we present a simple derivation. We need to determine both the direction of the boost and the $\gamma$ factor required for the boost. Starting with the rescaled jet (with the given jet mass $m_{0}$ and energy $E^{\prime}_J$),  we can split the situation into 2 cases:
\begin{enumerate}
\item	${E_{J}^{\prime}>E_{0}}$: the boost is along the three-momentum direction of the jet and
\item 	$E_{J}^{\prime}<E_{0}$: the boost is opposite to the three-momentum direction of the jet
\end{enumerate}
In order to determine the ${\gamma}$ factor required for the boost consider the first case $E_{J}^{\prime} > E_{0}$, where impose the condition that the boosted energy must be $E_0$.  
\begin{equation}
    E_{0}  \ = \  \gamma \: E_{J}^{\prime}  -
    \beta \gamma  \: \left| \vec{P}_{J}^{\prime} \right| \qquad
    \text{where:} \quad \beta^{2} \ = \
    1 - \frac{1}{\gamma^{2}}  \; .
\end{equation}
This gives:
\begin{equation}
        \left(\frac{m_{0}^{2}}
            {\left|\vec{P}_{J}^{\prime}\right|^{2}}\right)
        \gamma^{2} +
        \left(-\frac{2E_{J}^{\prime}E_{0}}
            {\left|\vec{P}_{J}^{\prime}\right|^{2}}\right)
        \gamma +
        \left(\frac{{E_{0}}^{2}}
            {\left|\vec{P}_{J}^{\prime}\right|^{2}}+1\right)
        \ = \ 0 \; , \label{eq:condgamma}
\end{equation}
where, we have used $ m_{0}^{2} = \left(E^{\prime }_{J}\right)^2 - \left| \vec{P}_{J}^{\prime} \right|^2$. Solving this quadratic equation in Eq.~\eqref{eq:gamma} and picking the smaller solution (so that we do not boost past the rest frame of the jet), we get:
\begin{equation}
  \gamma \ = \ \frac{1}{m_0^2} \ \left(  E_{J}^{\prime} E_{0}  -
            P_{0} \left| \vec{P}_{J}^{\prime}\right| \right)
      \label{eq:gamma} \; , 
\end{equation}
where $P_{0}^{2} =  E_{0}^{2} - m_{0}^{2}$ is the magnitude of three-momentum of the jet in the boosted frame.  

In case  $E_{J}^{\prime} < E_{0}$, we get the exact same solution for $\gamma$ but with the boost direction being opposite to the jet momentum. After boosting, we now have a jet with four-momentum $\mathbf{P}_{J}^{\mu}$ such that $\mathbf{E}_{J} = E_{0}$ and $\left|\mathbf{P}_{J}^{\mu}\right|^{2}={m_{0}}^{2}$.

%------------------------------------
\subsubsection{Forming the images} 
\label{sec:imageformation} 
%------------------------------------------------

A jet image is basically just a 2 dimensional histogram of the energy (or $p_T$) of its constituents with respect to the plane transverse to the jet axis. We use the  Gram-Schmidt procedure to determine the optimal set of basis vectors in the transverse plane while also removing the residual rotation symmetry still present in the constituents.

Denoting the set of all constituent momenta in the boosted frame $ \left\{ \mathbf{p}_{1}^{\mu},\mathbf{p}_{2}^{\mu},\dots, \mathbf{p}_{N}^{\mu} \right\} $ such that  $ \mathbf{p}_{1}^{0}  \ge   \mathbf{p}_{2}^{0}  \ge \dots \ge \mathbf{p}_{N}^{0} $,  we construct the new set of basis vectors  $\left\{ \hat{e}_{1}, \hat{e}_{2}, \hat{e}_{3}\right\}$ obtained by the Gram-Schmidt method:   
\begin{align}
    \hat{e}_{1}  &  \ \equiv \    \frac{\vec{\mathbf{P}}_{J}}
        {\left|{\vec{\mathbf{P}}_{J}}\right|} \; , \label{eq:e1} \\
    \hat{e}_{2}  &  \ \equiv \ \frac{ \vec{\mathbf{p}}_{1} -
        \left( \vec{\mathbf{p}}_{1}\cdot\hat{e}_{1} \right) \hat{e}_{1} }
        {\left| \vec{\mathbf{p}}_{1} - \left( \vec{\mathbf{p}}_{1} \cdot \hat{e}_{1} \right) \hat{e}_{1} \right|}  \label{eq:e2} \; ,\\
    \hat{e}_{3} & \ \equiv \ \frac{ \vec{\mathbf{p}}_{2} - 
        \hat{e}_{1} \left( \vec{\mathbf{p}}_{2} \cdot \hat{e}_{1} \right) -
            \hat{e}_{2} \left( \vec{\mathbf{p}}_{2}\cdot\hat{e}_{2} \right) }
        { \left| \vec{\mathbf{p}}_{2} -  \hat{e}_{1} 
            \left( \vec{\mathbf{p}}_{2} \cdot\hat{e}_{1} \right) -
              \hat{e}_{2} \left(\vec{\mathbf{p}}_{2}\cdot\hat{e}_{2}\right)
                 \right|} \; .
        \label{eq:e3}
 \end{align}
Noted here that we approximate the constituent four-momenta to be massless. The first step of the Gram-Schmidt ensures that the jet axis is along the origin, the second and third step ensure the major and minor axis of the jet is along the $x$ and $y$ axis. We are aware that the above method might not be IR safe for some situations. Therefore, we also use three-moments of $2$ hardest subjets (after the original jet constituents are reclustered to find 3 exclusive $k_{T}$-subjets) to instead of $\mathbf{p}_{1}$ and $\mathbf{p}_{2}$ -- we get similar results.

We convert the full set of constituents to a jet image using the new basis. Note that the jet image is basically a 2D histogram $\left\{ I, \ J, \ F_{IJ}  \right\}$, where $I$ and $J$ are integers, and represent the pixel coordinates. Also, for a image of $N_H \times N_H$ pixels, $1\le I,J\le N_{H}$. On the other hand, $F_{IJ}$ is a real number representing intensity of the  $\left(I,J\right)$-th pixel. In this work we derive $F_{IJ}$ from all of the constituents according to, 
\begin{equation}
\begin{split}
F_{IJ} \  = \   \sum_{i=1}^{N_J} \ \frac{\mathbf{p}_{i}^{0}}{E_0}  \: \delta_{IX_i} \: \delta_{JY_i} & \; , 
\qquad \qquad\text{ where} \\
& X_i \ = \ \left[ \frac{1}{2} \left(N_{H} + 1\right) + \frac{1}{2} \left(N_{H}-1\right) \left( \frac{\hat{e}_{2} \cdot \vec{\mathbf{p}}_{i} } {\mathbf{p}_{i}^{0}} \right) \right]  \\ 
& Y_i \ = \ \ \left[\frac{1}{2} \left(N_{H} + 1\right)  + \frac{1}{2} \left(N_{H}-1\right)\left(\frac{\hat{e}_{3}\cdot\vec{\mathbf{p}}_{i}} {\mathbf{p}_{i}^{0}}\right) \right] \; .
\label{eq:ImageDefn}
\end{split} 
\end{equation}
In the above $\left[f\right]$ is the usual floor function representing the greatest integer less than or equal to $f$. Further, it is easy to verify that by construction
\begin{equation}
F_{IJ} \ \ge \ 0  \quad \forall I,J \qquad \text{and} \qquad
\sum_{I =1}^{N_{H}} \ \sum_{J=1}^{N_{H}} \ F_{IJ}  \  = \  1 \; .
\end{equation}
and hence can be interpreted as the discrete probability distribution function (PDF). The 2D histogram (image) is used as input to an autoencoder (it is flattened to a 1D vector for the case of a fully connected autoencoder).

%----------------------------------------------------------
\subsection{The Autoencoder} 
\label{sec:autoencoder}
%------------------------------------------------

\begin{figure}[H]
    \begin{center}
        \includegraphics[width=0.5\textwidth]{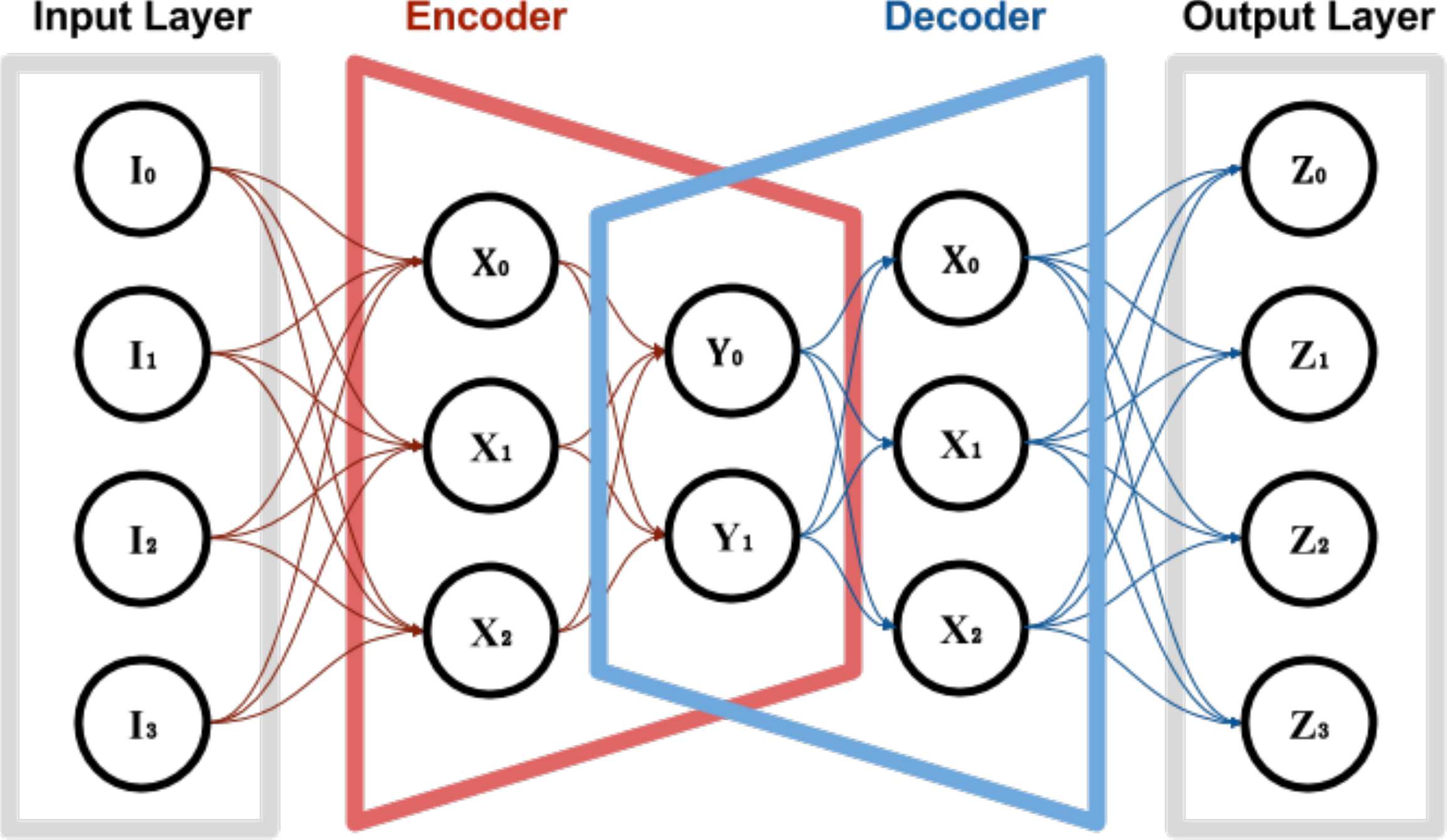}
        \caption{
            A schematic representation of an autoencoder (image was taken from Ref.~\cite{AutoEncoderImage}).
        }
        \label{fig:autoencoder}
    \end{center}
\end{figure}
An autoencoder consists of two segments of neural network. The first segment is called the encoder, which leads to a bottleneck with the number of nodes being less then the input. The second segment is a decoder which takes the output from the encoder bottleneck and tries to reconstruct the original input at the final output. In the simplest case, both encoder and decoder are fully connected (dense) neural networks, but can also be constructed by convolution layers. In this study we consider both fully connected and convolution autoencoders \cite{726791,Masci2011StackedCA}, note that the size of the bottleneck is an important characteristic of the autoencoder.

A schematic diagram representing a dense autoencoder is shown in \figref{fig:autoencoder}, it consists of multiple layers of fully connected neural networks with the output of one layer serving as input to the next layer:
\begin{equation}
O_{i}^{L} \ = \
\sigma\left(\sum_{j}W_{ij}^{L}I_{j}^{L}+B_{i}^{L}\right) \; ,
\end{equation}
where, $O_{i}^{L}$ and $I_{j}^{L}$  are the output and input vectors respectively, $W_{ij}^{L}$ is the weight matrix, and $B_{i}^{L}$ is the bias vector for the  neural network layer $L$. Note that, for internal layers $I_{j}^{L}$ is the output from the previous layer, or $I_{j}^{L} = O_{j}^{L-1}$. Finally, $\sigma$ is the activation function. We use ReLU~\cite{RIS_0} as the activation functions for all the intermediate layers and SoftMax for the final layer defined as 
\begin{equation}
    \text{ReLU}\left(x_{i}\right) \ = \
    \begin{cases}  
        0     & \mbox{if} \quad    x_{i} < 0 \\
        x_{i} & \mbox{otherwise}
        \end{cases}
    \; ,  \qquad \text{and} \qquad
   \text{SoftMax}\left(x_{i}\right) \ = \ \frac{e^{x_{i}}}{\sum_{j=1}^{N}e^{x_{j}}} \; .
    \label{eq:sigma_defn}
\end{equation}
Further note that the weight matrix $W_{ij}^{L}$ is of dimension $ M^{L} \times N^{L}$, where $M^{L}$ is the dimension of the output vector $O_{i}^{L}$ and $N^{L}$ is the dimension of the input vector $I_{j}^{L}$. For an example, in the first layer, $N^1 = {N_H}^2$. Given a network with $K$ layers, we additionally impose $M^K=N^1$, which ensures that the final output vector has the same dimensionality as the input and hence can be readily compared. Also, since we use SoftMax for the final layer, we ensure that the final output (namely, $O_{i}^{K}$), is also a discrete probability distribution function. This allows us to compare the final output to the initial image by a  simple $L_{2}$ norm function, which we use as the autoencoder loss function per input vector:
\begin{equation}
\epsilon \ \equiv \ 
\sqrt{	\sum_{i} \left(O_{i}^{K} - I_{i}^{1}\right)^{2}}
\label{eqn:loss}
\end{equation}
Similarly, a convolution autoencoder can be constructed by using convolution (Conv) and convolution transpose (ConvTrans) networks instead of dense networks in the encoder and decoder segments respectively. Numerically, the (discretized from the integral definition) convolution operation takes the form:
\begin{eqnarray*}
    O_{m,n} & = & \sum_{i,j}W_{i,j}I_{\left(m+i\right),\left(n+j\right)}
\end{eqnarray*}
where, the limits of the sum are determined based on the input and output dimensionality. Here, the indices of the weight matrix can also take negative values. Max pooling layers (MaxPool) are used to down-sample the image further in between the convolution layers, max pooling layer basically returns the maximum among all the input pixels.
Convolution transpose layers do the reverse of convolution layers in the sense that intensity of a single input pixel is added to all the output pixels.

We describe the details of the actual architecture of the convolutional autoencoders later in Appendix~\ref{sec:detailstraining}. Even though we focus on the simplest form of autoencoders, there have also been various improvements to the autoencoder brought in by variational decoding~\cite{2013arXiv1312.6114K} or adversarial training \cite{2017arXiv170104722M}. For further information see references and articles cited in Refs.~\cite{726791,Masci2011StackedCA,2011arXiv1106.1925P, Monk:2018zsb, Farina:2018fyg,Heimel:2018mkt,2013arXiv1312.6114K}.

%-------------------------------------------------
\subsection{Summary of the whole algorithm} 
%------------------------------------------------

Before concluding this section, we summarize the full algorithm in brief:
\begin{itemize}

\item Start with jets clustered using any of the standard IR-safe jet algorithms. We represent the jet momentum  by  $P_{J}^{\mu}$ and its constituent momenta by $p_{i}^{\mu}$ (see \eqnref{eqn:jetfoundation}).    

\item Rescale $p_{i}^{\mu}$ (and as a result $P_{J}^{\mu}$) such that the the new jet momentum $P_{J}^{\prime\mu}$ is characterized by a given mass $m_0$. We also denote the rescaled component momenta by $p_{i}^{\prime\mu}$ (see \eqnref{eq:rescale}).  

\item Lorentz boost either parallel or anti-parallel to the jet momenta as prescribed in \eqnref{eq:lorentzboostcondition} such that the energy of the jet in the new frame is given as a predetermined $E_0$. The boost factor required is given in \eqnref{eq:gamma}. We denote the jet momentum and the constituent momenta in the new frame by $\mathbf{P}_J^{\mu}$ and $\mathbf{p}_{i}^{\mu}$ respectively.

\item Find the optimal set of basis vectors  $\left\{ \hat{e}_{1}, \hat{e}_{2}, \hat{e}_{3} \right\}$ as described in Eqs.~(\ref{eq:e1} - \ref{eq:e3}). The vectors $\hat{e}_{2}$ and $ \hat{e}_{3}$ span the plane transverse
to the jet axis, which is given by $\hat{e}_{1}$. 

\item Use the constituent momenta $\mathbf{p}_{i}^{\mu}$, and the basis vectors $\hat{e}_{2}$ and $ \hat{e}_{3}$, to construct the final jet image (or rather a 2D histogram) of a given size $N_H \times N_H$ by using \eqnref{eq:ImageDefn}. 
    
\item
For the convolution autoencoder use the image (a 2D histogram) as input. On the other hand, for the dense autoencoder flatten the image to make  a one dimensional array and use it as the input.

\end{itemize}

%------------------------------------------------
\section{Results}
\label{sec:Results} 
%------------------------------------------------

Before proceeding, note that our tasks are divided into two main categories:
\begin{enumerate}
\item \label{task:1} The main task is to show that our procedure is robust. The robustness feature reflects the fact that the autoencoder loss function as defined in  \eqnref{eqn:loss} does not vary as jets from different $\rho$ bins are used to train the autoencoder.

\item \label{task:2} Also we benchmark the performance of our method as an anomaly finder. We carry this out in two parts:
    \begin{enumerate}[label=\textbf{\Alph*})]
        \item In the first part, we treat top-jets and $W$-jets as anomalies.
%        {\bf{\color{c1}{
		This also allows us to compare the performance of our method with tools based on physics information, as well as with previous proposals. 
%        }}}
            
        \item In the second part, we benchmark the performance of our method after introducing a NP particle decaying to di-$W$-jets giving rise to jets consisting of decay products of two $W$-bosons.  
    \end{enumerate}
\end{enumerate}
In the rest of this section we begin with the simulation details in \subsecref{sec:simulationdetails},  while results for task~\ref{task:1} and  task~\ref{task:2} are presented in \subsecref{sec:robustness} and \subsecref{sec:performance} respectively.

%------------------------------------------------
\subsection{Simulation Details} 
\label{sec:simulationdetails} 
%------------------------------------------------

In this study we generate all events using {\pythia} and use {\tt Delphes-3.4.1} \cite{deFavereau:2013fsa} to simulate detector environment. We compare results before and after adding various levels of complexity, such as turning on  multi parton interaction (MPI) or detector effects.  

We begin by clustering final state detectable particles from the output of {\pyth} into jets {\akjet} of $R=1.0$ using {\fastjet}. We only consider the hardest jet from an event in the central region $\left|{\eta}\right|<2.0$. On the other hand, we use  {\tt eflow} elements such as {\tt tracks}, {\tt neutrals} and {\tt photons} from the output of {\tt Delphes} to find jets, when we simulate detector effects.

The jet mass and transverse momentum we have considered in our study vary over a wide range as discussed later. Using these jets as input, jet images are formed using the method detailed in \subsecref{sec:procedure} with the parameters $m_0 = 1/2$ GeV, $E_0 = 1$ GeV and $N_H=40$. The full structure of the network can be read from the source code linked at the end of the document.
\begin{itemize}

    \item {\archA}:
    This is a large, dense autoencoder. Basically the input consists of a $40\times40$ pixel flattened image, each layer decreases both the row and column by 7 until the narrowest bottleneck of $5\times5$ is reached. The network expands in the same way again by 7 for both column and row, each layer has the ReLU activation, only the final layer has a softmax activation. We try to give a short hand notation for the structure of our network in \figref{fig:autoencoderstructure}. Most of the results in this article are presented for this architecture of autoencoder.

    \item {\archB}:
    This is a shallow dense autoencoder, again the  number of inputs is $40 \times 40$ pixel flattened image. The structure is shown in \figref{fig:autoencoderstructureshort}, as before, SoftMax activation is used in the final layer and ReLU activation is used for every other layer. We have mainly used this architecture of autoencoder to benchmark and verify that the results we observe are also reproducible using a much simpler autoencoder.

    \item {\archC}:
    This is a deep convolution autoencoder, the input image shape is $\left(1,40,40\right)$. Its architecture is described in Appendix.\ref{sec:archc}. We mainly use this architecture of autoencoder to benchmark the robustness of our method and find that, even though it performs slightly worse than the above two architectures, it is also more robust than both of them.

\end{itemize}

\begin{center}
	
\begin{minipage}{0.49\textwidth}
\begin{figure}[H]
    \begin{center}
        \includegraphics[width=1.0\textwidth]{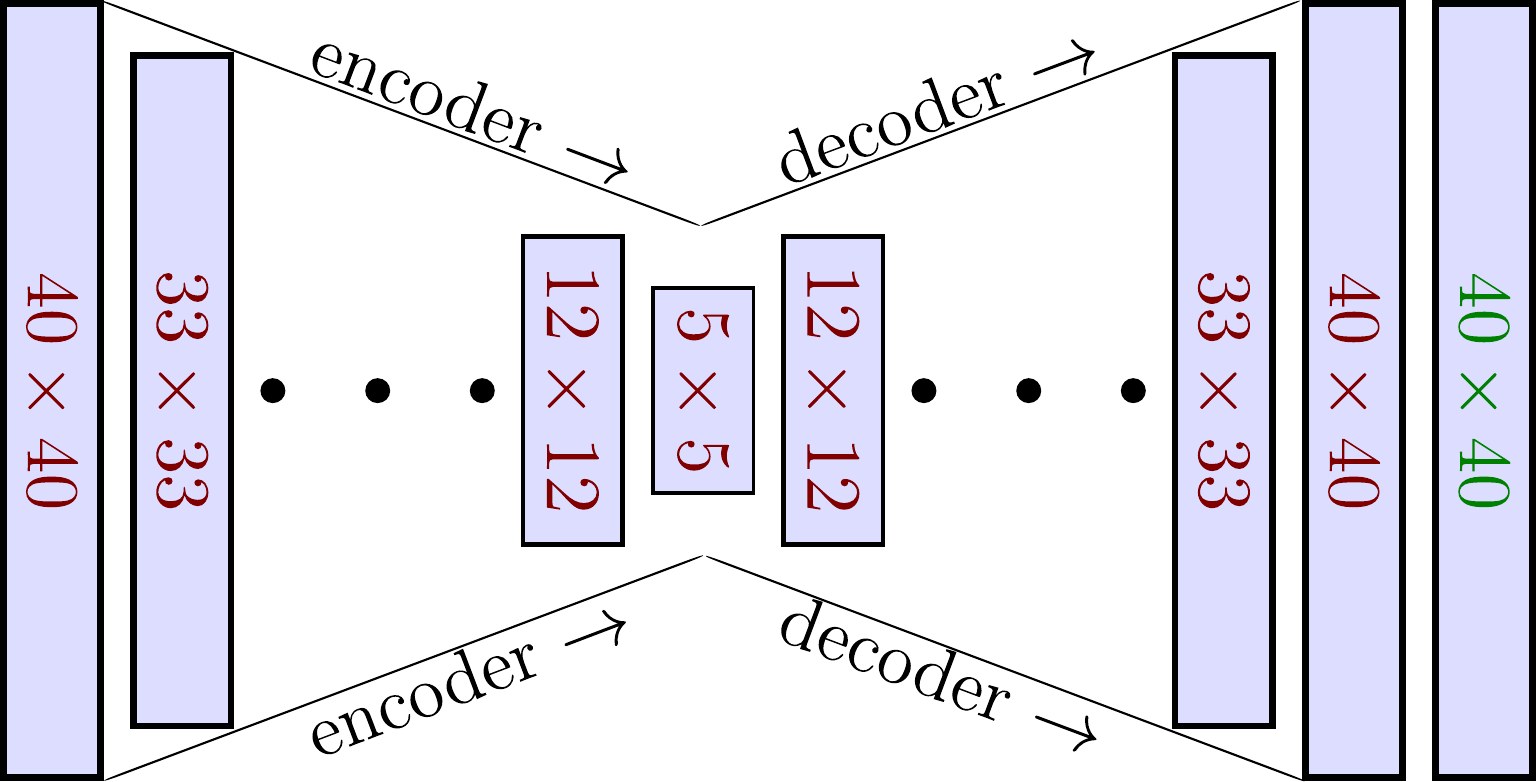}
        \caption{
            A schematic of the autoencoder structure used in our work ({\archA}). Layers with text in {\color{mred}{red}} have ReLU activation while the last layer with text in {\color{mgreen}{green}} has SoftMax activation (see \eqnref{eq:sigma_defn}).
        }
        \label{fig:autoencoderstructure}
    \end{center}
\end{figure}
\end{minipage}\hspace{0.01\textwidth}
\begin{minipage}{0.49\textwidth}
\begin{figure}[H]
    \begin{center}
        \includegraphics[width=0.8\textwidth]{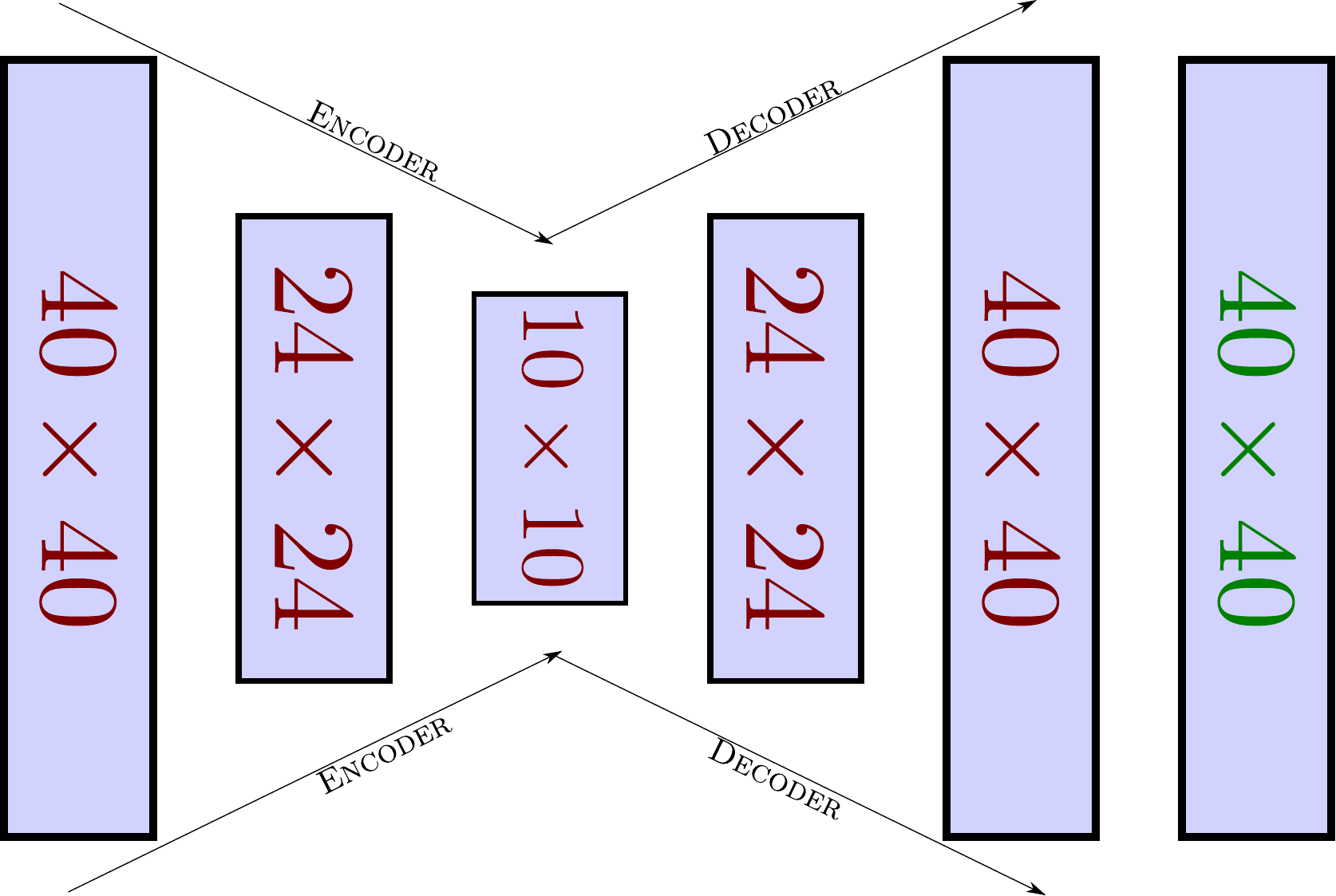}
        \caption{
            A schematic of the autoencoder structure used in our work ({\archB}). Layers with text in {\color{mred}{red}} have ReLU activation while the last layer with text in {\color{mgreen}{green}} has SoftMax activation (see \eqnref{eq:sigma_defn}).
        }
        \label{fig:autoencoderstructureshort}
    \end{center}
\end{figure}
\end{minipage}
\end{center}

Note that the jet image used as input to the autoencoder is normalized to 1 and each bin (pixel) is non-negative, \textit{i.e.}, it represents a discrete PDF. These properties might be lost when the image is propagated through the neural network. The softmax activation is used in the last layer to restore the normalization of energy clusters and also to make them non-negative. This forces the network to learn the main features of the image instead of just reproducing the magnitude and improves the overall training speed.

We arrive at these structures by checking through a few configurations mainly determined by the size of the  bottleneck and settled on these architectures, as increasing it further did not yield any significant improvement on reconstruction (loss function) of QCD in the first few training epochs. Also, for training this network, we use the {\tt Adam}~\cite{DBLP:journals/corr/KingmaB14} gradient descent with a batch size of around 100 to 200 and learning rate (step size of the gradient descent) of  around 0.001 to 0.0001.  

We have written and trained the network using \href{https://github.com/apache/incubator-mxnet} {the gluon API of Apache MXNet} \cite{2015arXiv151201274C} in python. This was chosen over \href{https://www.tensorflow.org/}{\tt TensorFlow} {\cite{DBLP:journals/corr/AbadiABBCCCDDDG16}} because of superior training performance on the CPU and easier imperative API, although we verified that the results are similar between this and {\href{https://www.tensorflow.org/}{\tt TensorFlow}} using both the {\href{http://tflearn.org/}{\tt TFLearn}} {\cite{DBLP:journals/corr/Tang16d}} andx \href{https://keras.io/}{\tt{Keras}} wrapper API.

%-------------------------------------------
\subsection{Robustness} 
\label{sec:robustness} 
%-------------------------------------------

The main goal of this section (as stated before) is to establish that the autoencoder loss function obtained using our method is rather insensitive to  the jet masses of the control sample. We demonstrate this is by training the  autoencoder network on QCD-jets from a particular region of the phase space and by testing the trained network on QCD-jets from a different region of the phase space, we divide our result into three parts:
\begin{enumerate}

\item  We train the autoencoder network on QCD-jets with  $800\gev <{p_T}_J < 900\gev$  produced at $\sqrt{s} = 13\tev$  collision energy  (with no cuts on the jet mass).  The data is then divided into bins depending on jet masses. The loss function (from the autoencoder trained in the previous step) is  evaluated on events from each bin and compared with each other as well as with the distribution of the whole sample  (\subsecref{sec:13TeV}).

\item We use QCD-jets produced at  $\sqrt{s}=100\tev$ with ${p_T}_J > 10\tev $, and then divide the full sample into bins depending on jet masses. We train an autoencoder network using QCD-jets from one of these mass bins and test with the others (\subsecref{sec:extremelimit}).

\item As extreme comparions, we train on jets from $\sqrt{s} = 13\tev$ collider, and test on data from $\sqrt{s}=100\tev$ collider, and vice versa. The results from this study is shown in Appendix~\ref{sec:crossvalidation}.

\end{enumerate}

%-----------------------------------------------------
\subsubsection{Robustness at $\sqrt{s}=13$ TeV}
\label{sec:13TeV} 
%-----------------------------------------------------

\begin{figure}[H]
\begin{center}
\begin{tabular}{cc}
\includegraphics[width=0.45\textwidth] {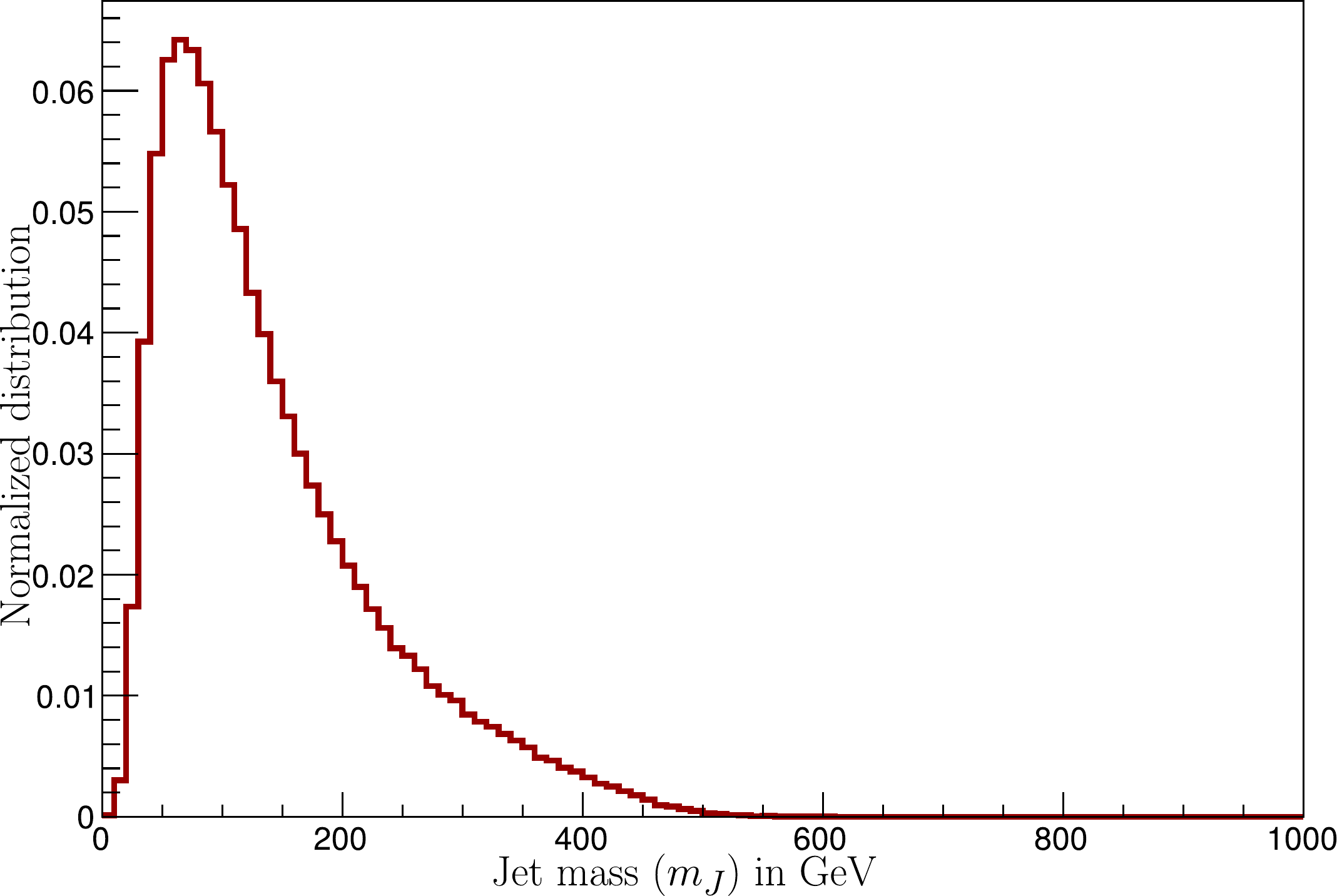}
            &
\includegraphics[width=0.45\textwidth] {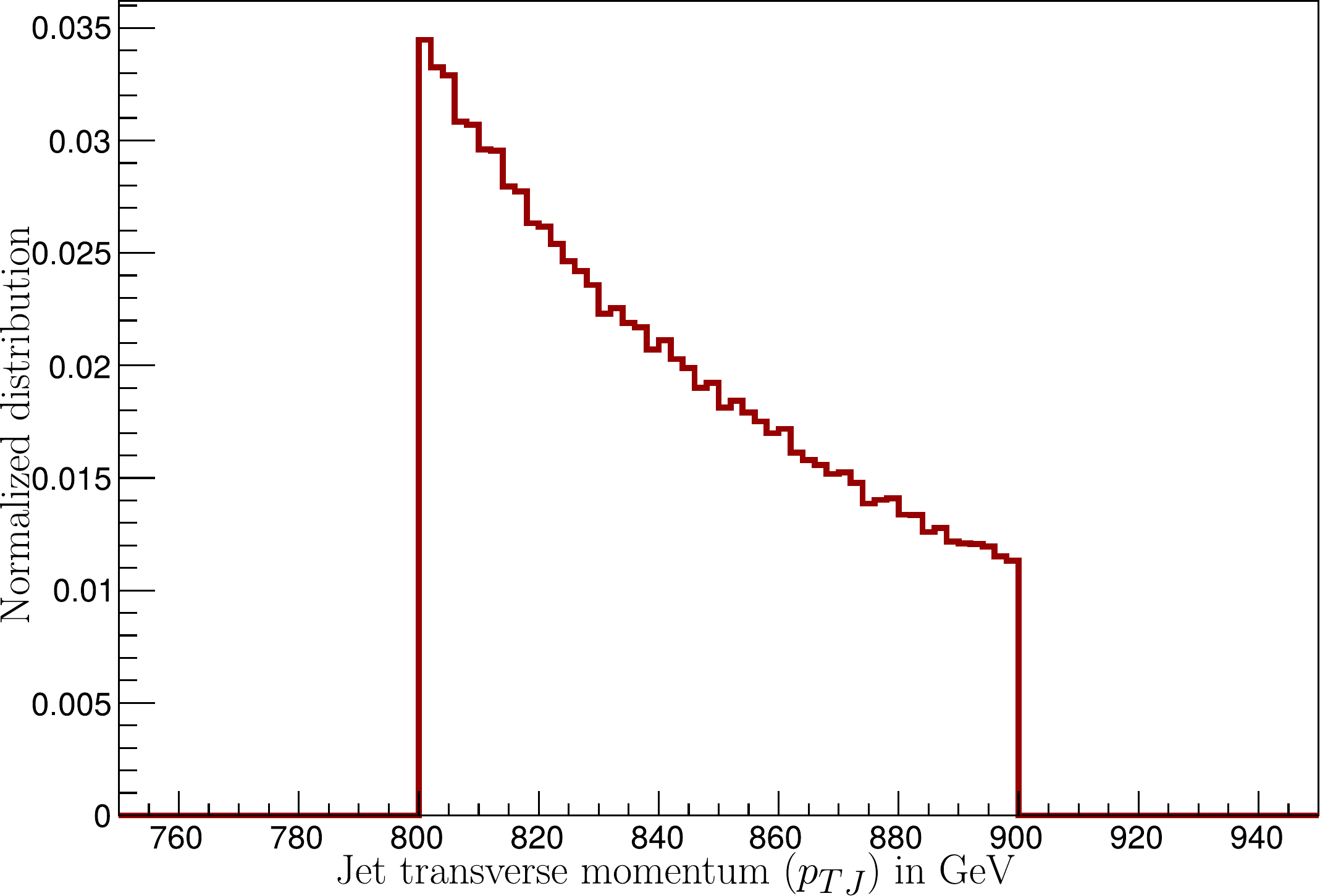}
\end{tabular}
\caption{ The mass (left) and transverse momentum (right) distribution of jets we use for our study. }
\label{fig:massandpt}
\end{center}    
\end{figure}

In this part, we study QCD-jets at $\sqrt{s}=13$ TeV. To be particular, we consider QCD di-jet events and study the jet hardest in $p_T$ from each event. We begin by studying jets with $800\gev< {p_T}_J < 900\gev$  generated without MPI. The jet mass and transverse momentum distribution of the sample is shown in \figref{fig:massandpt}.  The transverse momentum distribution is, as expected, lying in the window of 800 GeV to 900 GeV due to the imposed phase  space cuts and the jet mass gives the expected falling  distribution.

The jet images (as explained in \subsecref{sec:imageformation}) are basically a map from a  set of two dimensional integers (representing the pixels or the bins $\left(I ,J \right)$) to the set of real numbers between  0 and 1  (representing the intensities, $F_{IJ}$) given by \eqnref{eq:ImageDefn}.
\begin{figure}[H]
    \begin{center}
        \begin{tabular}{ccc}
\includegraphics[width=6.8cm,height=7.0cm]{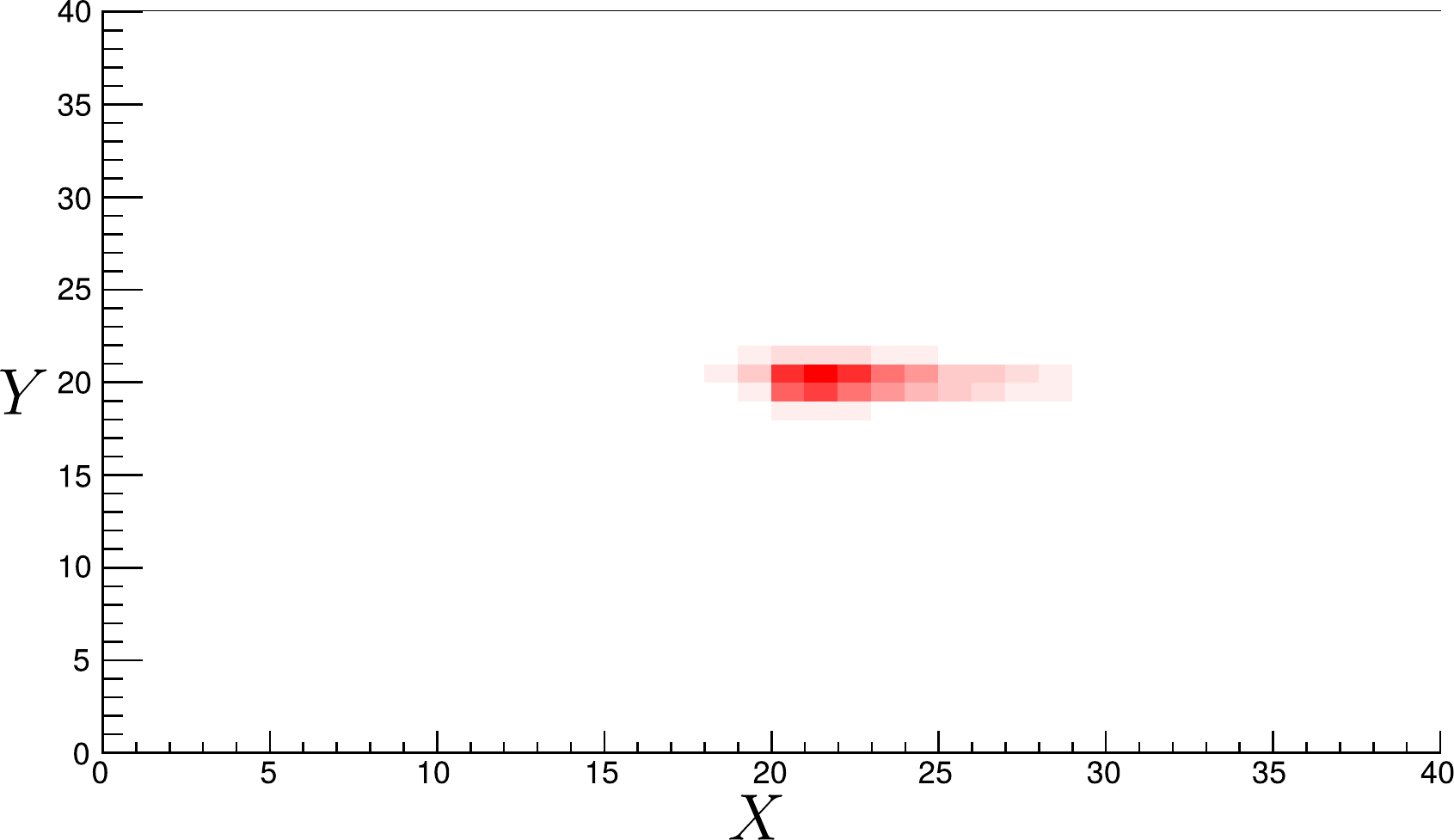}
            &
\includegraphics[width=0.71cm,height=7.0cm]{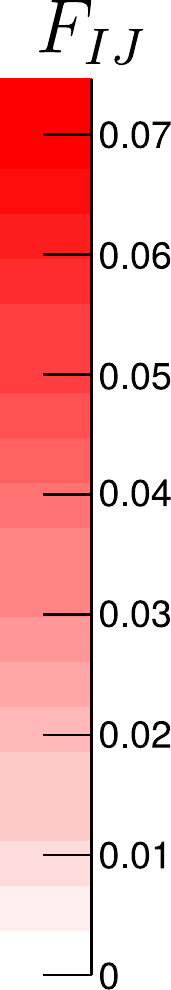}
            &
\includegraphics[width=6.8cm,height=7.0cm]{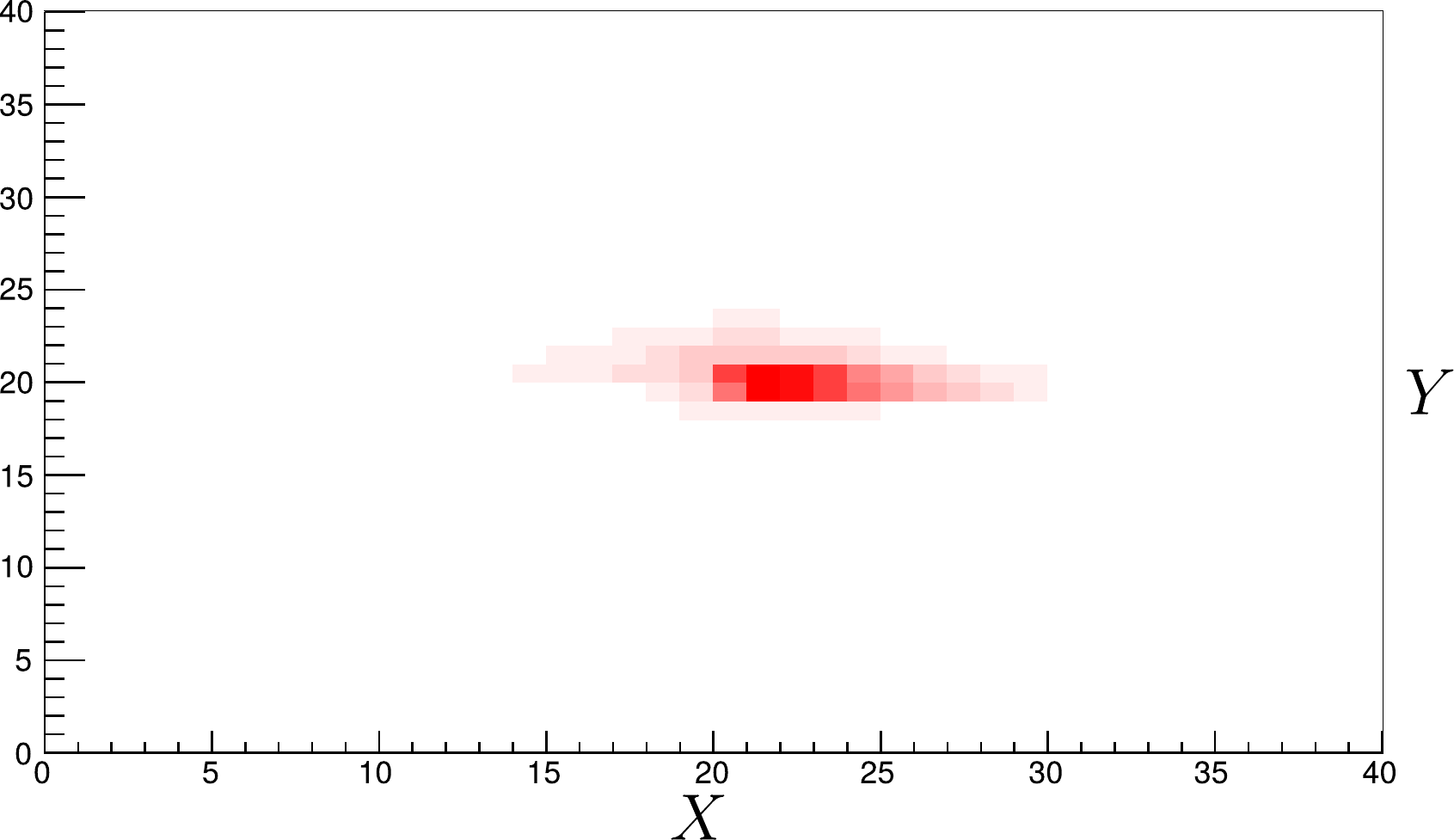}
        \end{tabular}
        \caption{
            The image of a QCD-jet (averaged over $\approx 500000$ jets) obtained after our pre-processing method  without{\bf(left)} and with{\bf(right)} detector effects (using {\tt Delphes}). The image is a 3D plot with the color density (the $z$ axis) representing intensity. To be specific, the axis represent:  $\left( x,y,z \right)\rightarrow \left(I ,J, \langle F_{IJ} \rangle \right)$, the axis labels follow the convention in \eqnref{eq:ImageDefn}.
        }
\label{fig:QCDIMAGE}
    \end{center}
\end{figure}
In \figref{fig:QCDIMAGE}, we present the averaged  (over $500000$ jets) jet image of QCD-jets  using our method. The color intensity of a bin (pixel) with coordinate $\left(I,J\right)$,  represents  the average $\langle F_{IJ} \rangle$  (the intensity of each pixel or bin is averaged across many jets).  The averaging procedure captures the broad features of the event type while smoothing over sharp fluctuations from individual events. 

The image faithfully reproduces the simple dipole emission structure of QCD-jets, where the hardest emission from the  parton initiating the jet decides the major (elongated) axis    of the jet  (which is brought along the $x$ axis of the image \figref{fig:QCDIMAGE} due to the Gram-Schmidt procedure in \eqnref{eq:e2})  while most subsequent emissions tend to lie along this axis (owing to the color connected \cite{Gallicchio:2010sw} structure of QCD interactions).  We also observe that the main impact of detector simulation  seems to smear the energy distribution of the jet  constituents.

%At this point, we divide our task into two parts:
%\begin{enumerate}[label=\textbf{(\Alph*)}]
%    \item \label{task:A}
%    Train the autoencoder network on the full set without any
%    cut on the jet mass and then test the variation of the
%    autoencoder loss function on various mass bins.
%    \item \label{task:B}
%    Train the autoencoder network on jets with
%     $m_J <  200\gev$ and test the trained network on
%    jets with $300\gev < m_J <  500\gev$.
%\end{enumerate}

\begin{figure}[h]
    \begin{center}
        \begin{tabular}{cc}
            \includegraphics[width=0.45\textwidth]{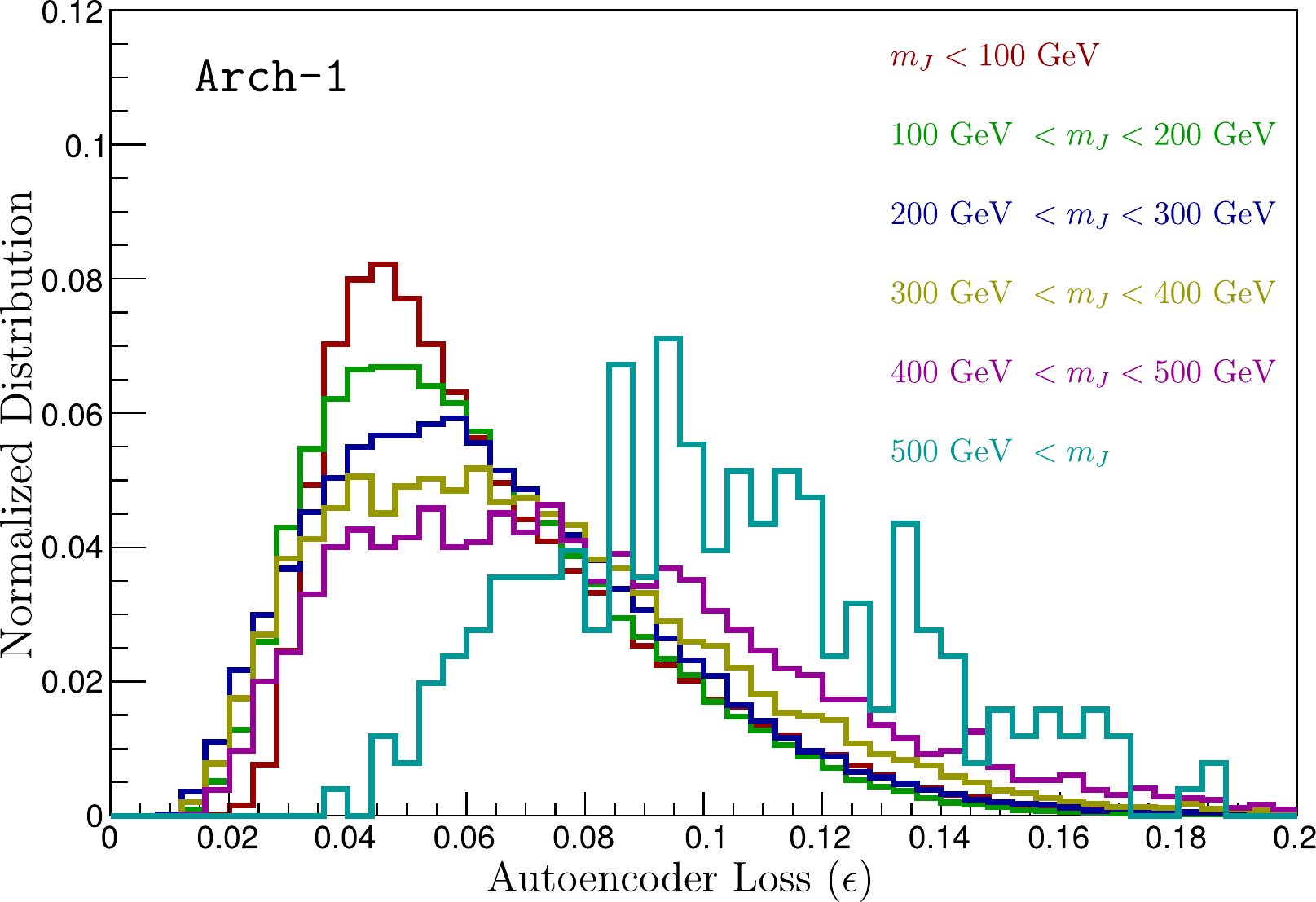}
            &
            \includegraphics[width=0.45\textwidth]{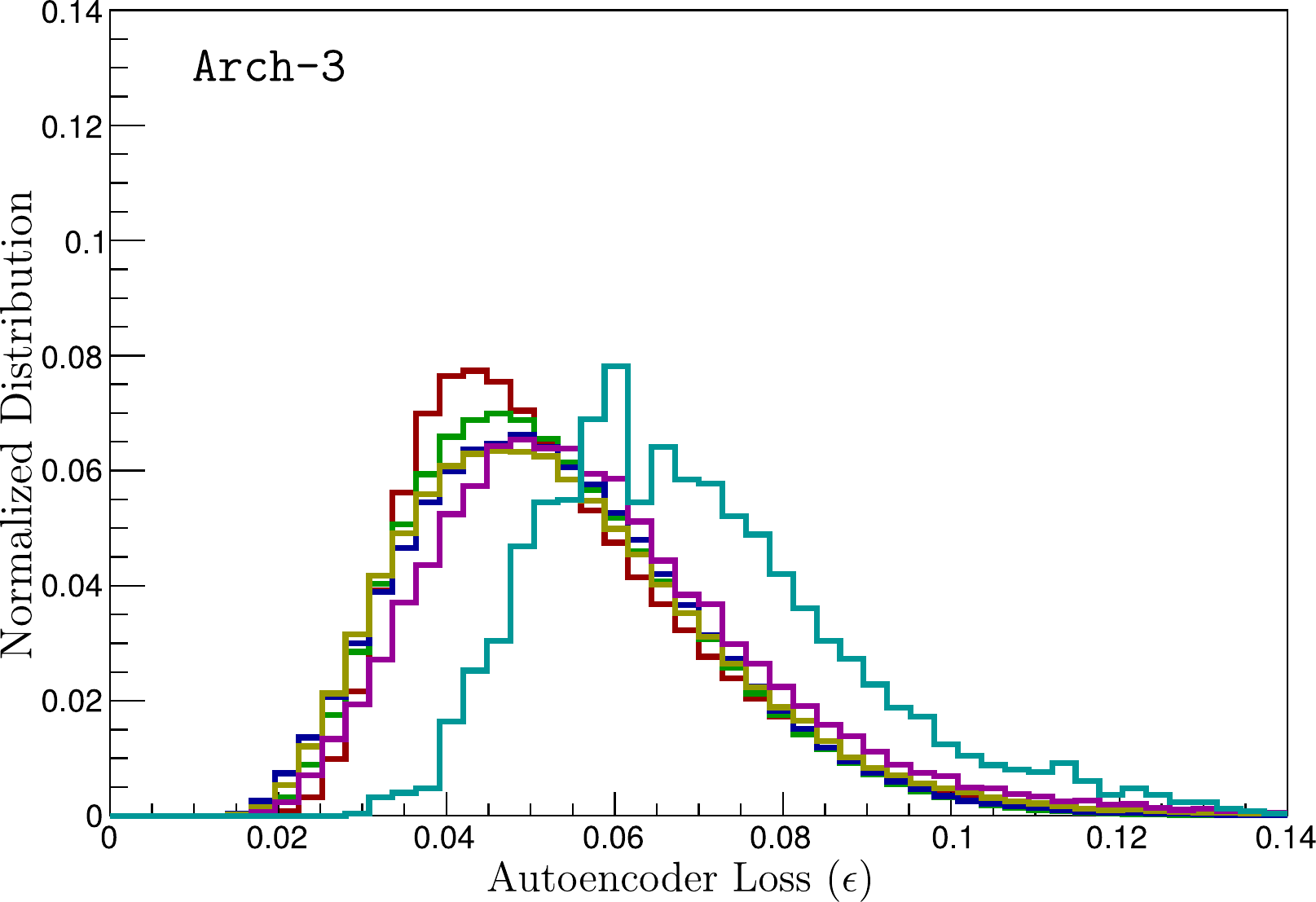}
            \\
            \includegraphics[width=0.45\textwidth]{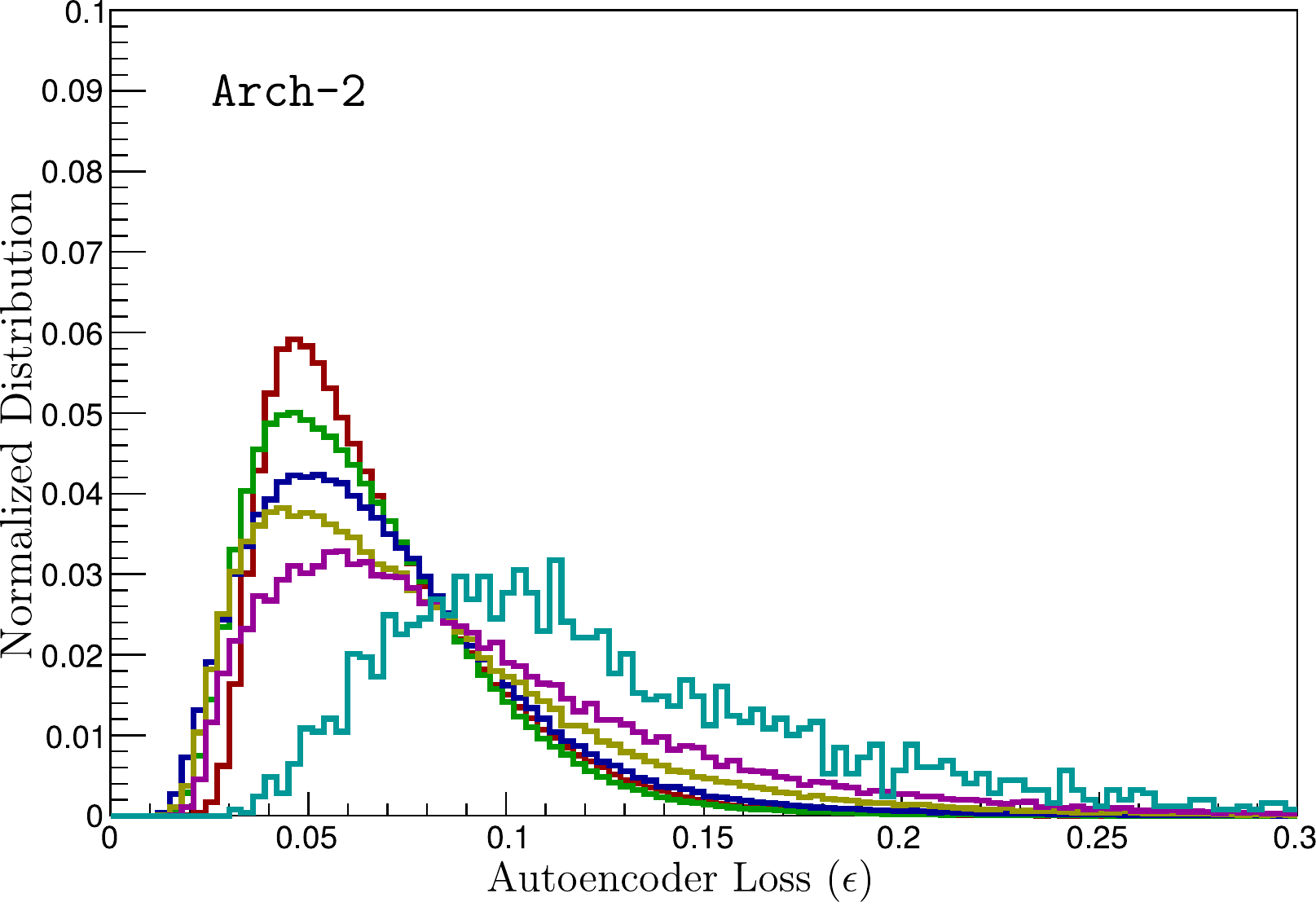}
            &
            \includegraphics[width=0.45\textwidth]{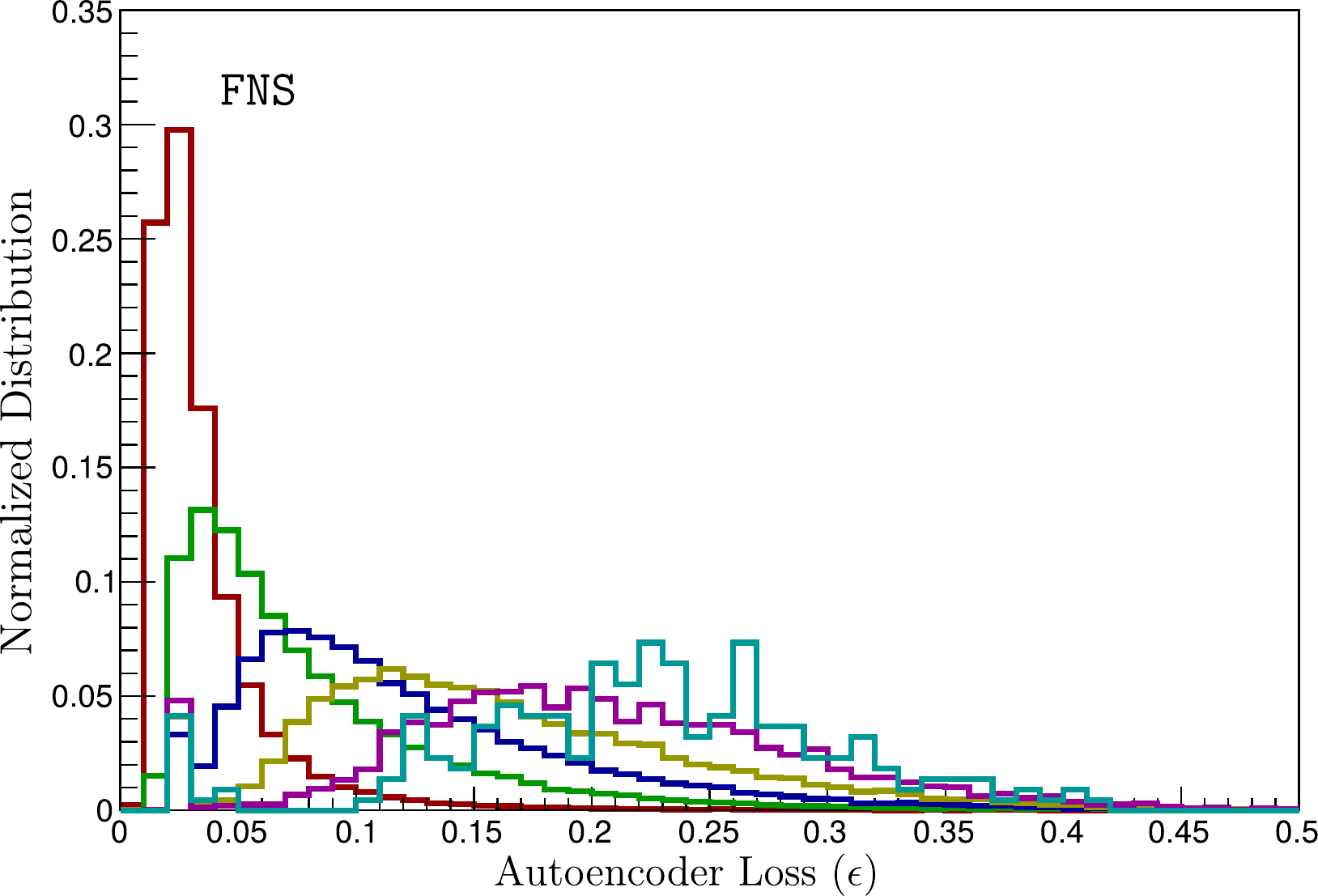}
            \\
        \end{tabular} 
        \caption{
            The effect of mass cuts on the autoencoder response (for QCD-jets) using our method for {\archA}  (top-left), {\archB} (bottom-left), {\archC} (top-right) and using the dense autoencoder from Ref.~\cite{Farina:2018fyg}  (lower-right).
        }
        \label{fig:masssculpt}
    \end{center}
\end{figure}

At this point, we train the autoencoder network on the full set of jets without any cut on the jet mass and then test the variation of the autoencoder loss function on various mass bins. We present the result of these studies  in \figref{fig:masssculpt}, where  we plot the autoencoder loss functions  for all the three architectures as jets are analyzed using various mass bins. We  compare these with responses of  dense autoencoder from Ref.~\cite{Farina:2018fyg} to calculate loss function (bottom-right plot of \figref{fig:masssculpt}).  Clearly, our method yields better robustness. The change in the peak position of the autoencoder loss  distribution for different jet masses using our method  (for {\archA}) is  within $10 \% - 15 \%$. On the other hand,  when we use the method from Ref.~\cite{Farina:2018fyg}, the peak position shifts by  factors of $4$ to $5$. Similar trends are also seen in the widths of the distribution which hardly changes for different mass bins when using our method while changing significantly  (by several factors) for the method proposed in Ref.~\cite{Farina:2018fyg}. Although, we have compared only the dense autoencoder from Ref.~\cite{Farina:2018fyg} here, the robustness of convolution autoencoders are found to be only slightly better (order $30 \%$ for $m_J>300$ GeV)~\cite{Farina:2018fyg}. However, as one can see from the lower-right plot of \figref{fig:masssculpt}, 20\% to 30\% improvement is not sufficient. Our simple method is significantly more robust. 

A comparison of the three architectures used in this work is warranted.  The convolutional autoencoder gives significantly more robust performance than any of the other autoencoders using dense networks. As noted before, this observation  was also mentioned in Ref.~\cite{Farina:2018fyg}. Note also that the shallower dense network employed in  {\archB} yields substantially more robust features than that of  Ref.~\cite{Farina:2018fyg}, even though it is worse than the deeper dense network in {\archA}. 

\begin{figure}[h]
	\begin{center}
		\includegraphics[width=\textwidth]{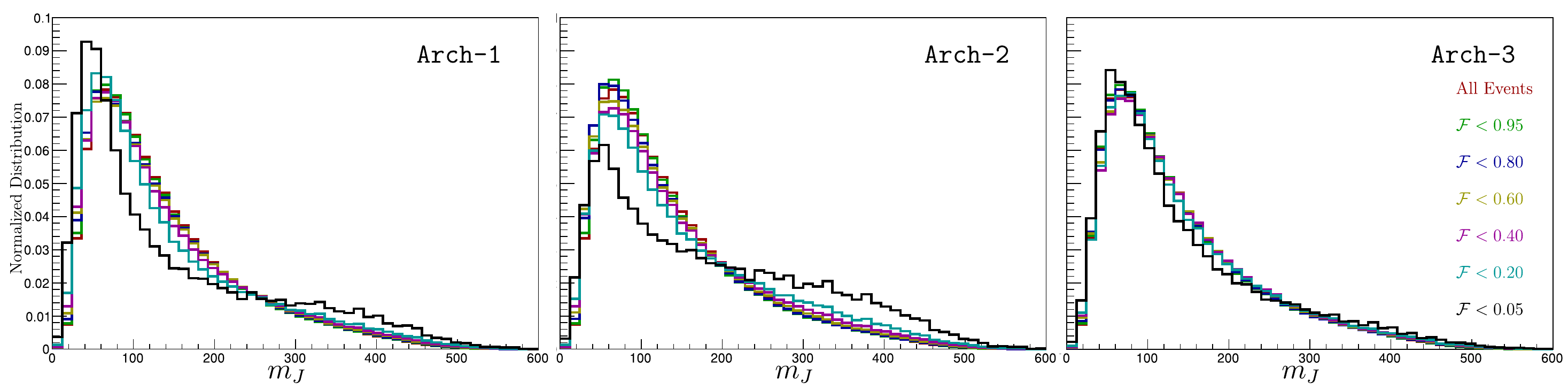}
		\caption{The distribution of jet mass in bins of autoencoder loss.}
		\label{fig:jet massinlossbins}
	\end{center}    
\end{figure}

This de-correlation of loss and jet mass is an extremely interesting phenomenon and deserves more attention. This feature can also be demonstrated in the distributions of jet masses where jets are binned in loss $\epsilon$. In \figref{fig:jet massinlossbins}, we show these jet mass distributions in bins of $\epsilon$.  To be specific, we follow the procedure as outlined in \cite{Heimel:2018mkt} and choose a cut on $\epsilon$ dynamically such that only a given fraction (say, $\mathcal{F}$) of QCD-jets pass this cut. The jet mass distribution for these  passed jets are then plotted, The procedure is repeated for different values of $\mathcal{F}$. We show the results of these studies  in  \figref{fig:jet massinlossbins}, where jet mass distributions remain robust for all the three architectures. Only for $\mathcal{F} < 0.05$ we start seeing deviations and the three architectures  show slightly different results.  Not surprisingly, the convolutional autoencoder ({\archC}) produces the best result in that even for  $\mathcal{F} < 0.05$ the jet mass distributions are virtually identical.  The $\mathcal{F} < 0.05$ plot of deeper dense network in  {\archA} shows some variation from other $\mathcal{F}$ plots, and the difference is more enhanced for the shallower dense network in {\archB}. 
It is tempting to compare the robustness presented in Ref.~\cite{Heimel:2018mkt}, where the authors use  additional adversarial training  to decorrelate QCD-jets from jet masses. Our method, relying largely on physics driven preprocessing gives as  robust a  performance as the proposal  in  Ref.~\cite{Heimel:2018mkt}.  We perform additional checks to pin down the exact reason behind the observed de-correlation effect between the autoencoder loss and the jet mass.  In order to quantify it, we use the usual   linear correlation coefficient. Given two variables $A$ and $B$, the  linear correlation coefficient, denoted by $\rho(A,B)$, is defined by the following equation 
%\begin{equation}
%\rho(A,B) \ \equiv \ \frac{E(AB) \ - \ E(A)E(B)}{E(A) \ E(B)}  \; , 
%\label{Eq_Correlation}
%\end{equation}
\begin{eqnarray}
\rho\left(A,B\right) & \equiv & \frac{\left\langle AB\right\rangle -\left\langle A\right\rangle \left\langle B\right\rangle }{\sqrt{\left\langle A^{2}\right\rangle -\left\langle A\right\rangle ^{2}}\sqrt{\left\langle B^{2}\right\rangle -\left\langle B\right\rangle ^{2}}}
\label{Eq_Correlation}
\end{eqnarray}
where $E(A)$, $E(B)$, and $E(AB)$  represents the expectation values of the variables $A$, $B$,  and $AB$ respectively. In order to demonstrate how the loss becomes de-correlated as we perform various stages during preprocessing of jets, we calculate $\rho(\epsilon, m_J)$ in four different studies: 
\begin{enumerate}
	\item Preprocessing without Lorentz boost + a fully connected network {\archA} for the autoencoder.
	\item Preprocessing without using  Gram-Schmidt axis + a fully connected network  {\archA}  for the autoencoder. Note that, unfortunately, due to the (arbitrary) rescaling and boosting steps, it no longer makes sense to use the usual $\eta$ and $\phi$ coordinates. Gram-Schmidt remains the only sensible way to pick the basis in the transverse plane (to the jet axis). Here we attempt to estimate the effect of using  Gram-Schmidt by using an  arbitrary basis vectors instead of the hardest constituent (or subjet) vectors.
	\item Full preprocessing + a fully connected network {\archA} for the autoencoder. 
	\item Full preprocessing + a convolutional autoencoder {\archC} for the autoencoder. 
\end{enumerate} 

\begin{table}[H]
    \begin{center}
    \begin{tabular}{|c|c|}
        \hline 
        CASE & Correlation between $\epsilon$ and $m_{J}$\tabularnewline
        \hline 
        1: No Lorentz Boost         + {\archA} & 75 \\
        2: Random Gram-Schmidt Axis + {\archA} & 32 \\
        3: Full pre-processing      + {\archA} & 7  \\
        4: Full pre-processing      + {\archC} & 2  \\
        \hline 
    \end{tabular}
		\caption{
            The linear correlation coefficients $\rho(\epsilon, m_J)$ (in $\%$) for the four cases mentioned in the text: preprocessing without Lorentz boost + {\archA} (first from left), preprocessing without using Gram-Schmidt axis + {\archA} (second), full preprocessing + {\archA} (third), and finally, full preprocessing + {\archC} (fourth).
        }
        \label{fig:correlation}
    \end{center}
\end{table}
%\begin{figure}[h]
%	\begin{center}
%		\includegraphics[width=\textwidth]{CorrelationMatrices_compare.pdf}
%		\caption{
%            The  linear correlation coefficients $\rho(\epsilon, m_J)$ (in $\%$)  for  the four cases mentioned in the text:  preprocessing without Lorentz boost + {\archA} (first from left),  preprocessing without using  Gram-Schmidt axis + {\archA} (second), full preprocessing + {\archA}   (third), and finally, full preprocessing  + {\archC} (fourth).
%        } 
%		\label{fig:correlation}
%	\end{center}
%\end{figure}
The results are shown In Table~\ref{fig:correlation}. It is clear that the largest impact of de-correlation comes from performing the Lorentz boost during preprocessing. Using {\archA} for the autoencoder, one improves $\rho(\epsilon, m_J)$  from $75\%$ to $7\%$ if the Lorentz boost step is performed (compare the first and third rows in Table~\ref{fig:correlation}). The effect of using Gram-Schmidt axes is rather mild. Using arbitrary axes, instead of Gram-Schmidt axes, deteriorates  $\rho(\epsilon, m_J)$  from $7\%$ to $32\%$ (compare the third and second rows in  Table~\ref{fig:correlation}). Finally, as seen in previous figures, we see modest improvement as we change from {\archA}  to {\archC} -- a  gain in  $\rho(\epsilon, m_J)$  from $7\%$ to $2\%$ is seen when we compare the last two rows in Table~\ref{fig:correlation}. For the full 2-dimensional distribution of autoencoder loss $\epsilon$ as a function of jet mass $m_J$ observed in these studies, see Appendix~\ref{sec:lossvsm}.

Note that throughout \subsecref{sec:13TeV} we have shown results after dividing the testing sample  in bins of jet masses, whereas the training sample had no such binning. In other words, the training dataset, in principle, contain information of jets of all masses. A rather more stringent test is to train using jets from a particular bin of jet masses, and test with jets from different bins. Operationally, it is more convenient to use jets produced at much higher center of mass energy collisions since, as shown in  \figref{fig:massandpt}, the sample we work with (at $\sqrt{s}=13$~TeV) is characterized by a rather shallow range in jet masses. We, therefore, leave this study for the next section where jets are produced at $\sqrt{s}=100$~TeV collisions. 

%----------------------------------------------------
\subsubsection{Robustness at $\sqrt{s}=100$ TeV} 
\label{sec:extremelimit} 
%----------------------------------------------------

\begin{figure}[h]
    \begin{center}
        \begin{tabular}{cc}
\includegraphics[width=0.48\textwidth]
{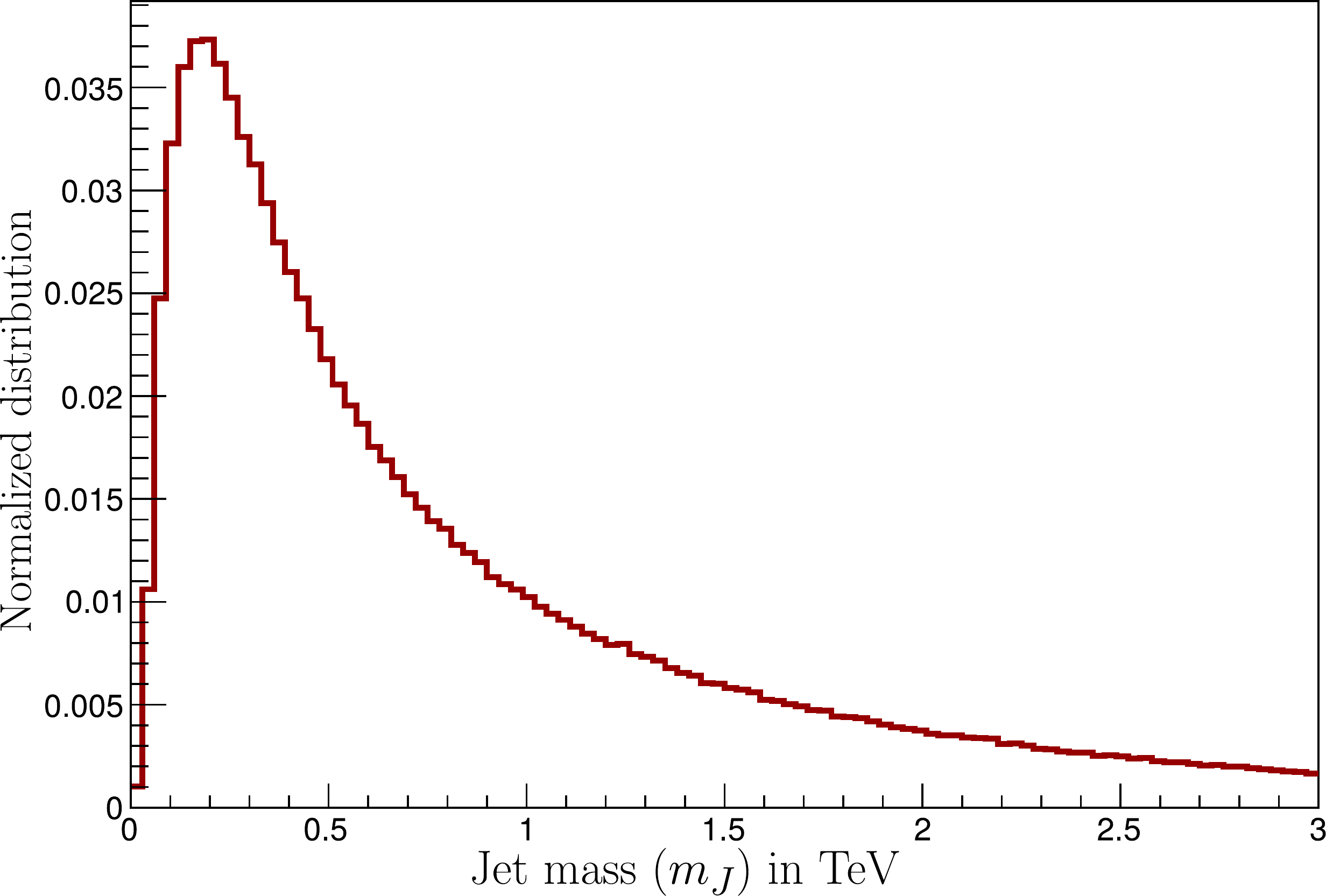}
            &
\includegraphics[width=0.48\textwidth]
{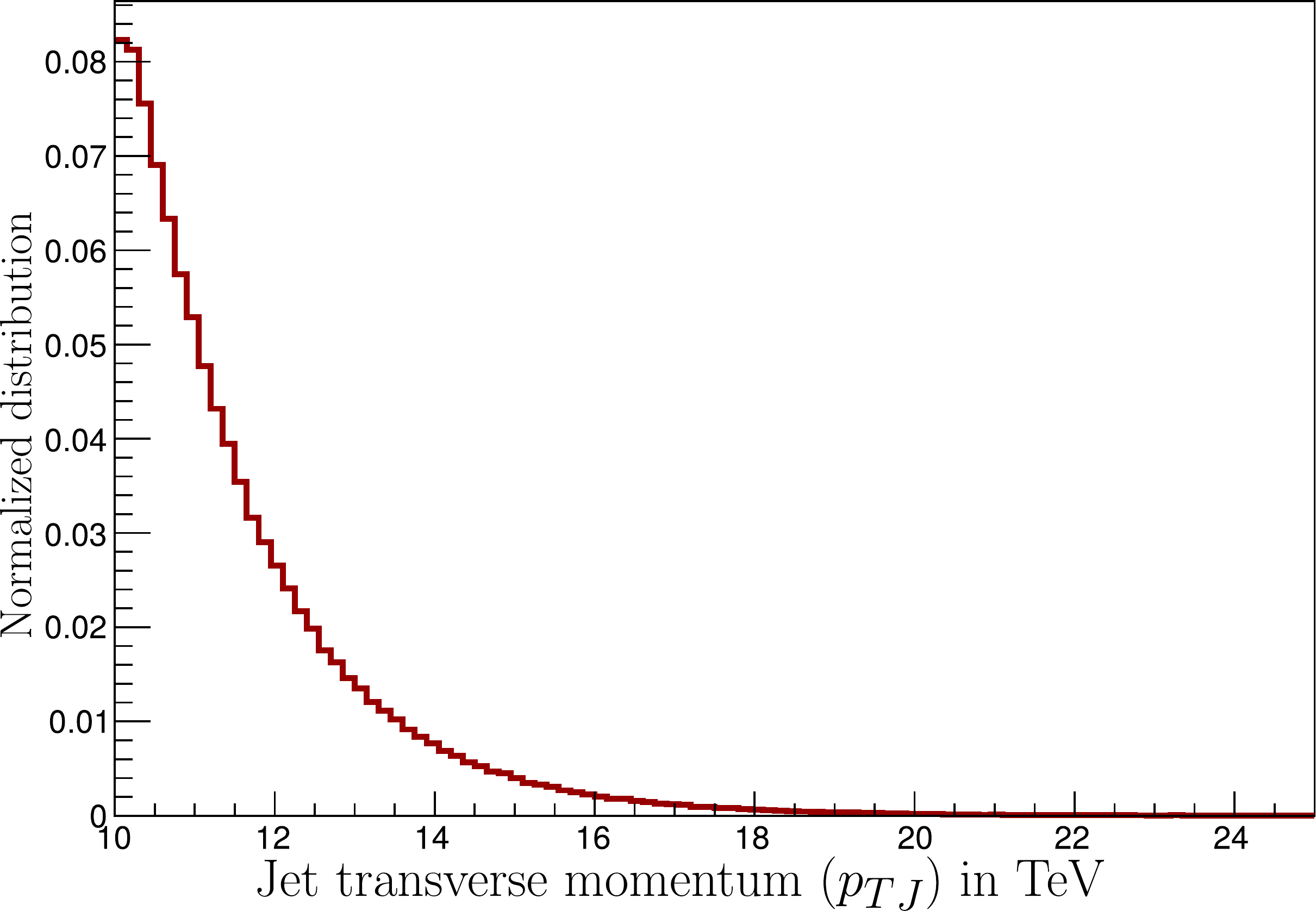}
        \end{tabular}
        \caption{
            The mass (left) and transverse momentum (right)
            distribution of jets we have used for our study
            at $\sqrt{s}=100$ TeV.
        }
        \label{fig:massandpt100tev}
    \end{center}    
\end{figure}

In this subsection we study the robustness with  QCD-jets of  high transverse momentum and jet mass. For this purpose, we consider jets with ${p_T}_J>10\tev$ from QCD-dijet events at a $\sqrt{s}=100\tev$ collider, generated without MPI. For these event we do not simulate any detector effects. The $m_J$ and ${p_T}_J$ distributions of these jets are  presented in \figref{fig:massandpt100tev}. Note the broad range of jet mass distribution of the dataset. It allows us to divide the training sample into a large number of bins of different jet masses while still keeping statistics under control. 

\begin{figure}[H]
    \begin{center} 
\includegraphics[width=0.6\textwidth]
{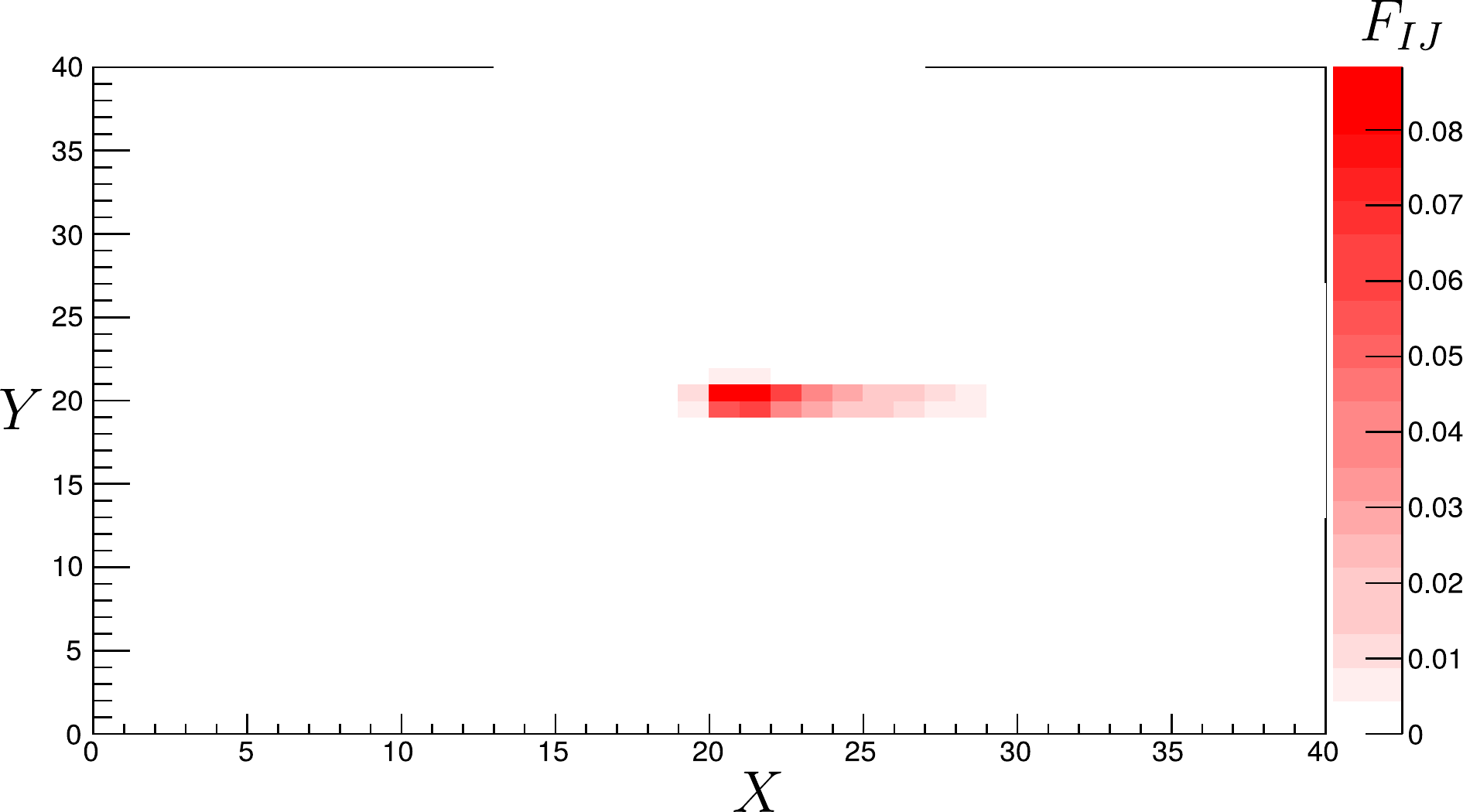}
\caption{
    The image of QCD-jets at $\sqrt{s}=100\tev$ obtained  using our method without MPI or detector effects (the axis labels follow the convention in \eqnref{eq:ImageDefn}).
} 
 \label{fig:highptimage}
    \end{center}
\end{figure}
Before beginning, we show the average image of these jets obtained after our preprocessing steps in \figref{fig:highptimage}. Clearly, the image is almost identical to the image in \figref{fig:QCDIMAGE} obtained for QCD-jets from  $\sqrt{s}=13\tev$ collisions further fortifying our claims. Note that the most energetic part of the image (as measured using the scales on the $x$ and $y$ axis) is identical between \figref{fig:QCDIMAGE} and \figref{fig:highptimage}. This demonstrates that our method always resolves the jet at the same length scales irrespective of the jet mass or momentum (boost).

\begin{figure}[t]
	\begin{center}
        \includegraphics[width=1.0\textwidth]{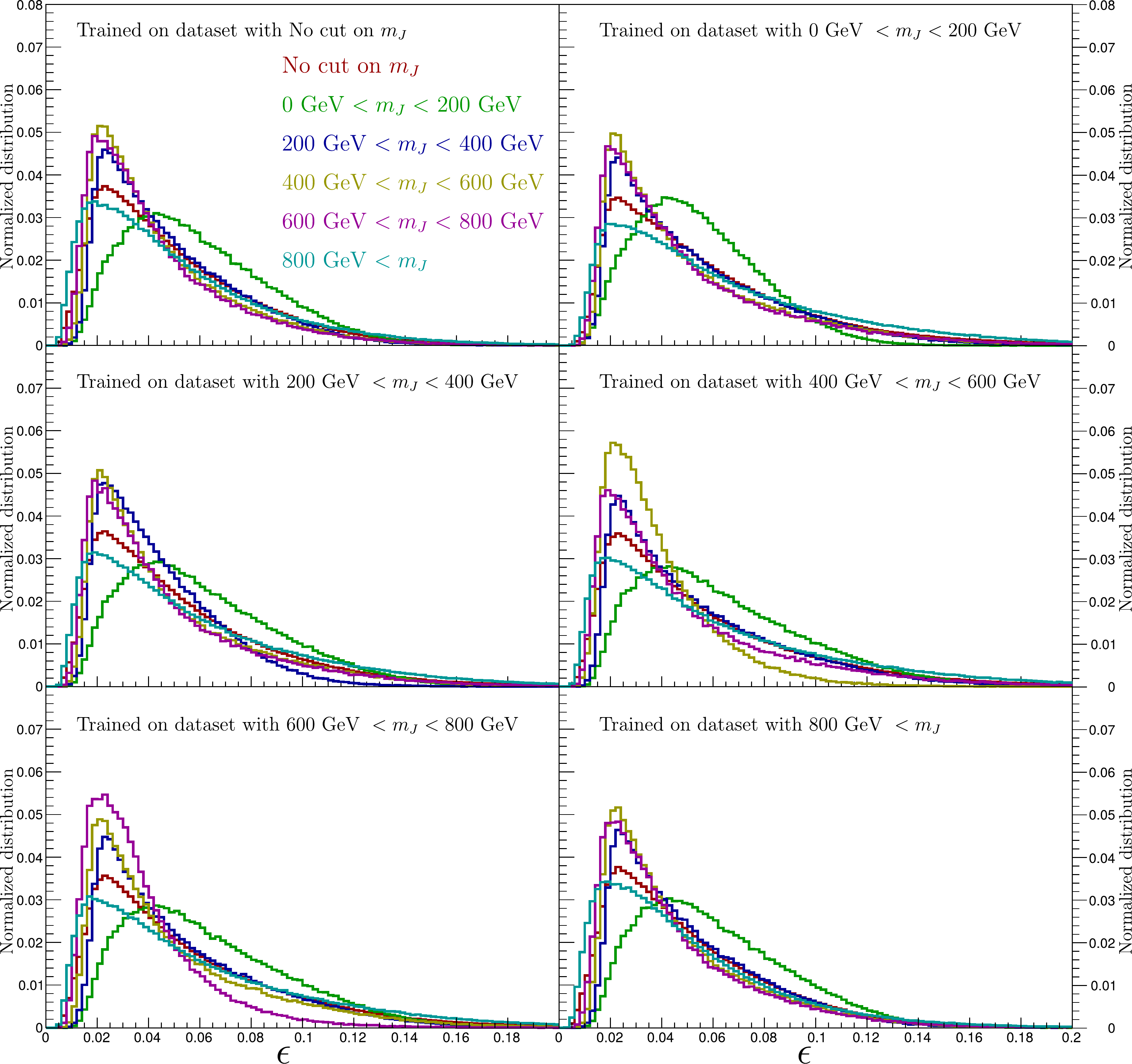}
        \caption{
            The (normalized) distribution of autoencoder loss function for QCD 	jets (${p_T}_J>10\tev$) in various mass bins, the autoencoder with jets in a particular mass bin (indicated in each of the figures) and tested on all the other mass bins. The colors of the histogram represent the mass bins as labeled.
        }
		\label{fig:highptkoreanmedicine}
	\end{center}
\end{figure}

Note that the purpose of this section is to check the robustness of our methodology, where we train jets from a given bin of jet masses while we test it with jets from different bins. We begin by dividing the full set of jets into subcategories according to:
\begin{multicols}{2}
    \begin{enumerate}
        \item
        No cut on $m_J$ (all jets are considered)
        \item
        $m_{J}<200\text{ GeV}$
        \item
        $200\text{ GeV }<m_{J}<400\text{ GeV}$
        \item
        $400\text{ GeV }<m_{J}<600\text{ GeV}$
        \item
        $600\text{ GeV }<m_{J}<800\text{ GeV}$
        \item
        $800\text{ GeV }<m_{J}$
    \end{enumerate}
\end{multicols}
In all of these cases, we form images using our method and analyze further using the autoencoder in various permutations. In order to check for robustness, we use the following  strategy:
\begin{enumerate}[label=\textbf{\Alph*})]
\item Train an autoencoder on jets from each of the mass bins mentioned above.
\item Evaluate the performance of these networks on jets from each of the mass bins listed above.
\end{enumerate}
We only use the dense network in {\archA} for the autoencoder. The comparative studies between different architectures as well as with/without various stages of preprocessing is already presented in \subsecref{sec:13TeV} and we do not repeat these here. We show the results from this study  in \figref{fig:highptkoreanmedicine}, where we plot distributions of loss-functions for jets (in various bins of jet masses). All of these figures again strongly support the claim that our method mostly eliminates the dependence of jet mass on the autoencoder loss function. As can be observed from the distributions  \figref{fig:highptkoreanmedicine},  most of the peak  position lie within $\approx 20\%$ of each other and the distributions largely overlap. 

At this point, we do not show any more plots (such as distributions of jet masses in bins of loss) to demonstrate the robustness feature in our proposal. As shown in \subsecref{sec:13TeV}, robustness in the distribution of $\epsilon$ for various $m_J$ bins immediately implies robustness in the distribution of $m_J$  for various $\epsilon$ bins.  

The lesson from all these different studies in  \subsecref{sec:13TeV} and  \subsecref{sec:extremelimit} is rather straight-forward:  our (physics-driven) proposal to preprocess jets before it is subjected to an autoencoder, manages to de-correlate jet masses from jets successfully, yielding  a robust methodology that can look for jets with high masses as anomalies while only being trained on  jets of low masses.

%-----------------------------------------------------
\subsection{Performance} 
\label{sec:performance} 
%-----------------------------------------------------

\begin{figure}[h]
	\begin{center}
		\includegraphics[width=0.99\textwidth]{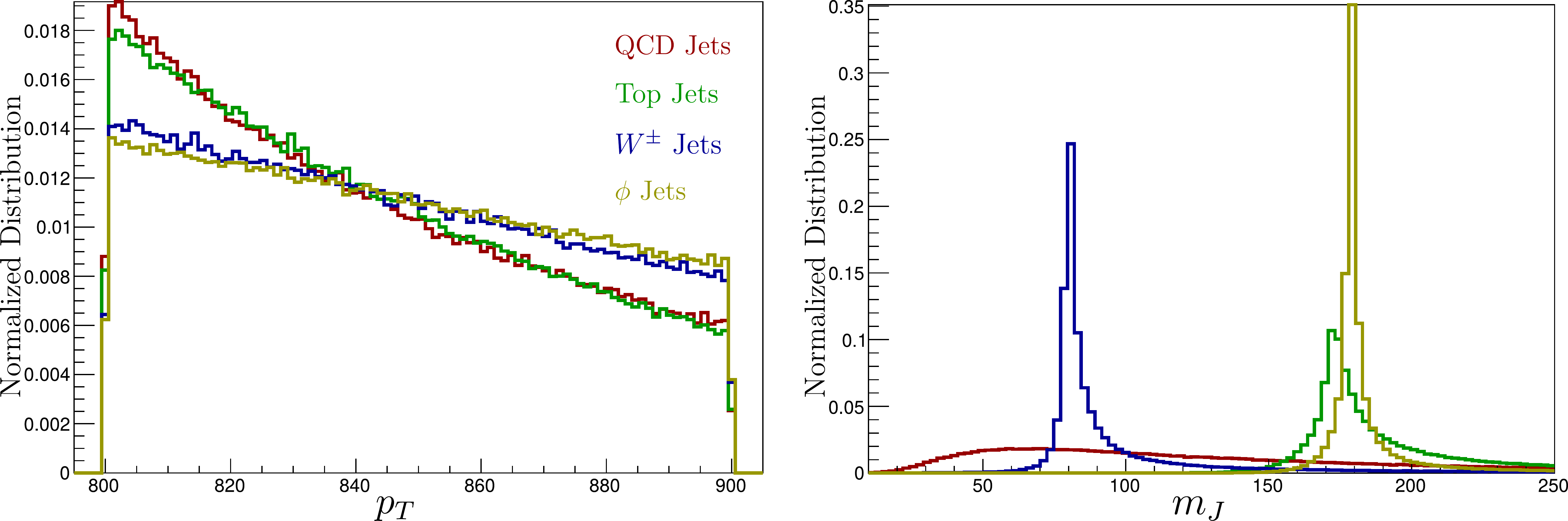}
		\caption{
            The $p_T$ and jet mass distributions for jets in different datasets.
        }
		\label{fig:jetphasespace}
	\end{center}
\end{figure}

The task in this subsection is to examine and benchmark the performance of our method  as an anomaly finder for various types of standard jets. As stated before, the autoencoder loss function can be used as a measure to determine how different a jet ``looks like'' as compared to the QCD-jets on which the autoencoder was trained on. Apart from considering top-jets and $W$-jets as anomalies, we also examine the performance of our method when these anomalous jets are  due to new physics. In particular, we study jets containing decay products of two hadronically decaying $W$-particles (namely, di-$W$ jet). The $p_T$ and jet mass distributions of jets in these datasets are shown in \figref{fig:jetphasespace}. As shown in   \figref{fig:jetphasespace}, all jets in this discussion are characterized by a $p_T$ in the range $800\gev \leq p_T \leq 900\gev$. The jet mass distributions for   top jets, $W$-jets, and di-$W$-jets  are peaked nicely near top-mass, $W$-mass, and the di-$W$ resonance (we take it to be $180\gev$).  

\begin{figure}[h]
    \begin{center}
\includegraphics[width=0.6\textwidth]
{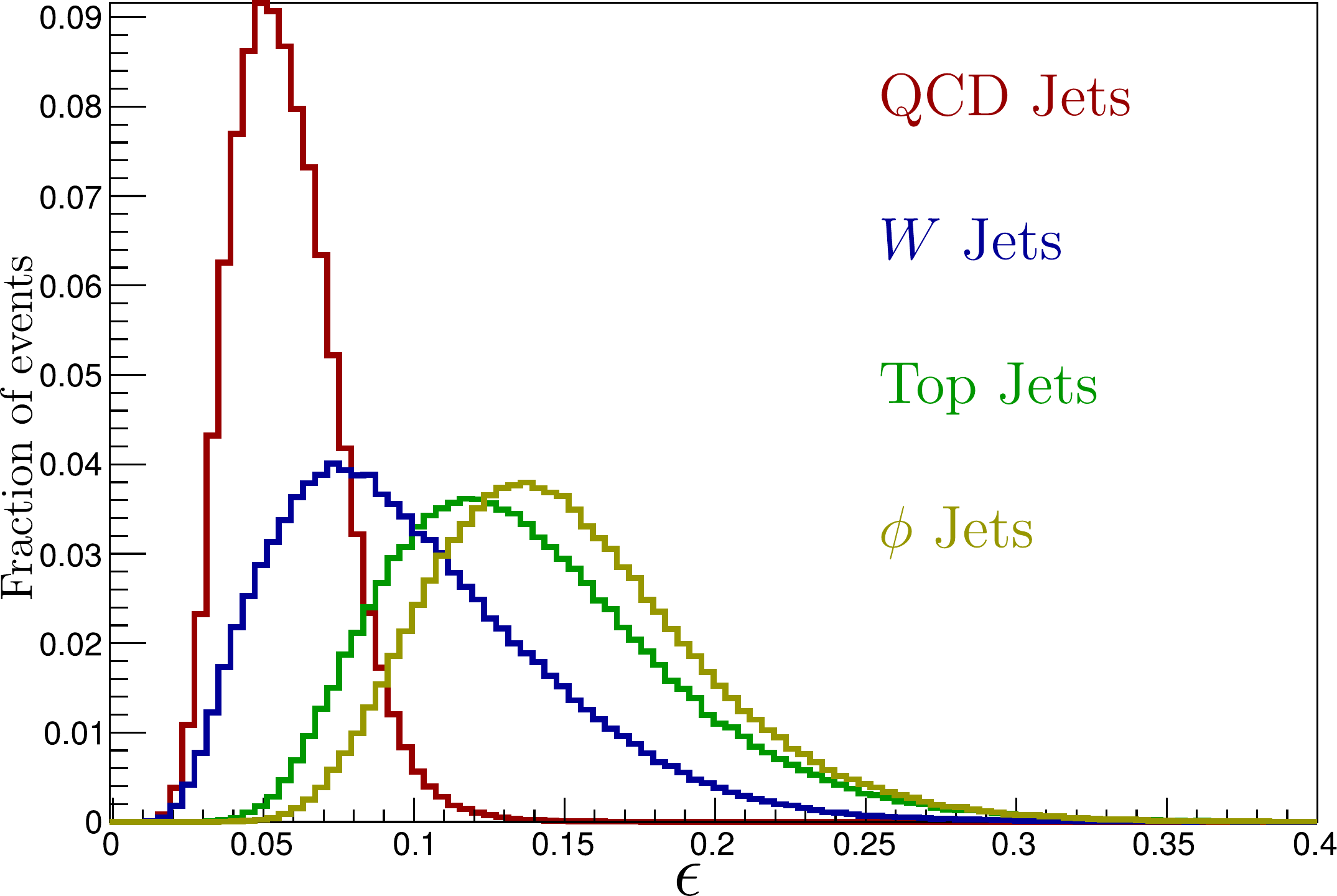}
        \caption{
            The autoencoder loss function
            ($\epsilon$, \eqnref{eqn:loss}) obtained from the
            autoencoder using our method of pre-processing for
            the case with no detector effects or multi parton
            interaction.
        }
        \label{fig:Compare_Error_5}
    \end{center}
\end{figure}

We first train the autoencoder using  QCD-jets.  Note that once the training on QCD-jets are done, the trained autoencoder acts as a  so-called ``black-box" which, given any jet simply associates  with it a  loss $\epsilon$.  
The loss function for these three different jet types (we use an autoencoder with the dense network in {\archA}) is presented in \figref{fig:Compare_Error_5}. We do not use MPI or detector effects in order to produce these plots, the purpose of which are simply to demonstrate that the loss $\epsilon$, by itself, can identify jets of any of these three types  as  anomalous jets.   It is clearly seen that the autoencoder loss increases as more complex structure is added to the jet. QCD-jets (on which the autoencoder was trained on and also have the simplest 1-prong dipole emission substructure) have, on average, the smallest loss function. $W$-jets, on the other hand,  have a 2-prong structure which is slightly more  ``complex'' than the 1-prong QCD-jet structure, and therefore have larger loss functions on average. Top jets come next in the hierarchy as they mostly have a 3-prong structure (with an onshell $W$) making it more ``complex'' than a $W$-jet. Finally, we show jets originating due to boosted $\phi \rightarrow W^+ W^-$. This gives rise to jets with 4-prong structure (containing 2 on-shell $W$ resonances in the same jet). Not surprisingly, we find these jets with the highest loss functions. 
In order to quantify the comparison between the performance of various methods, we compare the receiver operating  characteristic (ROC) curves. In our study, these curves are presented with the convention  that the $x$ axis represent the signal selection efficiency   and the $y$ axis represent the reciprocal of the background  fake rate (in log scale). Therefore, the algorithm producing curves closer towards the top right region of the graph would mean a better performing algorithm. It is also possible to compare two algorithms based on the  area under the ROC curves of the two algorithms, larger area  would generally mean better performance. 
    
Next, we examine each of the cases in greater detail, benchmarking our method against some standard techniques  which are well established and also against some recent  techniques found in literature.

%--------------------------------------------------
\subsubsection{Benchmark with top jets} 
\label{sec:topperformance} 
%--------------------------------------------------

The problem of discriminating jets originating from boosted tops against jets from QCD is extremely well studied theoretically~\cite{Dasgupta:2018emf, Dasgupta:2013ihk, Dasgupta:2013via, Marzani:2019hun}, numerically~\cite{Plehn:2011tg, Kasieczka:2015jma, Plehn:2010st, Kaplan:2008ie, Plehn:2009rk, Thaler:2011gf} and experimentally~\cite{CMS:2014fya,Aaboud:2018psm}.
Let us reiterate that the purpose of the subsection is \emph{not} to promote the loss of an autoencoder, which is trained entirely on QCD jets, to be an efficient top-tagger.
Since the autoencoder loss has no information about top jets, it is expected to under perform  when compared to any decent top-tagger, which uses information specific to top decays.
By treating top-jets as anomalies in this exercise, we rather hope to measure qualitatively the efficiency of this proposal if indeed we encounter anomalous jets (for an example, a jet consisting of decay products of a new beyond the standard model particle).
Indeed, earlier work on anomaly detection~\cite{Farina:2018fyg,Heimel:2018mkt} also used top jets as anomalies in order to benchmark.
In this subsection we take an identical approach to establish the performance of the dense and convolution autoencoder loss functions and compare it with the one in Ref.~\cite{Farina:2018fyg}, while also showing the performance of {\heptoptagger} in parallel as it is a well understood technique.

Top jets are obtained from $t \bar{t}$ events as detailed in  \subsecref{sec:simulationdetails}. We present the results for three main cases:
\begin{enumerate}[label=\textbf{(\Alph*)},ref=\textbf{(\Alph*)}]
    \item
    Top jets with $800\gev < {p_T}_J <  900\gev$ simulated without MPI or detector effects. \label{case:toppsA}

    \item
    Top jets with $800\gev <{p_T}_J<900\gev$  simulated without MPI but with detector effects. \label{case:toppsB}
    
    \item
    Top jets with $ {p_T}_J > 400\gev$ simulated with MPI and  detector effects. \label{case:toppsC}
\end{enumerate}
While generating these events in cases.~\ref{case:toppsA} and \ref{case:toppsB}, we impose an additional  merge requirement between the jet and the partonic top along with its decay products  $\left({b, q, {\bar{q}}^{\prime}}\right)$ by demanding that they lie within the jet at {\pyth} level. 
\begin{equation}
\begin{split}
{\Delta}R{\left({p_{J}^{\mu},p_{\text{top}}^{\mu}}\right)}
\  <  \ 0.6 &
\qquad \qquad 
{\Delta}R{\left({p_{J}^{\mu},p_{q}^{\mu}}\right)}
\ < \ 0.6 \\
{\Delta}R{\left({p_{J}^{\mu},p_{b}^{\mu}}\right)}  \ <  \ 0.6 &
\qquad \qquad 
{\Delta}R{\left({p_{J}^{\mu},p_{\bar{q}^{\prime}}^{\mu}}\right)}
\ < \ 0.6  \; ,
\end{split}
\end{equation}
Which ensures a consistent comparison with Ref.~\cite{Farina:2018fyg}. The merge requirement is not imposed for {case.~\ref{case:toppsC}}.

\begin{figure}[t]
    \begin{center}
        \begin{tabular}{ccc}
\includegraphics[width=0.43\textwidth]{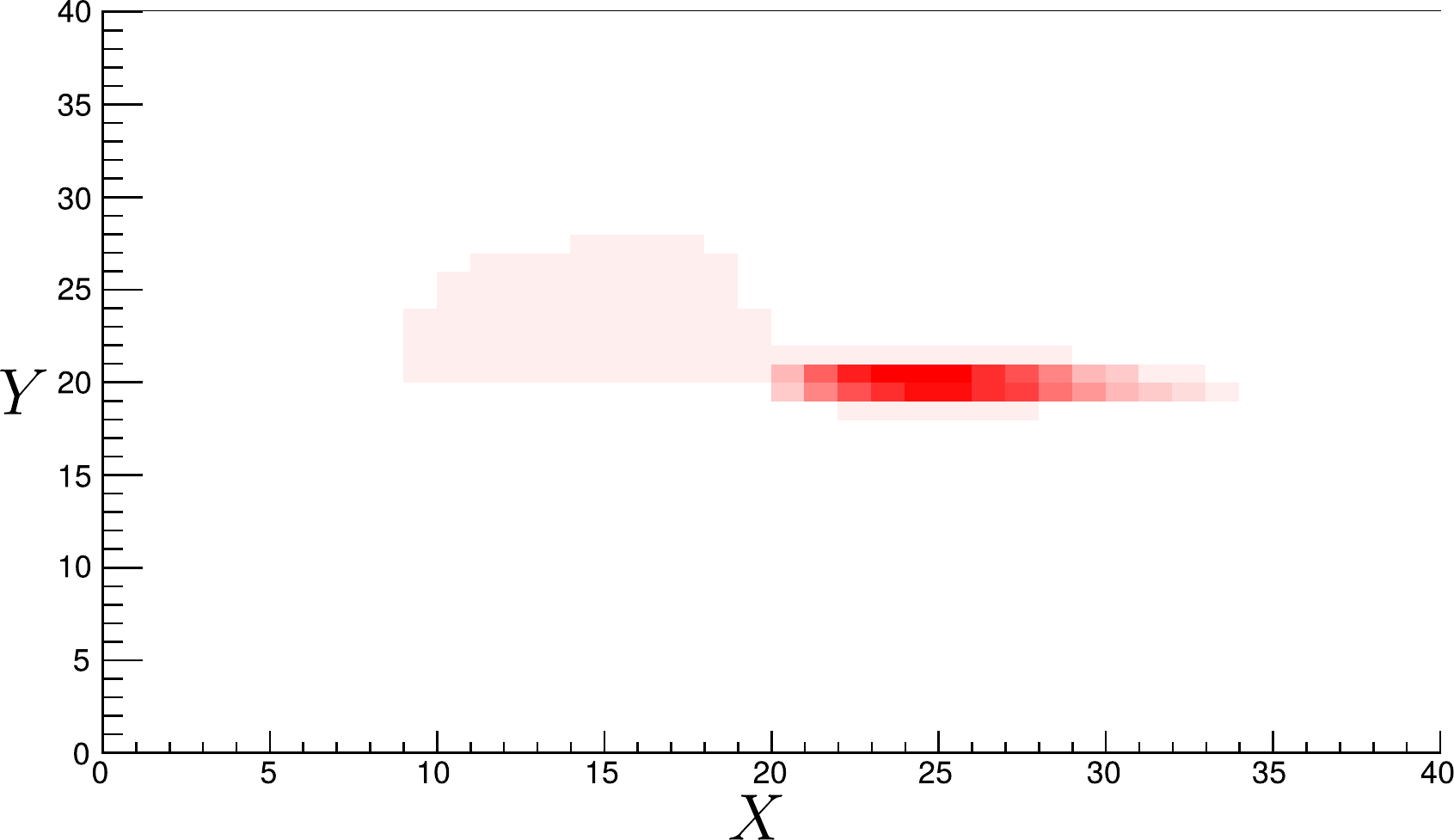}
            &
\includegraphics[width=0.053\textwidth]
{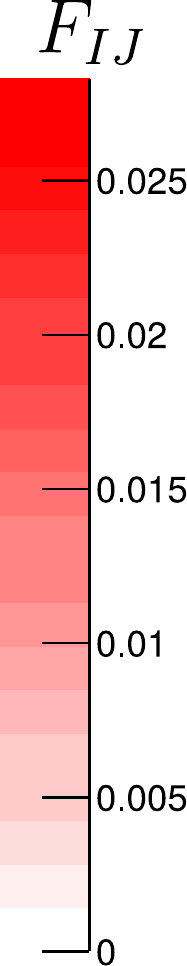}
            &
\includegraphics[width=0.43\textwidth]{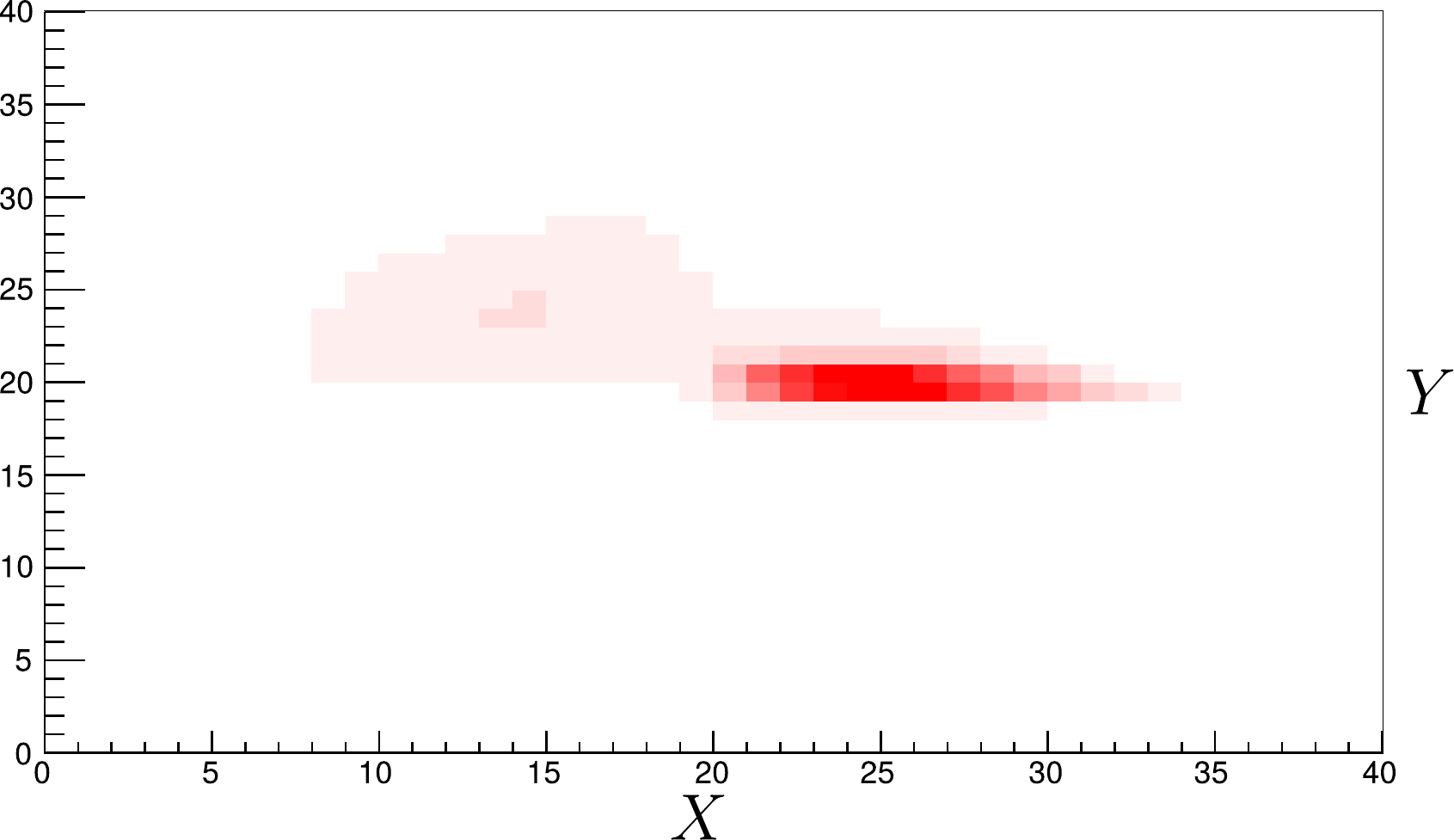}
        \end{tabular}
        \caption{
            The images of top-jets (averaged over  $\approx 500000$ events)  obtained after our preprocessing method  without (left) and  with (right) detector effects  (using {\tt Delphes}), the axis labels follow the convention in \eqnref{eq:ImageDefn}.
        }
        \label{fig:TOPIMAGE}
    \end{center}
\end{figure}

The jet images obtained using our method with and without detector effects are shown in \figref{fig:TOPIMAGE}, clearly  the structure is much more ``complex'' than the case of QCD  jets. Even though the three prong structure is not pronounced  (due to the three body decay) in the image, it is still visible in a subtle way. The extended horizontal lump and lobe approximately  constitute two prongs while the third is from the vertical  extension of the lobe towards the left.  Again, the effect of detector simulations seem just to smear the   momentum distribution of the jet constituents.

Note that the difference in distributions of top jets and QCD-jets suggest that simple cuts on the loss $\epsilon$, can be helpful in order to find top jets as anomalies. For example, if we consider  jets with $\epsilon > \epsilon_\text{cut} $, where $\epsilon_\text{cut} = 0.15$, to be signals (\textit{i.e.}, outliers as far as QCD-jets are concerned) we find that a negligible fraction of QCD-jets (namely $\epsilon_\text{QCD}$) and a rather large fraction of top-jets (namely $\epsilon_t$) pass the cut. Varying $\epsilon_\text{cut}$, one  obtains $\epsilon_t$ as a function of $\epsilon_\text{QCD}$ (the ROC for top jets).
\begin{figure}[h]
	\begin{center}
		\includegraphics[width=0.49\textwidth]
		{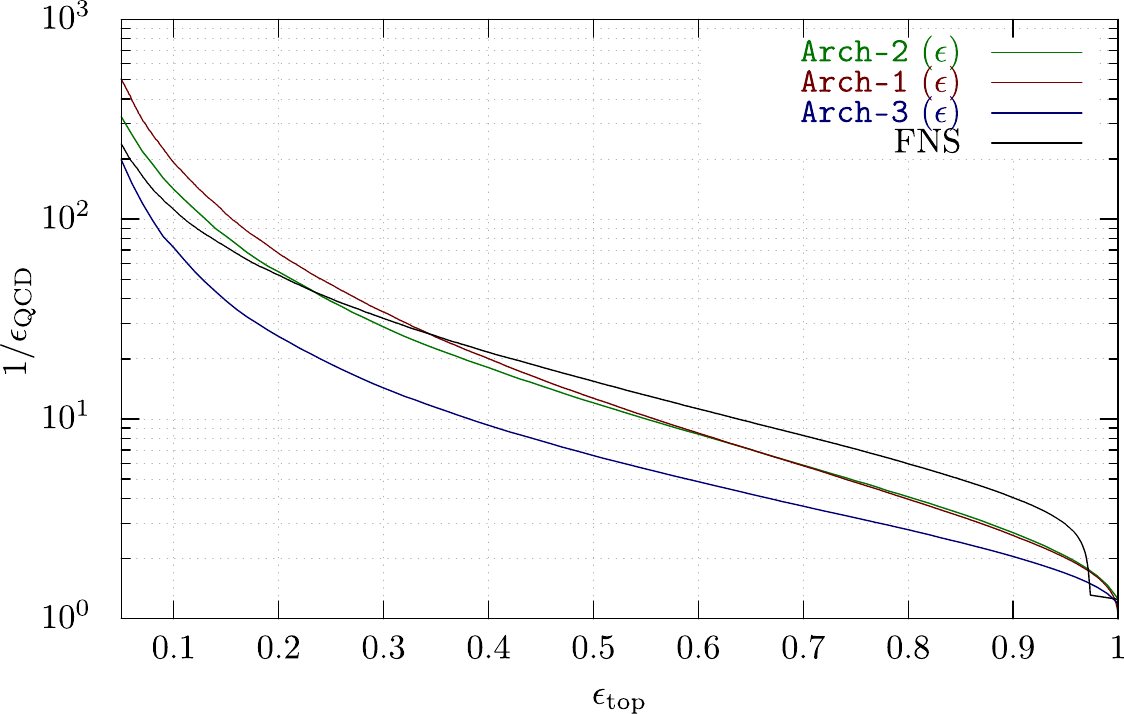}
		\includegraphics[width=0.49\textwidth]
		{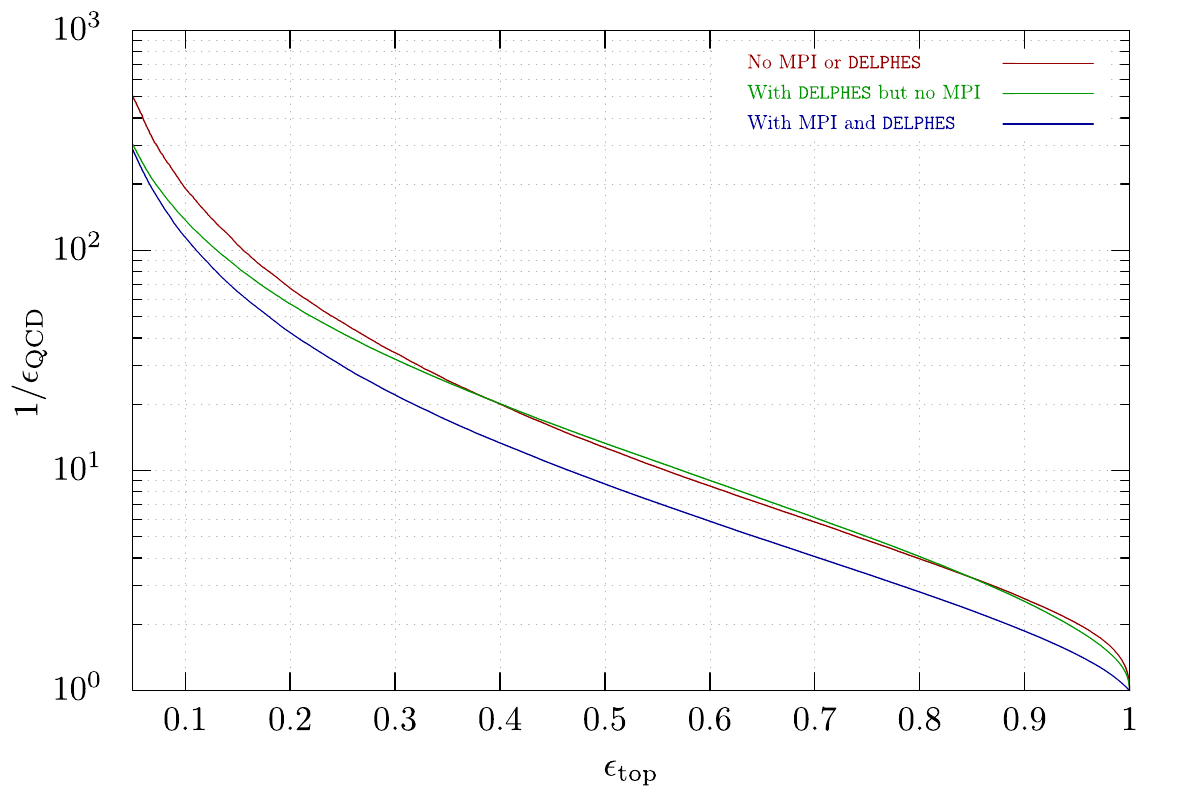}
		\caption{
			ROC Comparing the three architectures of the autoencoder along with the method from \cite{Farina:2018fyg} for QCD vs top jet discrimination using only autoencoder loss (left) and
			ROC comparing the effects of MPI and detector effects on {\archA} of the autoencoder (right).
		}
		\label{fig:roccomparearch}
	\end{center}
\end{figure}
In the left plot \figref{fig:roccomparearch}, we  compare the ROCs produced by autoencoders using network  architectures  in {\archA}, {\archB} and {\archC}.  Clearly, the dense network in  {\archA} offers the best performance but is closely followed by much shallower {\archB}. Even though the convolutional autoencoder in {\archC} is the most robust anomaly finder (as seen in \figref{fig:masssculpt}, \figref{fig:jet massinlossbins}, and  Table.~\ref{fig:correlation}), it does not offer as good a performance  to identify top-jets as anomalies. 
%{\tt{\color{c1}{(We probably have to re-write this whole paragraph?)}}}
%{\bf{\color{c1}{

\begin{figure}[h]
	\begin{center}
		\includegraphics[width=0.49\textwidth]
		{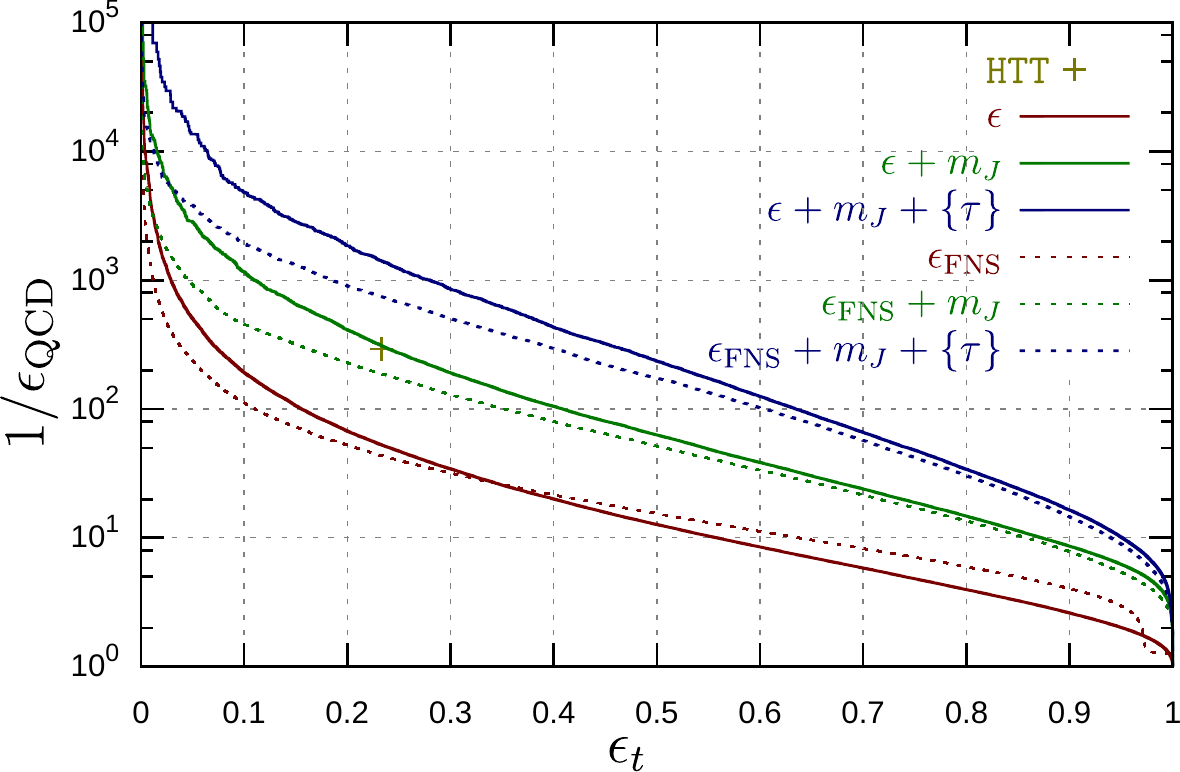}
		\includegraphics[width=0.49\textwidth]
		{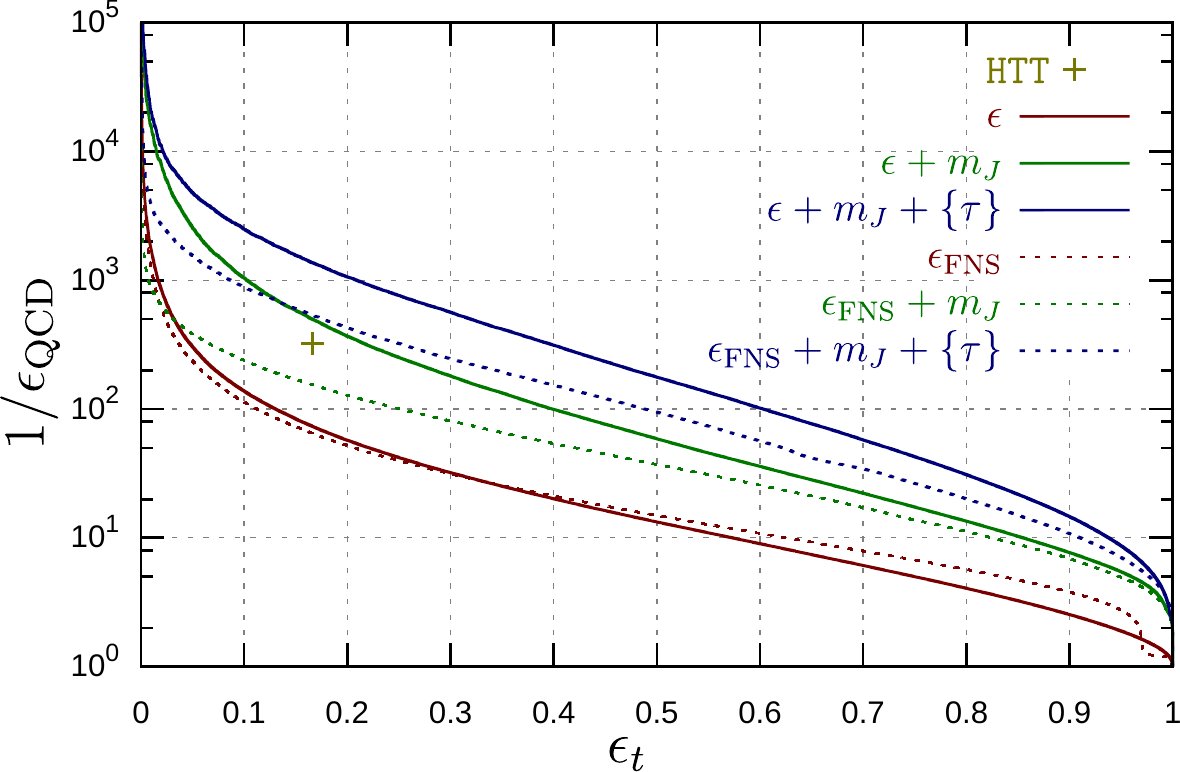}
		\caption{
			ROC for discriminating QCD vs.top using our auto encoder loss function $\epsilon$, as well as combining $\epsilon$ with jet mass and N-subjettiness variables $\{\tau_1, \dots, \tau_5 \}$.
			For comparison, we also produce the same but using $\epsilon_{\text{FNS}}$, loss functions trained using Ref.~\cite{Farina:2018fyg}.
			These results are presented without {\bf(left)} and with {\bf(right)} detector effects.
		}
		\label{fig:ROC_NOMPI_QCDTTBAR}
	\end{center}
\end{figure}

A potentially interesting example is to use the loss function $\epsilon$ in a supervised manner.
As mentioned before, after training on QCD-jets, the trained autoencoder can be taken to be a ``black-box"  where for any input jet the black-box gives an output  loss $\epsilon$.
One can use this $\epsilon$ as a jet-shape like any other physics based ones (such as mass or $N$-subjettiness), even though it is hard to interpret.
In the problem of finding top jets, therefore, we can combine $\epsilon$ and $m_J$ in the sense that we can introduce rectangular cuts on the $\epsilon$--$m_J$ plane to calculate $\epsilon_t$ and  $\epsilon_\text{QCD}$.
In order to perform this study on the  $\epsilon$--$m_J$ plane in a systematic manner and produce ROCs we use a Boosted Decision Tree (BDT) based multivariate analysis as implemented in Root~\cite{BRUN199781}, which simply finds signal rich hypercubes (here rich in top-jets) in multi-dimensional spaces of observables (here the $\epsilon$--$m_J$ plane), which are rich in signal (here top jets).
The results of these studies are presented in \figref{fig:ROC_NOMPI_QCDTTBAR}.
%}}}  
In order to demonstrate that our pre-processing step does not degrade the performance, we compare ROCs with and without pre-processing.
To be specific, we compare the ROC produced using cuts on  $\epsilon$, where we use the pre-processing stage and employ the architecture in {\archA}, with that of the dense network of Ref.~\cite{Farina:2018fyg} without preprocessing (we refer to it as $\epsilon_{\text{FNS}}$).
As shown in \figref{fig:ROC_NOMPI_QCDTTBAR}, both these ROCs are comparable.
%{\bf{\color{c1}{
			In the same figure, we also show the discriminating power of the well-studied, well-understood  {\heptoptagger} by the point marked ``$+$'' in the plot. Note that the default version functions without Qjets~\cite{Ellis:2014eya,Ellis:2012sn} or any other variables (like N-subjettiness \cite{Thaler:2011gf} or Energy Correlation~\cite{Larkoski:2013eya}), but uses the jet mass information in the mass drop step and also many other features from the physics of top quark decay kinetics. Clearly, {\heptoptagger} outperforms the autoencoder when we only use cuts on $\epsilon$. This is completely understandable since the autoencoder uses no physics information associated with top decays. The fact that one still achieves the impressive performance to find top jets as anomalies is impressive. As a further demonstration we additionally show ROCs where we combine $\epsilon$ with $m_J$ and $N$-subjettiness $\{\tau_1, \dots, \tau_5 \}$, by using BDTs. Not surprisingly, we find significantly better performance.
% -- it produces an improvement by a factor of $2.5$ over the {\toptag} in signal selection efficiency for a given background rejection.
%}}}
It shows that our technique is extremely general
%{\bf{\color{c1}{
			and its performance can be competitive with specialized taggers designed for each event topology.
%}}}

At this point, we also wish to note another important feature in the graph.
When just the autoencoder loss is used, $\epsilon_{\text{FNS}}$ does outperform our method in a large portion.
But it is worth pointing out that $\epsilon_{\text{FNS}}$ uses a large amount of information from the jet mass as seen from \figref{fig:masssculpt} and it is not possible to factor out this effect of jet mass from this method easily.
So instead of trying to factor out the effect of jet mass from $\epsilon_{\text{FNS}}$, we instead choose another way to make this comparison fair.
This is by incorporating the jet mass as an input observable for both the methods while making the comparison.
These arguments could be generalized to include other jet observables in addition to the autoencoder loss so that the fraction of information overlapping between these observables and the autoencoder loss can be gauged.

The detector simulation seems to affect the performance of all three methods by $10 \% - 20 \%$.  Using our method, for a signal selection efficiency of  $\approx 80 \%$, background fake rates as low as $5 \%$ to $10 \%$ can be achieved.
%{\color{c1}{\bf{
			The significant enhancement in the performance of the autoencoder loss when it is combined with  $m_J$ and $N$-subjettiness, suggests that $\epsilon$ is largely uncorrelated with  $m_J$ and $N$-subjettiness as well.
%}}}
We already have shown how the preprocessing stage de-correlates the loss from $m_J$. Since the $N$-subjettiness variables are usually strongly correlated with $m_J$, de-correlating mass should also result in de-correlating $N$-subjettiness. 
\begin{figure}[h]
	\begin{center}
			\includegraphics[width=0.9\textwidth]{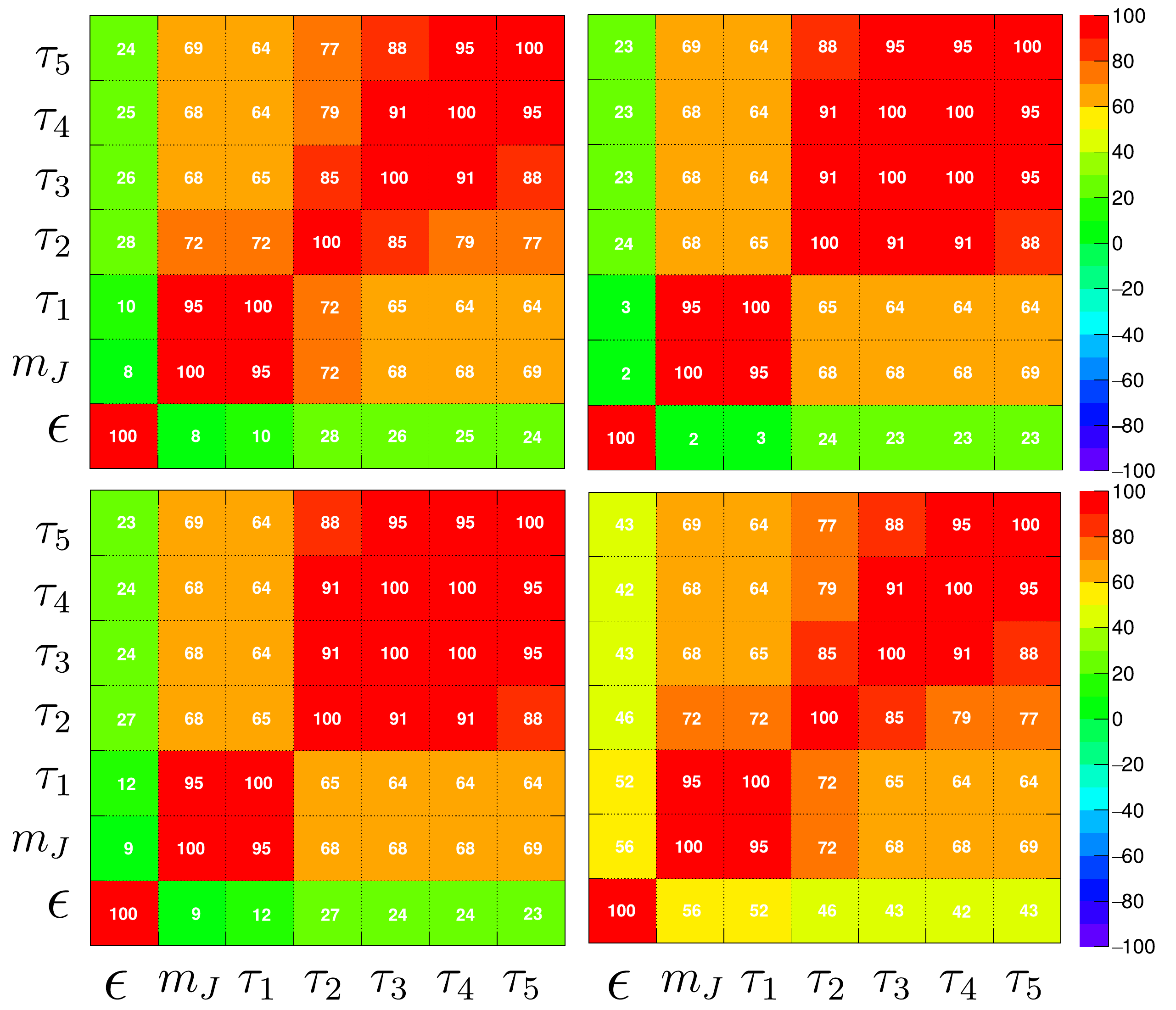}
		\caption{
			Linear correlation coefficients (in \%) among $\epsilon$, $m_J$ and $\left\{ \tau_1, \dots, \tau_5 \right\}$ for QCD-jets using {\archA} (top-left),  {\archB} (bottom-left),   {\archC} (top-right), and  $\epsilon_{\text{FNS}}$ (bottom-right).
		}
		\label{fig:Corrn}
	\end{center}
\end{figure}
In \figref{fig:Corrn} we plot the linear correlation coefficients (in \%) among $\epsilon$, $m_J$ and $\left\{ \tau_1, \dots, \tau_5 \right\}$ for QCD-jets using all the three networks discussed in this paper. We see what we expect --  correlation coefficients never reach even  $30\%$.  In the same figure we also show the same  correlation matrix where we combine $\epsilon_{\text{FNS}}$ with $m_J$ and $N$-subjettiness   variables. As shown in the bottom-right plot, the correlation coefficients for $\epsilon_{\text{FNS}}$ remains in the range  $\sim 45\% $ to $65\%$ for all these variables. 
%3abb6677af34ac57c0ca5828fd94f9d886c26ce59a8ce60ecf6778079423dccff1d6f19cb655805d56098e6d38a1a710dee59523eed7511e5a9e4b8ccb3a4686   
\begin{figure}[h]
	\begin{center}
		\includegraphics[width=0.6\textwidth]
		{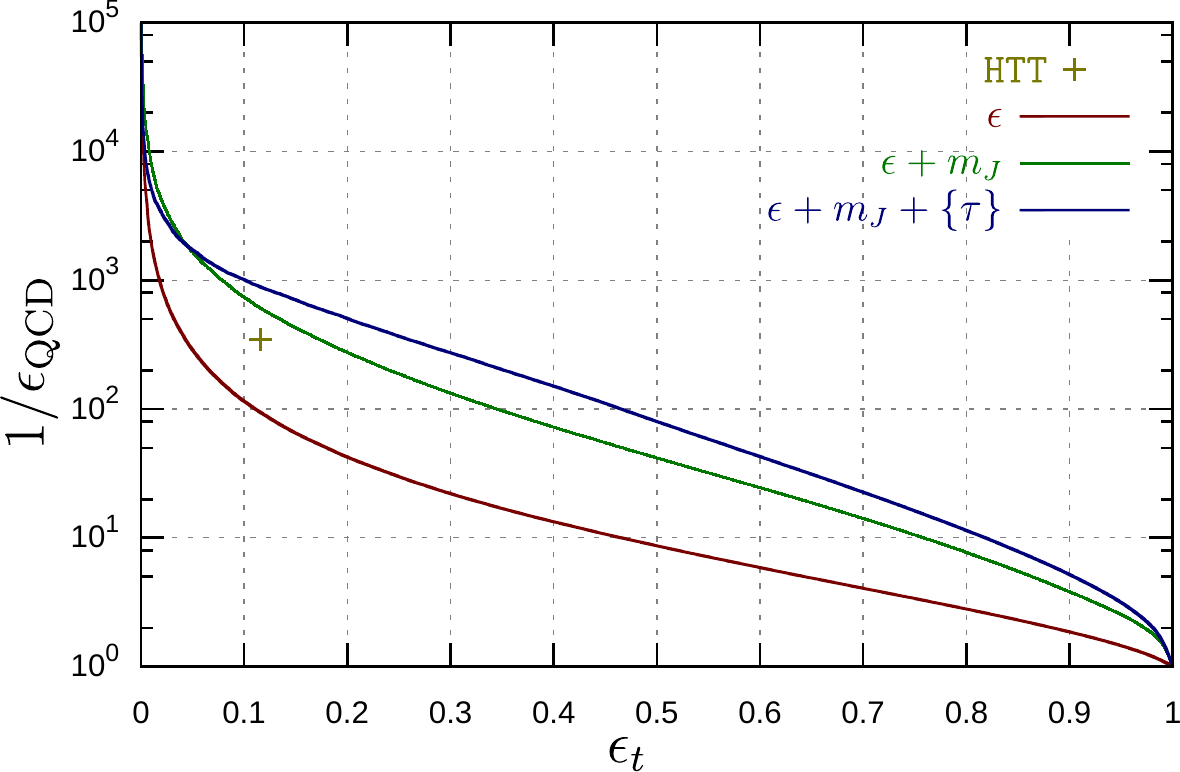}
		\caption{
			ROC for QCD vs top discrimination using our   method with MPI and detector effects.
		}
		\label{fig:ROC_MPI_DET_QCDTOP}
	\end{center}
\end{figure}
Finally, in order to demonstrate the fact that our method preserves  its performance in a more realistic situation, we benchmark its performance of QCD-jets vs top-jet discrimination on events  simulated with MPI and detector effects. We also discard all parton level merge requirements, imposed earlier, as explained in  case.~\ref{case:toppsC}. Additionally, we  use a much larger phase space region  with $ {p_T}_J > 400\gev$.  We show the results from this study in \figref{fig:ROC_MPI_DET_QCDTOP}.
%{\color{c1}{\bf{
			In the same plot we also show the performance of {\toptag} for comparison.
%		}}}
We see that much of the performance seen in the previous case is still retained even after MPI and detector effects are turned on (see, \figref{fig:ROC_NOMPI_QCDTTBAR} for comparison). 

%($\sim 4 \times 10^3$ background rejection for 20\% signal selection efficiency which is comparable to \figref{fig:ROC_NOMPI_QCDTTBAR}. 

%-------------------------------------------------
\subsubsection{Benchmark with $W$-jets} 
\label{sec:wperformance} 
%-------------------------------------------------

In this section, we consider the problem of discriminating $W$ jets against QCD-jets. This is again a well studied problem and variables like $N$-subjettiness~\cite{Stewart:2010tn,Thaler:2010tr} have been observed to work reasonably well. The CMS collaboration for LHC, for example, used $\tau_2/\tau_1$ for $W$-tagging~\cite{Khachatryan:2014vla}.
We use $W$-jets as anomalies and compare the performance of our methods  with that of  $\tau_2/\tau_1$.     

In this work we consider a sample of boosted $W$ jets with $800\gev <{p_T}_J<  900\gev$, generated without MPI or detector effects.
We also impose a merge requirement (as was the case for top) that the parent $W$ and its decay products lie within the jet  $ \left[ \Delta R \left( p_{J}^{\mu} , p_{a}^{\mu}  \right)  < 0.6  \right] $, where $a$ stands for decaying $W$, as well as quarks from the decay. 

\begin{figure}[H]
	\begin{center}
		\includegraphics[width=0.6\textwidth]{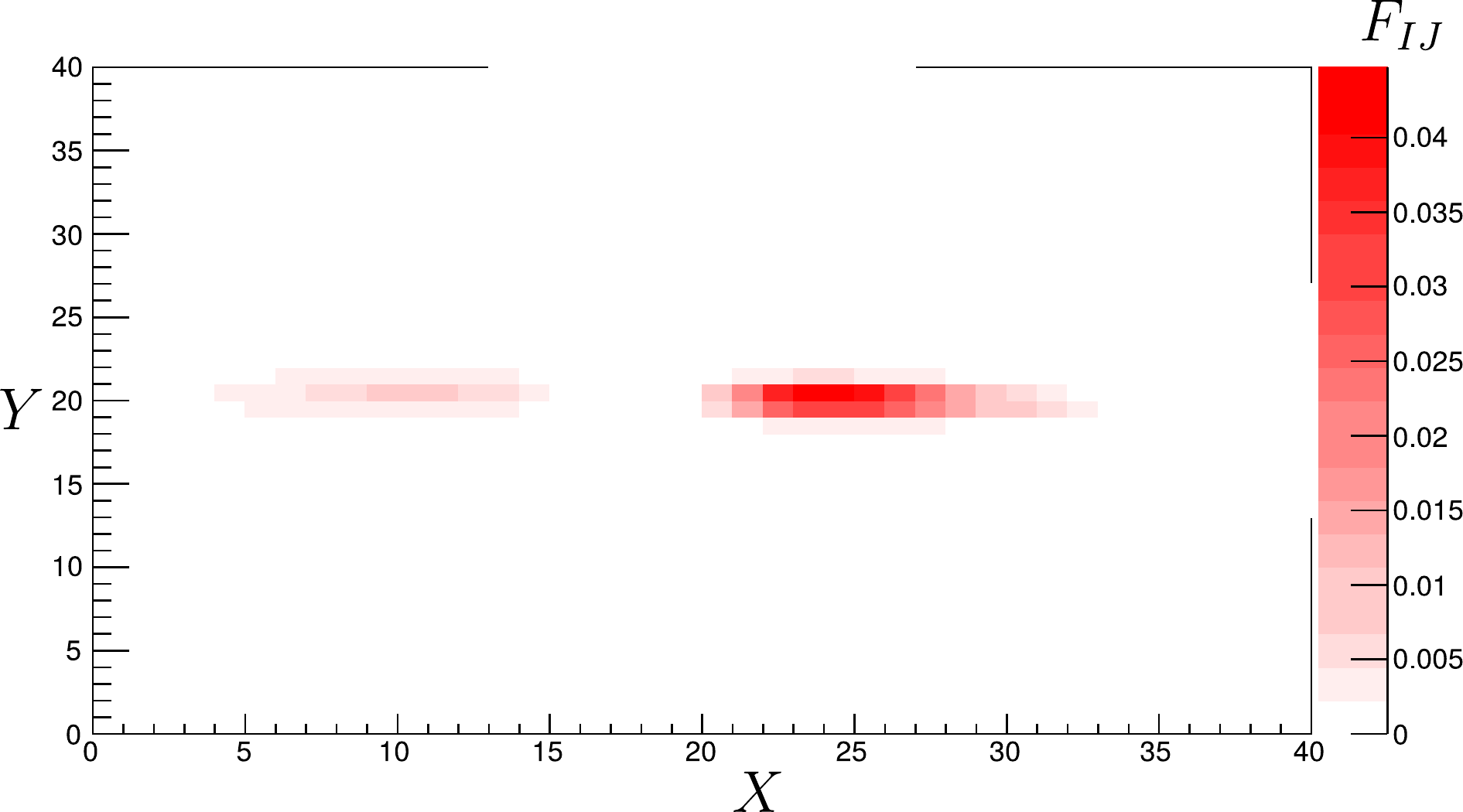}
		\caption{
			The average jet image of $W$-jet obtained after our pre-processing method (the axis labels follow the convention in \eqnref{eq:ImageDefn}).
		}
		\label{fig:WBSIMGGEN}
	\end{center}
\end{figure}
\begin{figure}[h]
	\begin{center}
		\includegraphics[width=0.6\textwidth]
		{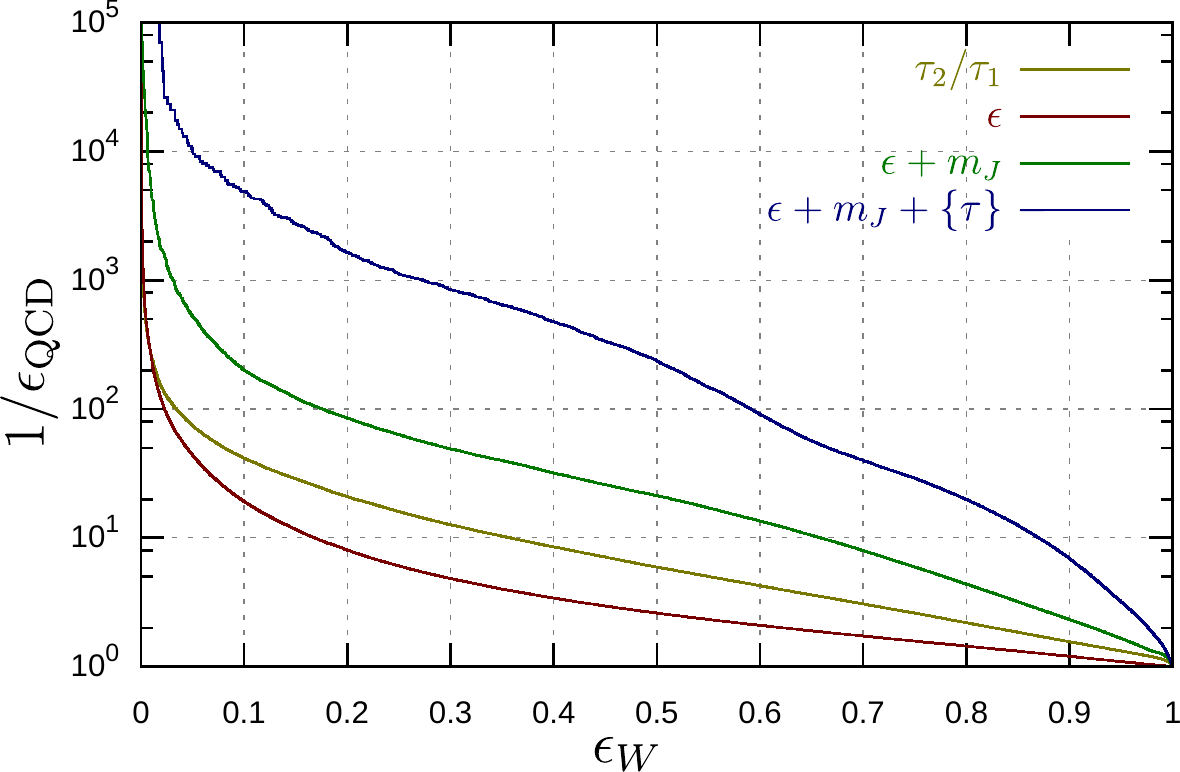}
		\caption{
			ROCs for discriminating QCD-jets vs $W$-jets using our anomaly finder and a standard $N$-subjettiness variable.
		}
		\label{fig:ROC_NOMPI_NODET_QCDW}
	\end{center}
\end{figure}
The jet images obtained using our method is presented in \figref{fig:WBSIMGGEN}.
Notice that the two prong structure of the jet is clearly visible.
The asymmetry in the intensities of the two prongs is due to the Gram-Schmidt procedure \eqnref{eq:e2}, which always   brings the harder prong along the right side of the  horizontal axis by construction. 

As before, we use an ROCs to quantify the performance of our method and compare it against an existing standard. The ROC in \figref{fig:ROC_NOMPI_NODET_QCDW} shows the achievable background rejection efficiency as a function of signal efficiency, when we use autoencoder loss function alone as well as when we combine it with $m_J$ and N-Subjettiness variables.
We find that our method performs reasonably well as an anomaly finder only when it is combined with the other variables.
At the level of $1\%$ QCD-jet reduction it manages to yield a signal selection efficiency of $\sim 20 \%$.
On the other hand, using $\epsilon$ along with $m_J$ and N-Subjettiness as well, yields signal selection efficiency as high as $\approx 60\%$ for the same QCD-jet rejection.

%-------------------------------------------------
\subsubsection{Benchmark with di-$W$ jets from new physics} 
\label{sec:NPperformance}
%-------------------------------------------------

In this section, we study the efficiency of finding jets consisting of decay products a NP particle (namely $\phi$, a di-$W$ resonance) as anomalies. To be specific, we use a scalar  with mass $180\gev$. We use the effective two Higgs doublet model, where $\phi$ is the heavy scalar Higgs, for generating events. The actual process we consider is given below 
\begin{equation}
p \ + \ p \ \rightarrow \  Z \left( \nu_e \bar{\nu}_e \right) \
      + \ \phi \left( W^+ W^- \right) \; . 
\end{equation}    
The events are generated at $\sqrt{s}=13$ TeV without MPI or detector effects. We consider jets with $800\gev <{p_T}_J< 900\gev$ that satisfy the usual merge requirement that the heavy parent $\phi$, $W^+$, $W^-$ and the four final partons lie inside the jet, namely $\left[ \Delta R \left( p_{J}^{\mu} , p_{a}^{\mu} \right) < 0.6 \right]$, where $a$ stands for decaying $\phi$, as well as all the partons from $W$-decays.

\begin{figure}[H]
	\begin{center}
		\includegraphics[width=0.6\textwidth]{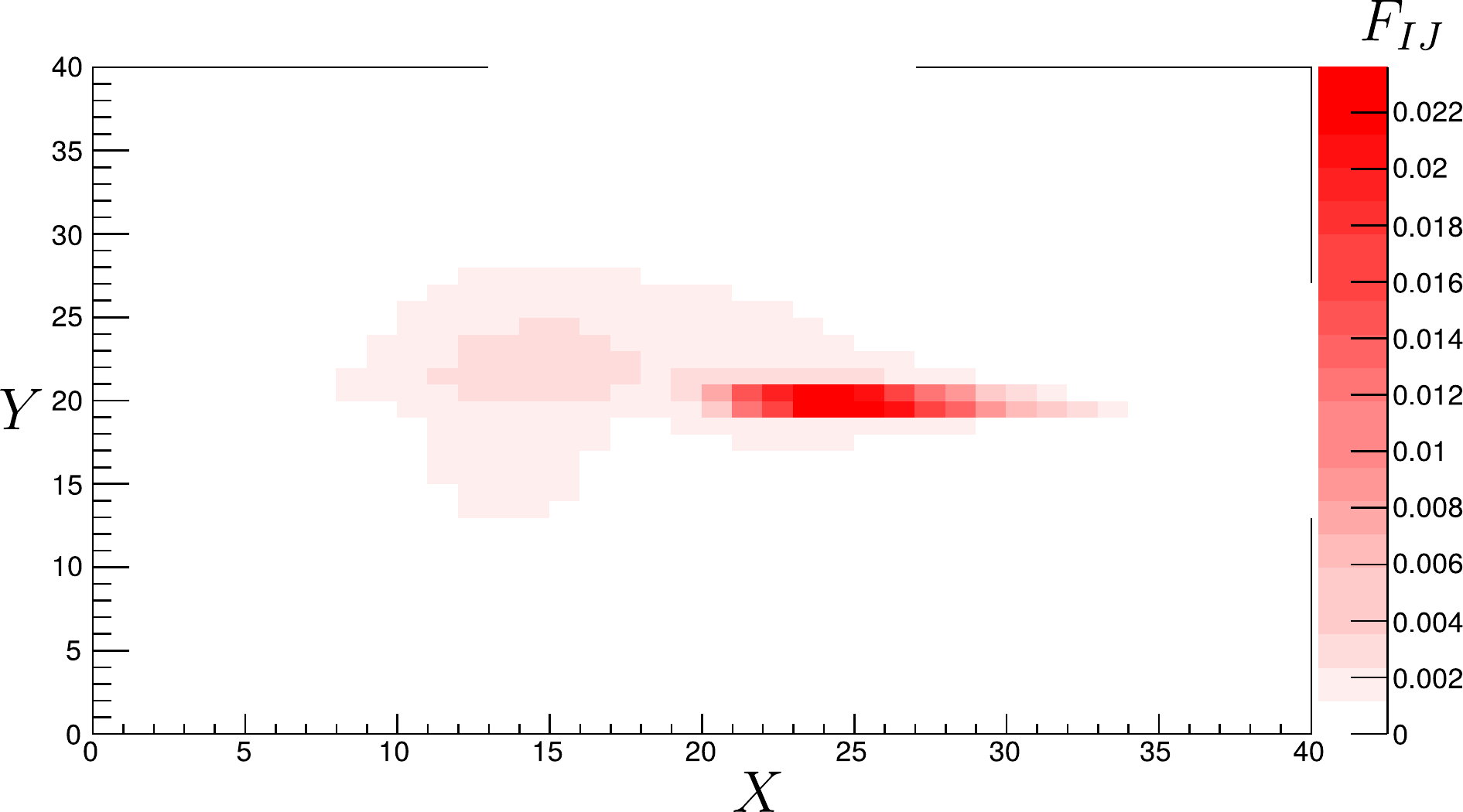}
		\caption{
			The average jet image for a di-$W$ resonance with mass $180\gev$ after our pre-processing method (the axis labels follow the convention in \eqnref{eq:ImageDefn}).
		}
		\label{fig:H4JIMGGEN}
	\end{center}
\end{figure}
The jet image formed using our method is presented in \figref{fig:H4JIMGGEN}. Even though the four prong structure is not pronounced, it is still visible in a subtle  way. As in the case of \figref{fig:TOPIMAGE}, the  horizontal lump and the extended lobe towards the left constitute two prongs while the third and fourth prongs overlap with the first two leading to the halo around the lump towards the right and a bright spot on the left lobe.

\begin{figure}[H]
    \begin{center}
\includegraphics[width=0.6\textwidth]
{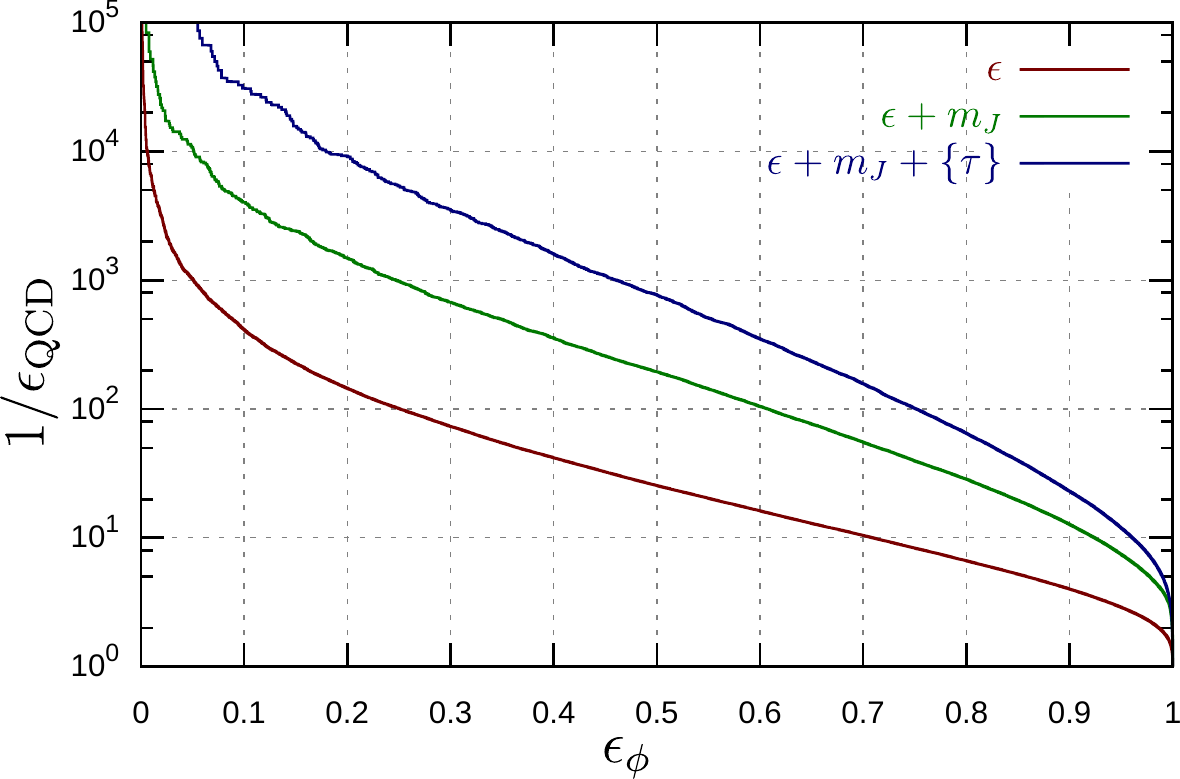}
        \caption{
            ROC for discriminating QCD vs di-$W$ jets using our anomaly finder.
        }
        \label{fig:ROC_NOMPI_NODET_QCDH4J}
    \end{center}
\end{figure}
Finally, we show the ROCs for discriminating these $\phi$-jets as  anomalies in \figref{fig:H4JIMGGEN}. The  performance of our method is highly promising. Even with just the autoencoder loss, for a signal selection efficiency of $\approx 20\%$, it suffers a fake rate of only $\approx 0.5 \%$ due to QCD-jets.
%{\color{c1}{\bf{
%	When combined with mass and N-subjettiness information the acceptance rate for the di-$W$ jets can be increased to order $70\%$ keeping the same fake rate. 
%}}}

%----------------------------------------------------
\section{Conclusion} 
\label{sec:Conclusion} 
%----------------------------------------------------

In this paper, we propose a new anomaly finder or rather an anti-QCD tagger based on autoencoder. While the proposal uses a version of an autoencoder (without variational or adversarial training), its novelty lies in the preprocessing stage, where we exploit the full Lorentz symmetry apart from  symmetries of shift, rotation and rescaling (approximate conformal symmetry of QCD). Our method gives comparable performances to some existing techniques in the literature, while, at the same time, removes the dependence of autoencoder loss function on the jet mass.  Consequently, it gives a straight forward way to find pure control samples to train. Our anomaly finder can be trained using jets in low $\rho$ bins, which is rich in QCD-jets, and can be used to analyze jets from high $\rho$ bins to find anomalies (or, signature of NP). The lack of jet mass dependency of the autoencoder loss  brings in additional advantages -- it can be readily combined with existing discriminating variables to improve performances of even supervised learning to some effect.

Apart from this robustness feature, we also find that detector effects or multi parton interactions have little effects in the performance of the anomaly finder.  We have not explicitly tested our method against pileup in this study. However, we see no reason why this method can not be made pileup robust by simply  using the standard pileup mitigation  techniques~\cite{Krohn:2013lba, Cacciari:2014gra, Bertolini:2014bba}  before forming the jet image, or even after the jet image is formed as shown in Ref.~\cite{Komiske:2017ubm}.

We were unable to optimize on the structure of the autoencoder network due to lack of sufficient computational power. However, even with our naive optimization methods, we manage to get remarkable results as compared to  existing techniques in literature. Our method provides an extremely simple yet a powerful technique with great potential for generalization as it can be readily incorporated into any existing neural network based algorithms such as image based top tagging~\cite{Almeida:2015jua,Macaluso:2018tck,Kasieczka:2017nvn,Cogan:2014oua,Baldi:2016fql,deOliveira:2015xxd}, Lorentz layers~\cite{Butter:2017cot}, or recurrent networks~\cite{Louppe:2017ipp}  with minimal effort. Because of its simplicity, the preprocessing stage can even be extended to existing boosted jet substructure classification algorithms that uses only physics information. With more improvements and optimizations, it might even be possible to use the autoencoder loss obtained with our method (combined with the jet mass) as a spectroscopy like observable to find anomalous events. 

\begin{flushleft}
Source code for most of the content discussed here can be found at:
[\url{https://github.com/aravindhv10/CPP_Wrappers/tree/master/AntiQCD4}]
\end{flushleft}

\acknowledgments

We are grateful to Tanmoy Bhattacharya and Michael Graesser who originally suggested using Auto-encoders for anomaly hunting at colliders. This work also benefitted from stimulating conversations with Harrison Prosper and Deepak Kar. 
TSR was supported in part by the Early Career Research Award by Science and Engineering Research Board, Dept. of Science and Technology, Govt. of India (grant no. ECR/2015/000196). We also acknowledge the workshop ``Beyond the Standard Model: where do we go from here?'' hosted at the Galileo Galilei Institute for Theoretical Physics (GGI), as well as the workshop ``International Meeting on High Energy Physics" hosted at Institute of Physics (IOP) where parts of this work was completed. 
AHV is extremely grateful to Brij Jashal for lending access to a GPU on the CMS T2 at TIFR without which some parts of this work would have been impossible.

\appendix

%-----------------------------------------------------
\section{Cross validation} 
\label{sec:crossvalidation} 
%-----------------------------------------------------
The purpose of this appendix is to demonstrate that even if the autoencoder is trained using jets in a given $p_T$ range, it can still be used to find anomalies in jets with largely different $p_T$.  To bring home the point we present results from two studies:
\begin{enumerate}
	\item Train the autoencoder on $\sqrt{s}=13$ TeV QCD-jet data and test it on QCD-jets at $\sqrt{s}=100$ TeV.
	\item Train the autoencoder on QCD-jets from $\sqrt{s}=100\tev$ collisions and test it  on the QCD-jets produced at $\sqrt{s}=13$ TeV.
\end{enumerate}
\begin{figure}[h]
	\begin{center}
		\includegraphics[width=1.0\textwidth]
		{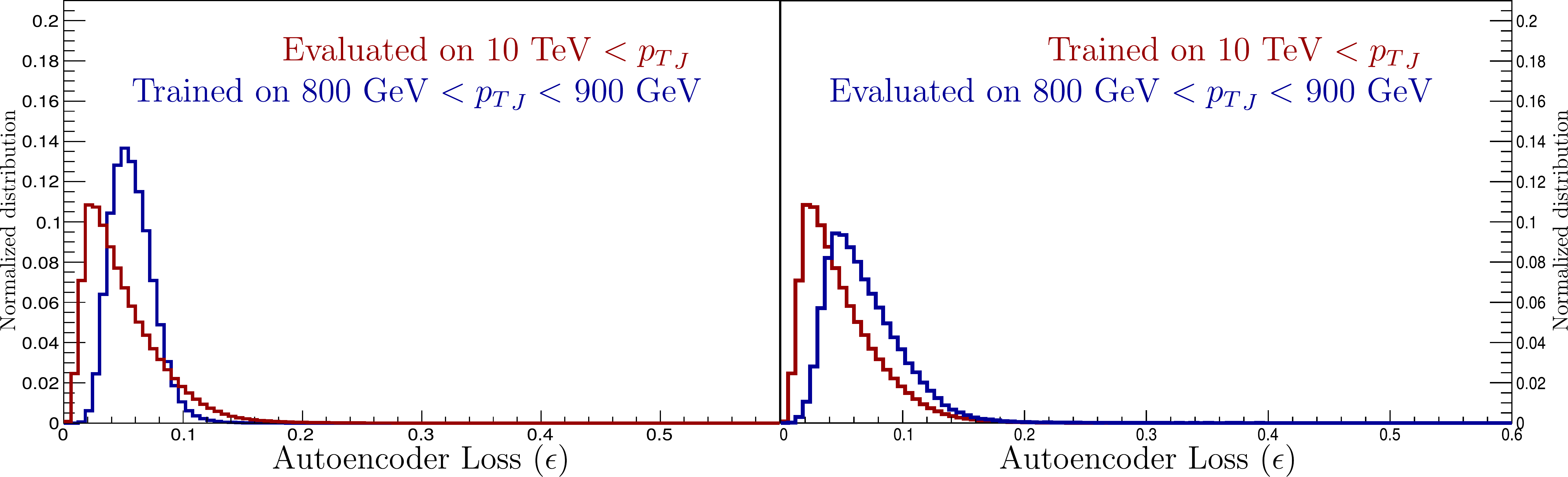}
		\caption{ Dependence of autoencoder loss function on jet mass when the network is trained on low energy jets and  evaluated on high energy jets (left) and when  network is trained on high energy jets and  evaluated on low energy jets (right). } 
		\label{fig:LIGHT_HEAVY_ROBUST}
	\end{center}
\end{figure}
Even with these extreme comparisons of jets with drastic differences in kinematics, our method yields largely insensitive results as can be seen in \figref{fig:LIGHT_HEAVY_ROBUST}. Note that the shift in peak positions in both cases are less than 50\% of the width. Most of the actual distributions overlap and, consequently, widths and averages in all cases are within $20\%$ of each other.

%-----------------------------------------------------
\section{Autoencoder loss and jet mass}
\label{sec:lossvsm}
%-----------------------------------------------------

As mentioned in the text, we show the full 2-dimensional distribution of autoencoder loss $\epsilon$ as  a function of jet mass $m_J$ for QCD-jets.
In particular we show results for all the four cases mentioned in \subsecref{sec:13TeV} in order to understand the variation of the joint probability distribution as we remove various stages in preprocessing. 
\begin{figure}[H]
	\begin{center}
		\includegraphics[width=1.0\textwidth]{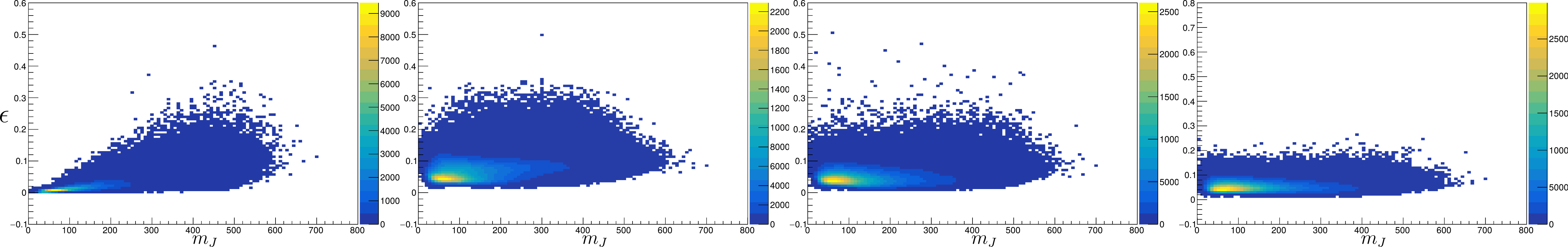}
		\caption{
			2D Histogram of jet mass vs autoencoder loss  for the four cases mentioned in the text:  preprocessing without Lorentz boost + {\archA} (first from left),  preprocessing without using  Gram-Schmidt axis + {\archA} (second), full preprocessing + {\archA}   (third), and finally, full preprocessing + {\archC} (fourth).
		}        
		\label{fig:2dcorrelation}
	\end{center}
\end{figure}

%-----------------------------------------------------
\section{The Convolutional Autoencoder}
\label{sec:archc}
%-----------------------------------------------------
Here, we give a the pseudo-code describing the convolutional autoencoder ({\archC}):
{\small \color{mred}
	\begin{verbatim}
	Conv2D(channels=50,kernel_size=5,activation='relu')
	MaxPool2D(pool_size=2, strides=2)
	Conv2D(channels=50,kernel_size=5,activation='relu')
	MaxPool2D(pool_size=2, strides=2)
	Conv2D(channels=50,kernel_size=5,activation='relu')
	Flatten()
	Dense(100,activation="relu")
	Dense(450,activation="relu")
	reshape(50,3,3)
	Conv2DTranspose(channels=50,kernel_size=9,stride=2,padding=0,activation="relu")
	Conv2DTranspose(channels=50,kernel_size=9,stride=2,padding=0,activation="relu")
	Conv2DTranspose(channels=1,kernel_size=8,stride=1,padding=0,activation="relu")
	Flatten()
	softmax()
	reshape(1,40,40)
	\end{verbatim}
}

%-----------------------------------------------------
\section{Details of training}
\label{sec:detailstraining} 
%-----------------------------------------------------

\begin{figure}[H]
	\begin{center}
		\includegraphics[width=0.5\textwidth]{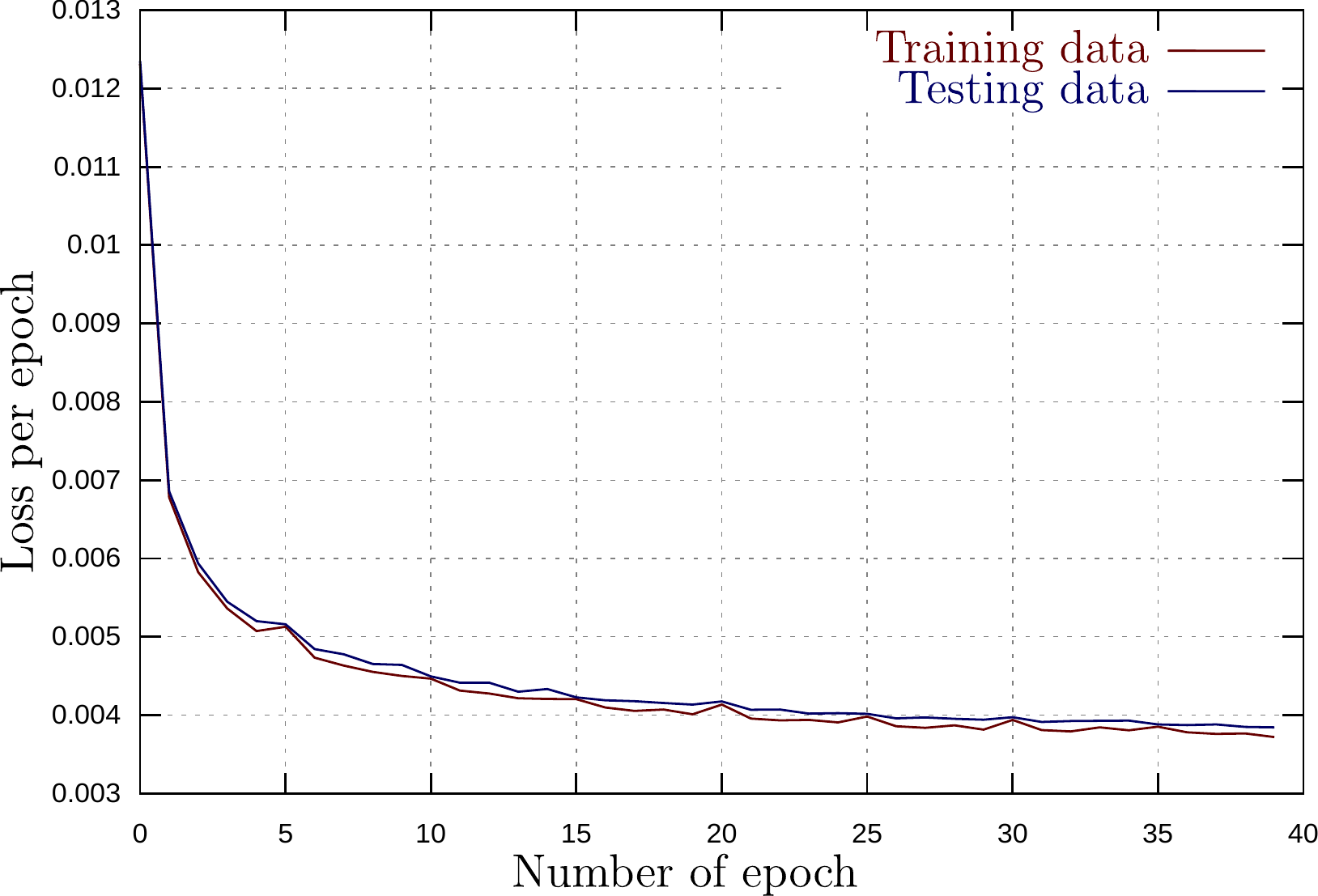}
		\caption{
			The variation of training and testing loss curves with number of training epochs for the CNN autoencoder ({\archC})
		}
		\label{fig:traininglosscurve}
	\end{center}
\end{figure}

We have used 2400000 QCD-jet images for training and 400000 QCD-jet images for testing. We generally trained for 20 epochs, since   as can be seen  in  \figref{fig:traininglosscurve},  there is no significant difference in the loss per epoch even if one goes beyond 20. 
\begin{figure}[H]
	\begin{center}
		\includegraphics[width=1.0\textwidth]{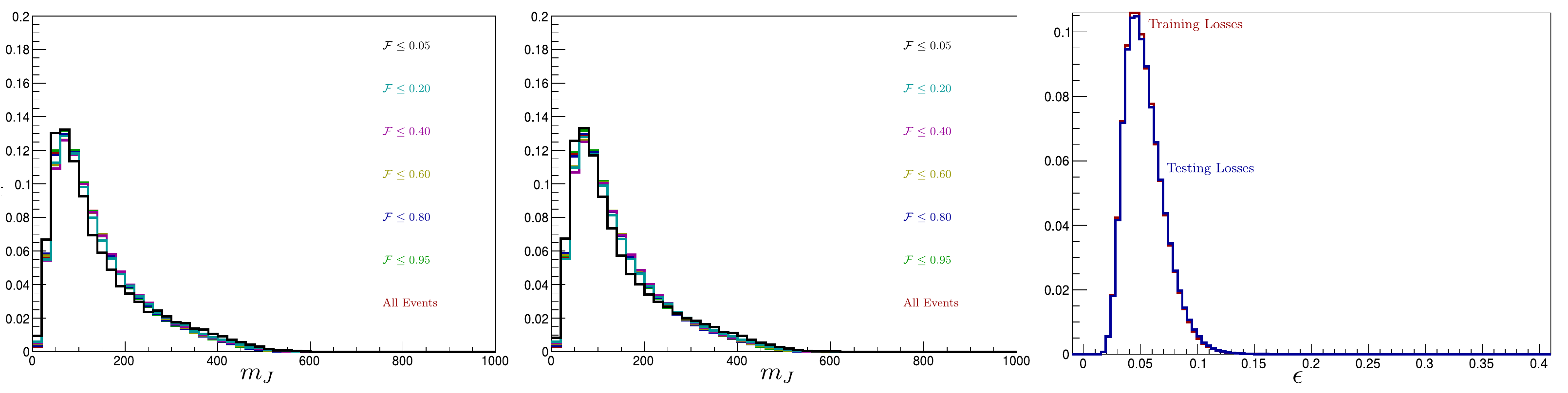}
		\caption{ 
			The distributions of $m_J$ in bins of  $\epsilon$ for QCD-jets using the training (left) and testing (middle) datasets, as well as the distributions of $\epsilon$ (right), when we use the convolutional autoencoder from {\archC}.
		}
		\label{fig:comparetrainingtesting}
	\end{center}
\end{figure}

We consider additional checks to make sure that we do not over-train. For example, we provide  comparison plots for training and testing datasets in \figref{fig:comparetrainingtesting}.  Clearly, the difference between the distributions for training and testing are generally negligible.

%-----------------------------------------------------
\section{The choice of $m_0/E_0$}
\label{sec:m0e0sensitivity} 
%-----------------------------------------------------

\begin{figure}[h]
	\begin{center}
		\includegraphics[width=0.5\textwidth]{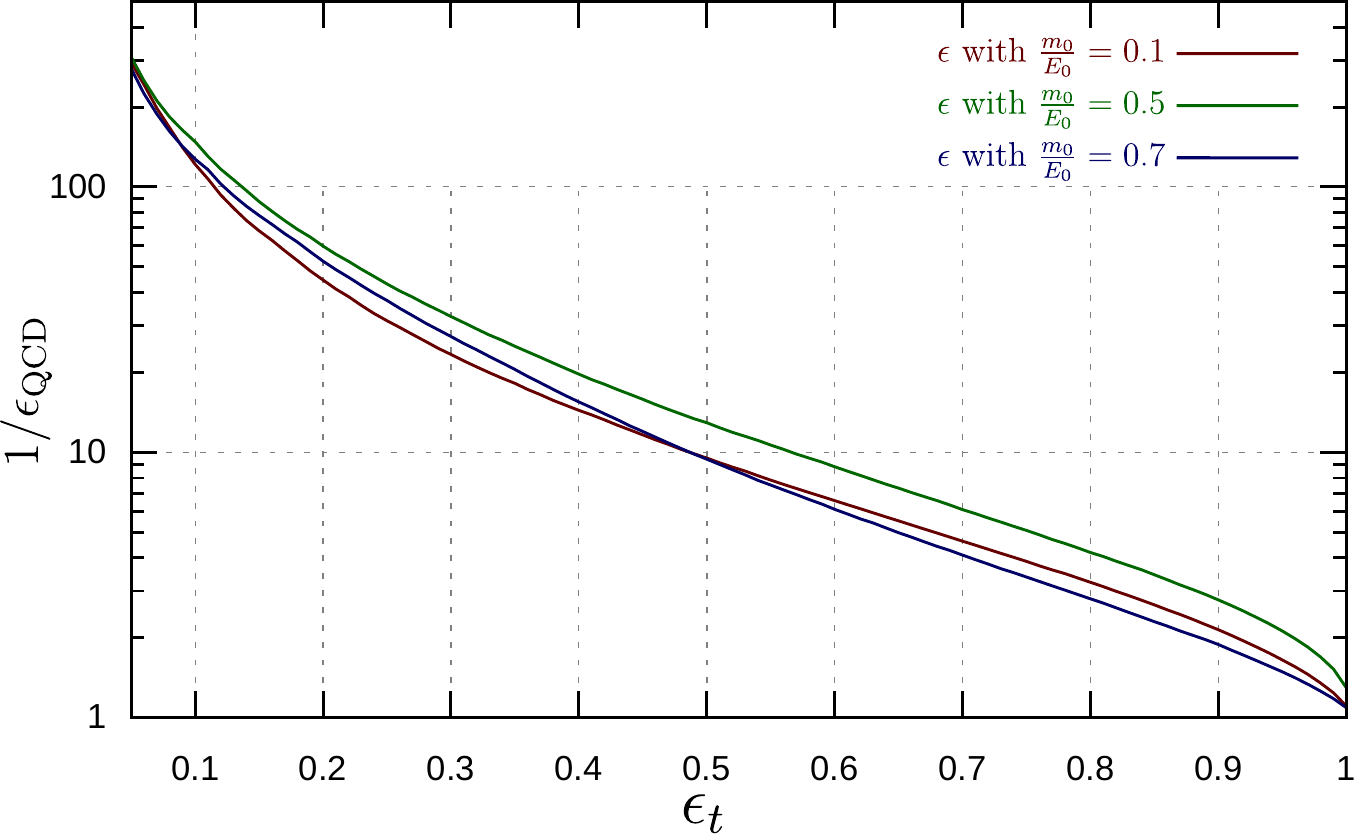}
		\caption{
			A comparison of QCD vs top jet discrimination ROC (using only the autoencoder loss) for the 3 cases of $\frac{m_0}{E_0}=0.1$, $\frac{m_0}{E_0}=0.5$ and $\frac{m_0}{E_0}=0.7$ for the shallow dense network ({\tt{Arch-2}}).
		}
		\label{fig:e0m0compare}
	\end{center}
\end{figure}
Throughout this work we use $m_0 = 0.5\gev$ and $E_0 = 1\gev$, in order to prepare jet images.  One expects that the information content in these images and, therefore, the performances of auto-encoders to be sensitive on the choice of $m_0$ and $E_0$, or rather on the ratio $m_0/E_0$. Clearly, one does not take arbitrary $m_0/E_0$ and expects to get identical performance. A tiny $m_0/E_0$ implies that the most energetic components of the final boosted jet occupy a narrow cone inside the jet, and get mapped to a few pixels at the center of the image by construction.   Similarly, a large $m_0/E_0$, sends all energetic particles to the periphery. In both these extreme limits, a large number of constituents may get mapped to the same pixels, and consequently the image looses substructure information of the jet. 

We demonstrate how the performance of the auto-encoder changes  as we  vary $m_0/E_0$ in Figure~\ref{fig:e0m0compare}. Here we compare ROC curves for QCD-jets vs top jet discrimination. The red, green, and blue  lines in Figure~\ref{fig:e0m0compare} represent performances when images are produced using $m_0/E_0$ to be $0.1$, $0.5$ and $0.7$ respectively.  As can be seen in the figure, the change in performance is rather reasonable. For the same background efficiency, we see that  the signal efficiency varies by order $10\%$,  even when we change $m_0/E_0$ by order $40\%$ or more around  $m_0/E_0= 0.5$ (the operating point in this work).

%----------------------------------------------------
\section{Benchmarking auto-encoder loss for supervised learning with standard dataset}
\label{sec:comparison_supervised} 
%-----------------------------------------------------

\begin{figure}[H]
	\begin{center}
		\includegraphics[width=0.5\textwidth]{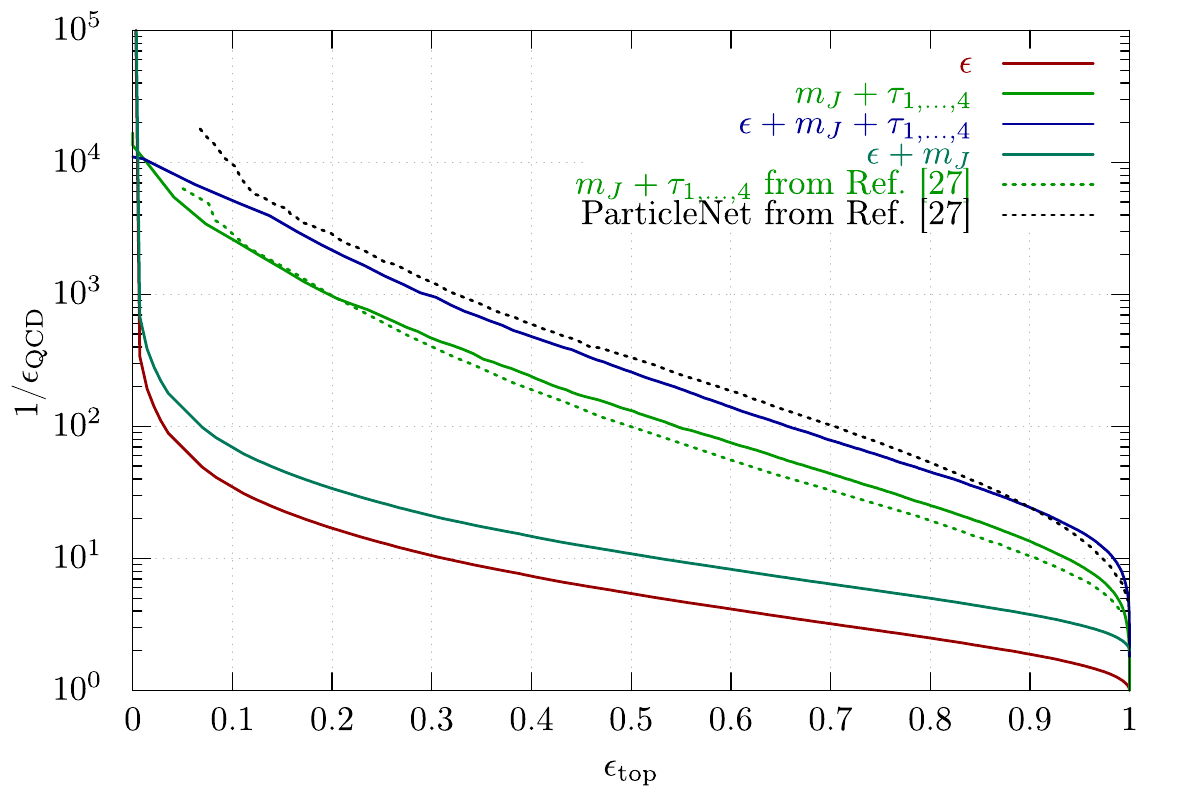}
		\caption{
			ROC for QCD vs top jet discrimination compared with results from Ref.~\cite{Kasieczka:2019dbj}.}
		\label{fig:supervisecomparisons}
	\end{center}    
\end{figure}

In Section~\ref{sec:topperformance} we suggested that the loss function $\epsilon$ can be used in a supervised manner as well.
It is tempting to benchmark this performances with respect to the garden variety of supervised algorithms available.
Of course, the exact dataset one uses to draw comparisons always provides grounds for contentions.
In order to mitigate these concerns, we provide the plot in \figref{fig:supervisecomparisons}, where we repeat our analyses using the data sample\footnote{the dataset was downloaded from \url{https://desycloud.desy.de/index.php/s/llbX3zpLhazgPJ6}} used in Ref.~\cite{Kasieczka:2019dbj}.
For this analysis, we do not perform any new optimization but just use the {\archB} model that was already described.
In this plot, the solid curves represent the result of our analyses.
One of the advantage of using the same dataset as in  Ref.~\cite{Kasieczka:2019dbj}, implies that we can readily compare these performances with the stack of ROCs (using an ensemble of supervised learners) given in  Ref.~\cite{Kasieczka:2019dbj}.
For the sake of reference, we also show the ROC for supervised learning using ParticleNet, as reported in Ref.~\cite{Kasieczka:2019dbj}.
As clearly demonstrated in \figref{fig:supervisecomparisons}, the discrimination performance of a BDT (to separate top-jets from QCD-jets) which combines high level jet observables like the auto-encoder loss, jet-mass, and $N$-subjettiness variables compares favorable well to the performance of the ``state-of-the-art" supervised algorithms from Ref.~\cite{Kasieczka:2019dbj}.

\bibliography{main.bib}

\providecommand{\href}[2]{#2}\begingroup\raggedright\begin{thebibliography}{10}

\bibitem{Chatrchyan:2012xdj}
{\bfseries CMS Collaboration} Collaboration, S.~Chatrchyan {\em et~al.},
  ``{Observation of a new boson at a mass of 125 GeV with the CMS experiment at
  the LHC},'' \href{http://dx.doi.org/10.1016/j.physletb.2012.08.021}{{\em
  Phys.Lett.} {\bfseries B716} (2012) 30–61},
  \href{http://arxiv.org/abs/1207.7235}{{\ttfamily arXiv:1207.7235 [hep-ex]}}.
\url{https://arxiv.org/abs/1207.7235}.
%\%CITATION = ARXIV:1207.7235;\%\%.

\bibitem{Aad:2012tfa}
{\bfseries ATLAS Collaboration} Collaboration, G.~Aad {\em et~al.},
  ``{Observation of a new particle in the search for the Standard Model Higgs
  boson with the ATLAS detector at the LHC},''
  \href{http://dx.doi.org/10.1016/j.physletb.2012.08.020}{{\em Phys.Lett.}
  {\bfseries B716} (2012) 1–29},
  \href{http://arxiv.org/abs/1207.7214}{{\ttfamily arXiv:1207.7214 [hep-ex]}}.
\url{https://arxiv.org/abs/1207.7214}.
%\%CITATION = ARXIV:1207.7214;\%\%.

\bibitem{Butterworth:2008iy}
J.~M. Butterworth, A.~R. Davison, M.~Rubin, and G.~P. Salam, ``{Jet
  substructure as a new Higgs search channel at the LHC},''
  \href{http://dx.doi.org/10.1103/PhysRevLett.100.242001}{{\em Phys. Rev.
  Lett.} {\bfseries 100} (2008) 242001},
\href{http://arxiv.org/abs/0802.2470}{{\ttfamily arXiv:0802.2470 [hep-ph]}}.
%%CITATION = ARXIV:0802.2470;%%.

\bibitem{Salam:2009jx}
G.~P. Salam, ``{Towards Jetography},''
  \href{http://dx.doi.org/10.1140/epjc/s10052-010-1314-6}{{\em Eur. Phys. J.}
  {\bfseries C67} (2010) 637--686},
\href{http://arxiv.org/abs/0906.1833}{{\ttfamily arXiv:0906.1833 [hep-ph]}}.
%%CITATION = ARXIV:0906.1833;%%.

\bibitem{Krohn:2009th}
D.~Krohn, J.~Thaler, and L.-T. Wang, ``{Jet Trimming},''
  \href{http://dx.doi.org/10.1007/JHEP02(2010)084}{{\em JHEP} {\bfseries 02}
  (2010) 084},
\href{http://arxiv.org/abs/0912.1342}{{\ttfamily arXiv:0912.1342 [hep-ph]}}.
%%CITATION = ARXIV:0912.1342;%%.

\bibitem{Ellis:2009me}
S.~D. Ellis, C.~K. Vermilion, and J.~R. Walsh, ``{Recombination Algorithms and
  Jet Substructure: Pruning as a Tool for Heavy Particle Searches},''
  \href{http://dx.doi.org/10.1103/PhysRevD.81.094023}{{\em Phys. Rev.}
  {\bfseries D81} (2010) 094023},
\href{http://arxiv.org/abs/0912.0033}{{\ttfamily arXiv:0912.0033 [hep-ph]}}.
%%CITATION = ARXIV:0912.0033;%%.

\bibitem{Chakraborty:2017mbz}
A.~Chakraborty, A.~M. Iyer, and T.~S. Roy, ``{A Framework for Finding Anomalous
  Objects at the LHC},''
  \href{http://dx.doi.org/10.1016/j.nuclphysb.2018.05.019}{{\em Nucl. Phys.}
  {\bfseries B932} (2018) 439--470},
\href{http://arxiv.org/abs/1707.07084}{{\ttfamily arXiv:1707.07084 [hep-ph]}}.
%%CITATION = ARXIV:1707.07084;%%.

\bibitem{Aguilar-Saavedra:2017rzt}
J.~A. Aguilar-Saavedra, J.~H. Collins, and R.~K. Mishra, ``{A generic anti-QCD
  jet tagger},'' \href{http://dx.doi.org/10.1007/JHEP11(2017)163}{{\em JHEP}
  {\bfseries 11} (2017) 163},
\href{http://arxiv.org/abs/1709.01087}{{\ttfamily arXiv:1709.01087 [hep-ph]}}.
%%CITATION = ARXIV:1709.01087;%%.

\bibitem{Cohen:2017exh}
T.~Cohen, M.~Freytsis, and B.~Ostdiek, ``{(Machine) Learning to Do More with
  Less},'' \href{http://dx.doi.org/10.1007/JHEP02(2018)034}{{\em JHEP}
  {\bfseries 02} (2018) 034},
\href{http://arxiv.org/abs/1706.09451}{{\ttfamily arXiv:1706.09451 [hep-ph]}}.
%%CITATION = ARXIV:1706.09451;%%.

\bibitem{Qu:2019gqs}
H.~Qu and L.~Gouskos, ``{ParticleNet: Jet Tagging via Particle Clouds},''
\href{http://arxiv.org/abs/1902.08570}{{\ttfamily arXiv:1902.08570 [hep-ph]}}.
%%CITATION = ARXIV:1902.08570;%%.

\bibitem{Datta:2017rhs}
K.~Datta and A.~Larkoski, ``{How Much Information is in a Jet?},''
  \href{http://dx.doi.org/10.1007/JHEP06(2017)073}{{\em JHEP} {\bfseries 06}
  (2017) 073},
\href{http://arxiv.org/abs/1704.08249}{{\ttfamily arXiv:1704.08249 [hep-ph]}}.
%%CITATION = ARXIV:1704.08249;%%.

\bibitem{Moore:2018lsr}
L.~Moore, K.~Nordström, S.~Varma, and M.~Fairbairn, ``{Reports of My Demise
  Are Greatly Exaggerated: $N$-subjettiness Taggers Take On Jet Images},''
\href{http://arxiv.org/abs/1807.04769}{{\ttfamily arXiv:1807.04769 [hep-ph]}}.
%%CITATION = ARXIV:1807.04769;%%.

\bibitem{Komiske:2018cqr}
P.~T. Komiske, E.~M. Metodiev, and J.~Thaler, ``{Energy Flow Networks: Deep
  Sets for Particle Jets},''
  \href{http://dx.doi.org/10.1007/JHEP01(2019)121}{{\em JHEP} {\bfseries 01}
  (2019) 121},
\href{http://arxiv.org/abs/1810.05165}{{\ttfamily arXiv:1810.05165 [hep-ph]}}.
%%CITATION = ARXIV:1810.05165;%%.

\bibitem{Butter:2017cot}
A.~Butter, G.~Kasieczka, T.~Plehn, and M.~Russell, ``{Deep-learned Top Tagging
  with a Lorentz Layer},''
  \href{http://dx.doi.org/10.21468/SciPostPhys.5.3.028}{{\em SciPost Phys.}
  {\bfseries 5} no.~3, (2018) 028},
\href{http://arxiv.org/abs/1707.08966}{{\ttfamily arXiv:1707.08966 [hep-ph]}}.
%%CITATION = ARXIV:1707.08966;%%.

\bibitem{Komiske:2017aww}
P.~T. Komiske, E.~M. Metodiev, and J.~Thaler, ``{Energy flow polynomials: A
  complete linear basis for jet substructure},''
  \href{http://dx.doi.org/10.1007/JHEP04(2018)013}{{\em JHEP} {\bfseries 04}
  (2018) 013},
\href{http://arxiv.org/abs/1712.07124}{{\ttfamily arXiv:1712.07124 [hep-ph]}}.
%%CITATION = ARXIV:1712.07124;%%.

\bibitem{Louppe:2017ipp}
G.~Louppe, K.~Cho, C.~Becot, and K.~Cranmer, ``{QCD-Aware Recursive Neural
  Networks for Jet Physics},''
  \href{http://dx.doi.org/10.1007/JHEP01(2019)057}{{\em JHEP} {\bfseries 01}
  (2019) 057},
\href{http://arxiv.org/abs/1702.00748}{{\ttfamily arXiv:1702.00748 [hep-ph]}}.
%%CITATION = ARXIV:1702.00748;%%.

\bibitem{Erdmann:2018shi}
M.~Erdmann, E.~Geiser, Y.~Rath, and M.~Rieger, ``{Lorentz Boost Networks:
  Autonomous Physics-Inspired Feature Engineering},''
\href{http://arxiv.org/abs/1812.09722}{{\ttfamily arXiv:1812.09722 [hep-ex]}}.
%%CITATION = ARXIV:1812.09722;%%.

\bibitem{Hocker:2007ht}
A.~Hoecker, P.~Speckmayer, J.~Stelzer, J.~Therhaag, E.~von Toerne, and H.~Voss,
  ``{TMVA: Toolkit for Multivariate Data Analysis},'' {\em PoS} {\bfseries
  ACAT} (2007) 040,
\href{http://arxiv.org/abs/physics/0703039}{{\ttfamily arXiv:physics/0703039}}.
%%CITATION = PHYSICS/0703039;%%.

\bibitem{Pearkes:2017hku}
J.~Pearkes, W.~Fedorko, A.~Lister, and C.~Gay, ``{Jet Constituents for Deep
  Neural Network Based Top Quark Tagging},''
\href{http://arxiv.org/abs/1704.02124}{{\ttfamily arXiv:1704.02124 [hep-ex]}}.
%%CITATION = ARXIV:1704.02124;%%.

\bibitem{Almeida:2015jua}
L.~G. Almeida, M.~Backović, M.~Cliche, S.~J. Lee, and M.~Perelstein,
  ``{Playing Tag with ANN: Boosted Top Identification with Pattern
  Recognition},'' \href{http://dx.doi.org/10.1007/JHEP07(2015)086}{{\em JHEP}
  {\bfseries 07} (2015) 086},
\href{http://arxiv.org/abs/1501.05968}{{\ttfamily arXiv:1501.05968 [hep-ph]}}.
%%CITATION = ARXIV:1501.05968;%%.

\bibitem{Macaluso:2018tck}
S.~Macaluso and D.~Shih, ``{Pulling Out All the Tops with Computer Vision and
  Deep Learning},'' \href{http://dx.doi.org/10.1007/JHEP10(2018)121}{{\em JHEP}
  {\bfseries 10} (2018) 121},
\href{http://arxiv.org/abs/1803.00107}{{\ttfamily arXiv:1803.00107 [hep-ph]}}.
%%CITATION = ARXIV:1803.00107;%%.

\bibitem{Kasieczka:2017nvn}
G.~Kasieczka, T.~Plehn, M.~Russell, and T.~Schell, ``{Deep-learning Top Taggers
  or The End of QCD?},'' \href{http://dx.doi.org/10.1007/JHEP05(2017)006}{{\em
  JHEP} {\bfseries 05} (2017) 006},
\href{http://arxiv.org/abs/1701.08784}{{\ttfamily arXiv:1701.08784 [hep-ph]}}.
%%CITATION = ARXIV:1701.08784;%%.

\bibitem{Cogan:2014oua}
J.~Cogan, M.~Kagan, E.~Strauss, and A.~Schwarztman, ``{Jet-Images: Computer
  Vision Inspired Techniques for Jet Tagging},''
  \href{http://dx.doi.org/10.1007/JHEP02(2015)118}{{\em JHEP} {\bfseries 02}
  (2015) 118},
\href{http://arxiv.org/abs/1407.5675}{{\ttfamily arXiv:1407.5675 [hep-ph]}}.
%%CITATION = ARXIV:1407.5675;%%.

\bibitem{Baldi:2016fql}
P.~Baldi, K.~Bauer, C.~Eng, P.~Sadowski, and D.~Whiteson, ``{Jet Substructure
  Classification in High-Energy Physics with Deep Neural Networks},''
  \href{http://dx.doi.org/10.1103/PhysRevD.93.094034}{{\em Phys. Rev.}
  {\bfseries D93} no.~9, (2016) 094034},
\href{http://arxiv.org/abs/1603.09349}{{\ttfamily arXiv:1603.09349 [hep-ex]}}.
%%CITATION = ARXIV:1603.09349;%%.

\bibitem{deOliveira:2015xxd}
L.~de~Oliveira, M.~Kagan, L.~Mackey, B.~Nachman, and A.~Schwartzman,
  ``{Jet-images — deep learning edition},''
  \href{http://dx.doi.org/10.1007/JHEP07(2016)069}{{\em JHEP} {\bfseries 07}
  (2016) 069},
\href{http://arxiv.org/abs/1511.05190}{{\ttfamily arXiv:1511.05190 [hep-ph]}}.
%%CITATION = ARXIV:1511.05190;%%.

\bibitem{Collins:2019jip}
J.~H. Collins, K.~Howe, and B.~Nachman, ``{Extending the search for new
  resonances with machine learning},''
  \href{http://dx.doi.org/10.1103/PhysRevD.99.014038}{{\em Phys. Rev.}
  {\bfseries D99} no.~1, (2019) 014038},
\href{http://arxiv.org/abs/1902.02634}{{\ttfamily arXiv:1902.02634 [hep-ph]}}.
%%CITATION = ARXIV:1902.02634;%%.

\bibitem{Kasieczka:2019dbj}
A.~Butter {\em et~al.}, ``{The Machine Learning Landscape of Top Taggers},''
\href{http://arxiv.org/abs/1902.09914}{{\ttfamily arXiv:1902.09914 [hep-ph]}}.
%%CITATION = ARXIV:1902.09914;%%.

\bibitem{Collins:2018epr}
J.~H. Collins, K.~Howe, and B.~Nachman, ``{Anomaly Detection for Resonant New
  Physics with Machine Learning},''
  \href{http://dx.doi.org/10.1103/PhysRevLett.121.241803}{{\em Phys. Rev.
  Lett.} {\bfseries 121} no.~24, (2018) 241803},
\href{http://arxiv.org/abs/1805.02664}{{\ttfamily arXiv:1805.02664 [hep-ph]}}.
%%CITATION = ARXIV:1805.02664;%%.

\bibitem{Cerri:2018anq}
O.~Cerri, T.~Q. Nguyen, M.~Pierini, M.~Spiropulu, and J.-R. Vlimant,
  ``{Variational Autoencoders for New Physics Mining at the Large Hadron
  Collider},''
\href{http://arxiv.org/abs/1811.10276}{{\ttfamily arXiv:1811.10276 [hep-ex]}}.
%%CITATION = ARXIV:1811.10276;%%.

\bibitem{Andreassen:2018apy}
A.~Andreassen, I.~Feige, C.~Frye, and M.~D. Schwartz, ``{JUNIPR: a Framework
  for Unsupervised Machine Learning in Particle Physics},''
  \href{http://dx.doi.org/10.1140/epjc/s10052-019-6607-9}{{\em Eur. Phys. J.}
  {\bfseries C79} no.~2, (2019) 102},
\href{http://arxiv.org/abs/1804.09720}{{\ttfamily arXiv:1804.09720 [hep-ph]}}.
%%CITATION = ARXIV:1804.09720;%%.

\bibitem{Farina:2018fyg}
M.~Farina, Y.~Nakai, and D.~Shih, ``{Searching for New Physics with Deep
  Autoencoders},''
\href{http://arxiv.org/abs/1808.08992}{{\ttfamily arXiv:1808.08992 [hep-ph]}}.
%%CITATION = ARXIV:1808.08992;%%.

\bibitem{Heimel:2018mkt}
T.~Heimel, G.~Kasieczka, T.~Plehn, and J.~M. Thompson, ``{QCD or What?},''
\href{http://arxiv.org/abs/1808.08979}{{\ttfamily arXiv:1808.08979 [hep-ph]}}.
%%CITATION = ARXIV:1808.08979;%%.

\bibitem{Hajer:2018kqm}
J.~Hajer, Y.-Y. Li, T.~Liu, and H.~Wang, ``{Novelty Detection Meets Collider
  Physics},''
\href{http://arxiv.org/abs/1807.10261}{{\ttfamily arXiv:1807.10261 [hep-ph]}}.
%%CITATION = ARXIV:1807.10261;%%.

\bibitem{Dasgupta:2013ihk}
M.~Dasgupta, A.~Fregoso, S.~Marzani, and G.~P. Salam, ``{Towards an
  understanding of jet substructure},''
  \href{http://dx.doi.org/10.1007/JHEP09(2013)029}{{\em JHEP} {\bfseries 09}
  (2013) 029},
\href{http://arxiv.org/abs/1307.0007}{{\ttfamily arXiv:1307.0007 [hep-ph]}}.
%%CITATION = ARXIV:1307.0007;%%.

\bibitem{Plehn:2011tg}
T.~Plehn and M.~Spannowsky, ``{Top Tagging},''
  \href{http://dx.doi.org/10.1088/0954-3899/39/8/083001}{{\em J. Phys.}
  {\bfseries G39} (2012) 083001},
\href{http://arxiv.org/abs/1112.4441}{{\ttfamily arXiv:1112.4441 [hep-ph]}}.
%%CITATION = ARXIV:1112.4441;%%.

\bibitem{Kasieczka:2015jma}
G.~Kasieczka, T.~Plehn, T.~Schell, T.~Strebler, and G.~P. Salam, ``{Resonance
  Searches with an Updated Top Tagger},''
  \href{http://dx.doi.org/10.1007/JHEP06(2015)203}{{\em JHEP} {\bfseries 06}
  (2015) 203},
\href{http://arxiv.org/abs/1503.05921}{{\ttfamily arXiv:1503.05921 [hep-ph]}}.
%%CITATION = ARXIV:1503.05921;%%.

\bibitem{Plehn:2010st}
T.~Plehn, M.~Spannowsky, M.~Takeuchi, and D.~Zerwas, ``{Stop Reconstruction
  with Tagged Tops},'' \href{http://dx.doi.org/10.1007/JHEP10(2010)078}{{\em
  JHEP} {\bfseries 10} (2010) 078},
\href{http://arxiv.org/abs/1006.2833}{{\ttfamily arXiv:1006.2833 [hep-ph]}}.
%%CITATION = ARXIV:1006.2833;%%.

\bibitem{Plehn:2009rk}
T.~Plehn, G.~P. Salam, and M.~Spannowsky, ``{Fat Jets for a Light Higgs},''
  \href{http://dx.doi.org/10.1103/PhysRevLett.104.111801}{{\em Phys. Rev.
  Lett.} {\bfseries 104} (2010) 111801},
\href{http://arxiv.org/abs/0910.5472}{{\ttfamily arXiv:0910.5472 [hep-ph]}}.
%%CITATION = ARXIV:0910.5472;%%.

\bibitem{Sterman:2008kj}
G.~F. Sterman, ``{Some Basic Concepts of Perturbative QCD},'' {\em Acta Phys.
  Polon.} {\bfseries B39} (2008) 2151--2172,
\href{http://arxiv.org/abs/0807.5118}{{\ttfamily arXiv:0807.5118 [hep-ph]}}.
%%CITATION = ARXIV:0807.5118;%%.

\bibitem{Cacciari:2008gp}
M.~Cacciari, G.~P. Salam, and G.~Soyez, ``{The anti-$k_t$ jet clustering
  algorithm},'' \href{http://dx.doi.org/10.1088/1126-6708/2008/04/063}{{\em
  JHEP} {\bfseries 04} (2008) 063},
\href{http://arxiv.org/abs/0802.1189}{{\ttfamily arXiv:0802.1189 [hep-ph]}}.
%%CITATION = ARXIV:0802.1189;%%.

\bibitem{Dokshitzer:1997in}
Y.~L. Dokshitzer, G.~D. Leder, S.~Moretti, and B.~R. Webber, ``{Better jet
  clustering algorithms},''
  \href{http://dx.doi.org/10.1088/1126-6708/1997/08/001}{{\em JHEP} {\bfseries
  08} (1997) 001},
\href{http://arxiv.org/abs/hep-ph/9707323}{{\ttfamily arXiv:hep-ph/9707323
  [hep-ph]}}.
%%CITATION = HEP-PH/9707323;%%.

\bibitem{Ellis:1993tq}
S.~D. Ellis and D.~E. Soper, ``{Successive combination jet algorithm for hadron
  collisions},'' \href{http://dx.doi.org/10.1103/PhysRevD.48.3160}{{\em Phys.
  Rev.} {\bfseries D48} (1993) 3160--3166},
\href{http://arxiv.org/abs/hep-ph/9305266}{{\ttfamily arXiv:hep-ph/9305266
  [hep-ph]}}.
%%CITATION = HEP-PH/9305266;%%.

\bibitem{Catani:1993hr}
S.~Catani, Y.~L. Dokshitzer, M.~H. Seymour, and B.~R. Webber, ``{Longitudinally
  invariant $K_t$ clustering algorithms for hadron hadron collisions},''
\href{http://dx.doi.org/10.1016/0550-3213(93)90166-M}{{\em Nucl. Phys.}
  {\bfseries B406} (1993) 187--224}.
%%CITATION = NUPHA,B406,187;%%.

\bibitem{Blazey:2000qt}
G.~C. Blazey {\em et~al.}, ``{Run II jet physics},'' in {\em {QCD and weak
  boson physics in Run II. Proceedings, Batavia, USA, March 4-6, June 3-4,
  November 4-6, 1999}}, pp.~47--77.
\newblock 2000.
\newblock \href{http://arxiv.org/abs/hep-ex/0005012}{{\ttfamily
  arXiv:hep-ex/0005012 [hep-ex]}}.
\newblock
\url{http://lss.fnal.gov/cgi-bin/find_paper.pl?conf-00-092}.
\newblock
%%CITATION = HEP-EX/0005012;%%.

\bibitem{Cacciari2012}
M.~Cacciari, G.~P. Salam, and G.~Soyez, ``Fastjet user manual,''
  \href{http://dx.doi.org/10.1140/epjc/s10052-012-1896-2}{{\em The European
  Physical Journal C} {\bfseries 72} no.~3, (Mar, 2012) 1896}.
  \url{https://doi.org/10.1140/epjc/s10052-012-1896-2}.

\bibitem{AutoEncoderImage}
A.~Zucconi,
  ``\href{https://www.alanzucconi.com/2018/03/14/an-introduction-to-autoencoders/}{AN
  INTRODUCTION TO AUTOENCODERS}.''
\newblock
  \url{https://www.alanzucconi.com/2018/03/14/an-introduction-to-autoencoders/}.

\bibitem{726791}
Y.~{Lecun}, L.~{Bottou}, Y.~{Bengio}, and P.~{Haffner}, ``Gradient-based
  learning applied to document recognition,''
  \href{http://dx.doi.org/10.1109/5.726791}{{\em Proceedings of the IEEE}
  {\bfseries 86} no.~11, (Nov, 1998) 2278--2324}.

\bibitem{Masci2011StackedCA}
J.~Masci, U.~Meier, D.~C. Ciresan, and J.~Schmidhuber,
  ``\href{https://link.springer.com/chapter/10.1007/978-3-642-21735-7_7}{Stacked
  Convolutional Auto-Encoders for Hierarchical Feature Extraction},'' in {\em
  21st International Conference on Artificial Neural Networks, Espoo, Finland,
  June 14-17, 2011, Proceedings, Part I}, T.~Honkela, W.~Duch, M.~Girolami, and
  S.~Kaski, eds., {International Conference on Artificial Neural Networks
  (ICANN)}.
\newblock 2011.

\bibitem{RIS_0}
{Hahnloser Richard H. R.}, {Sarpeshkar Rahul}, {Mahowald Misha A.}, {Douglas
  Rodney J.}, and {Seung H. Sebastian}, ``{Digital selection and analogue
  amplification coexist in a cortex-inspired silicon circuit},''
  \href{http://dx.doi.org/https://doi.org/10.1038/35016072
  10.1038/35016072}{{\em Nature} {\bfseries 405} (Jun, 2000) 947}.
  \url{https://www.nature.com/articles/35016072\#supplementary-information}.

\bibitem{2013arXiv1312.6114K}
D.~P. {Kingma} and M.~{Welling}, ``{Auto-Encoding Variational Bayes},'' {\em
  arXiv e-prints} (Dec., 2013) arXiv:1312.6114,
  \href{http://arxiv.org/abs/1312.6114}{{\ttfamily arXiv:1312.6114 [stat.ML]}}.

\bibitem{2017arXiv170104722M}
L.~{Mescheder}, S.~{Nowozin}, and A.~{Geiger}, ``{Adversarial Variational
  Bayes: Unifying Variational Autoencoders and Generative Adversarial
  Networks},'' {\em arXiv e-prints} (Jan, 2017) arXiv:1701.04722,
  \href{http://arxiv.org/abs/1701.04722}{{\ttfamily arXiv:1701.04722 [cs.LG]}}.

\bibitem{2011arXiv1106.1925P}
R.~{Prescott Adams} and R.~S. {Zemel}, ``{Ranking via Sinkhorn Propagation},''
  {\em arXiv e-prints} (June, 2011) arXiv:1106.1925,
  \href{http://arxiv.org/abs/1106.1925}{{\ttfamily arXiv:1106.1925 [stat.ML]}}.

\bibitem{Monk:2018zsb}
J.~W. Monk, ``{Deep Learning as a Parton Shower},''
\href{http://arxiv.org/abs/1807.03685}{{\ttfamily arXiv:1807.03685 [hep-ph]}}.
%%CITATION = ARXIV:1807.03685;%%.

\bibitem{Sjostrand:2006za}
T.~Sjostrand, S.~Mrenna, and P.~Z. Skands, ``{PYTHIA 6.4 Physics and Manual},''
  \href{http://dx.doi.org/10.1088/1126-6708/2006/05/026}{{\em JHEP} {\bfseries
  05} (2006) 026},
\href{http://arxiv.org/abs/hep-ph/0603175}{{\ttfamily arXiv:hep-ph/0603175
  [hep-ph]}}.
%%CITATION = HEP-PH/0603175;%%.

\bibitem{Sjostrand:2014zea}
T.~Sjöstrand, S.~Ask, J.~R. Christiansen, R.~Corke, N.~Desai, P.~Ilten,
  S.~Mrenna, S.~Prestel, C.~O. Rasmussen, and P.~Z. Skands, ``{An Introduction
  to PYTHIA 8.2}'' \href{http://dx.doi.org/10.1016/j.cpc.2015.01.024}{{\em
  Comput. Phys. Commun.} {\bfseries 191} (2015) 159--177},
\href{http://arxiv.org/abs/1410.3012}{{\ttfamily arXiv:1410.3012 [hep-ph]}}.
%%CITATION = ARXIV:1410.3012;%%.

\bibitem{deFavereau:2013fsa}
{\bfseries DELPHES 3} Collaboration, J.~de~Favereau, C.~Delaere, P.~Demin,
  A.~Giammanco, V.~Lemaître, A.~Mertens, and M.~Selvaggi, ``{DELPHES 3, A
  modular framework for fast simulation of a generic collider experiment},''
  \href{http://dx.doi.org/10.1007/JHEP02(2014)057}{{\em JHEP} {\bfseries 02}
  (2014) 057},
\href{http://arxiv.org/abs/1307.6346}{{\ttfamily arXiv:1307.6346 [hep-ex]}}.
%%CITATION = ARXIV:1307.6346;%%.

\bibitem{DBLP:journals/corr/KingmaB14}
D.~P. Kingma and J.~Ba, ``Adam: {A} method for stochastic optimization,'' {\em
  CoRR} {\bfseries abs/1412.6980} (2014) ,
  \href{http://arxiv.org/abs/1412.6980}{{\ttfamily arXiv:1412.6980}}.
  \url{http://arxiv.org/abs/1412.6980}.

\bibitem{2015arXiv151201274C}
T.~{Chen}, M.~{Li}, Y.~{Li}, M.~{Lin}, N.~{Wang}, M.~{Wang}, T.~{Xiao},
  B.~{Xu}, C.~{Zhang}, and Z.~{Zhang}, ``{MXNet: A Flexible and Efficient
  Machine Learning Library for Heterogeneous Distributed Systems},'' {\em arXiv
  e-prints} (Dec., 2015) arXiv:1512.01274,
  \href{http://arxiv.org/abs/1512.01274}{{\ttfamily arXiv:1512.01274 [cs.DC]}}.

\bibitem{DBLP:journals/corr/AbadiABBCCCDDDG16}
M.~Abadi, A.~Agarwal, P.~Barham, E.~Brevdo, Z.~Chen, C.~Citro, G.~S. Corrado,
  A.~Davis, J.~Dean, M.~Devin, S.~Ghemawat, I.~J. Goodfellow, A.~Harp,
  G.~Irving, M.~Isard, Y.~Jia, R.~J{\'{o}}zefowicz, L.~Kaiser, M.~Kudlur,
  J.~Levenberg, D.~Man{\'{e}}, R.~Monga, S.~Moore, D.~G. Murray, C.~Olah,
  M.~Schuster, J.~Shlens, B.~Steiner, I.~Sutskever, K.~Talwar, P.~A. Tucker,
  V.~Vanhoucke, V.~Vasudevan, F.~B. Vi{\'{e}}gas, O.~Vinyals, P.~Warden,
  M.~Wattenberg, M.~Wicke, Y.~Yu, and X.~Zheng, ``Tensorflow: Large-scale
  machine learning on heterogeneous distributed systems,'' {\em CoRR}
  {\bfseries abs/1603.04467} (2016) ,
  \href{http://arxiv.org/abs/1603.04467}{{\ttfamily arXiv:1603.04467}}.
  \url{http://arxiv.org/abs/1603.04467}.

\bibitem{DBLP:journals/corr/Tang16d}
Y.~Tang, ``Tf.learn: Tensorflow's high-level module for distributed machine
  learning,'' {\em CoRR} {\bfseries abs/1612.04251} (2016) ,
  \href{http://arxiv.org/abs/1612.04251}{{\ttfamily arXiv:1612.04251}}.
  \url{http://arxiv.org/abs/1612.04251}.

\bibitem{Gallicchio:2010sw}
J.~Gallicchio and M.~D. Schwartz, ``{Seeing in Color: Jet Superstructure},''
  \href{http://dx.doi.org/10.1103/PhysRevLett.105.022001}{{\em Phys. Rev.
  Lett.} {\bfseries 105} (2010) 022001},
\href{http://arxiv.org/abs/1001.5027}{{\ttfamily arXiv:1001.5027 [hep-ph]}}.
%%CITATION = ARXIV:1001.5027;%%.

\bibitem{Dasgupta:2018emf}
M.~Dasgupta, M.~Guzzi, J.~Rawling, and G.~Soyez, ``{Top tagging : an analytical
  perspective},'' \href{http://dx.doi.org/10.1007/JHEP09(2018)170}{{\em JHEP}
  {\bfseries 09} (2018) 170},
\href{http://arxiv.org/abs/1807.04767}{{\ttfamily arXiv:1807.04767 [hep-ph]}}.
%%CITATION = ARXIV:1807.04767;%%.

\bibitem{Dasgupta:2013via}
M.~Dasgupta, A.~Fregoso, S.~Marzani, and A.~Powling, ``{Jet substructure with
  analytical methods},''
  \href{http://dx.doi.org/10.1140/epjc/s10052-013-2623-3}{{\em Eur. Phys. J.}
  {\bfseries C73} no.~11, (2013) 2623},
\href{http://arxiv.org/abs/1307.0013}{{\ttfamily arXiv:1307.0013 [hep-ph]}}.
%%CITATION = ARXIV:1307.0013;%%.

\bibitem{Marzani:2019hun}
S.~Marzani, G.~Soyez, and M.~Spannowsky, ``{Looking inside jets: an
  introduction to jet substructure and boosted-object phenomenology},''
\href{http://arxiv.org/abs/1901.10342}{{\ttfamily arXiv:1901.10342 [hep-ph]}}.
%%CITATION = ARXIV:1901.10342;%%.

\bibitem{Kaplan:2008ie}
D.~E. Kaplan, K.~Rehermann, M.~D. Schwartz, and B.~Tweedie, ``{Top Tagging: A
  Method for Identifying Boosted Hadronically Decaying Top Quarks},''
  \href{http://dx.doi.org/10.1103/PhysRevLett.101.142001}{{\em Phys. Rev.
  Lett.} {\bfseries 101} (2008) 142001},
\href{http://arxiv.org/abs/0806.0848}{{\ttfamily arXiv:0806.0848 [hep-ph]}}.
%%CITATION = ARXIV:0806.0848;%%.

\bibitem{Thaler:2011gf}
J.~Thaler and K.~Van~Tilburg, ``{Maximizing Boosted Top Identification by
  Minimizing N-subjettiness},''
  \href{http://dx.doi.org/10.1007/JHEP02(2012)093}{{\em JHEP} {\bfseries 02}
  (2012) 093},
\href{http://arxiv.org/abs/1108.2701}{{\ttfamily arXiv:1108.2701 [hep-ph]}}.
%%CITATION = ARXIV:1108.2701;%%.

\bibitem{CMS:2014fya}
{\bfseries CMS} Collaboration, C.~Collaboration,
``{Boosted Top Jet Tagging at CMS},''.
%%CITATION = CMS-PAS-JME-13-007;%%.

\bibitem{Aaboud:2018psm}
{\bfseries ATLAS} Collaboration, M.~Aaboud {\em et~al.}, ``{Performance of
  top-quark and $W$-boson tagging with ATLAS in Run 2 of the LHC},''
\href{http://arxiv.org/abs/1808.07858}{{\ttfamily arXiv:1808.07858 [hep-ex]}}.
%%CITATION = ARXIV:1808.07858;%%.

\bibitem{BRUN199781}
R.~Brun and F.~Rademakers, ``Root — an object oriented data analysis
  framework,''
  \href{http://dx.doi.org/http://dx.doi.org/10.1016/S0168-9002(97)00048-X}{{\em
  Nuclear Instruments and Methods in Physics Research Section A: Accelerators,
  Spectrometers, Detectors and Associated Equipment} {\bfseries 389} no.~1,
  (1997) 81 -- 86}.
  \url{http://www.sciencedirect.com/science/article/pii/S016890029700048X}. New
  Computing Techniques in Physics Research V.

\bibitem{Ellis:2014eya}
S.~D. Ellis, A.~Hornig, D.~Krohn, and T.~S. Roy, ``{On Statistical Aspects of
  Qjets},'' \href{http://dx.doi.org/10.1007/JHEP01(2015)022}{{\em JHEP}
  {\bfseries 01} (2015) 022},
\href{http://arxiv.org/abs/1409.6785}{{\ttfamily arXiv:1409.6785 [hep-ph]}}.
%%CITATION = ARXIV:1409.6785;%%.

\bibitem{Ellis:2012sn}
S.~D. Ellis, A.~Hornig, T.~S. Roy, D.~Krohn, and M.~D. Schwartz, ``{Qjets: A
  Non-Deterministic Approach to Tree-Based Jet Substructure},''
  \href{http://dx.doi.org/10.1103/PhysRevLett.108.182003}{{\em Phys. Rev.
  Lett.} {\bfseries 108} (2012) 182003},
\href{http://arxiv.org/abs/1201.1914}{{\ttfamily arXiv:1201.1914 [hep-ph]}}.
%%CITATION = ARXIV:1201.1914;%%.

\bibitem{Larkoski:2013eya}
A.~J. Larkoski, G.~P. Salam, and J.~Thaler, ``{Energy Correlation Functions for
  Jet Substructure},'' \href{http://dx.doi.org/10.1007/JHEP06(2013)108}{{\em
  JHEP} {\bfseries 06} (2013) 108},
\href{http://arxiv.org/abs/1305.0007}{{\ttfamily arXiv:1305.0007 [hep-ph]}}.
%%CITATION = ARXIV:1305.0007;%%.

\bibitem{Stewart:2010tn}
I.~W. Stewart, F.~J. Tackmann, and W.~J. Waalewijn, ``{N-Jettiness: An
  Inclusive Event Shape to Veto Jets},''
  \href{http://dx.doi.org/10.1103/PhysRevLett.105.092002}{{\em Phys. Rev.
  Lett.} {\bfseries 105} (2010) 092002},
\href{http://arxiv.org/abs/1004.2489}{{\ttfamily arXiv:1004.2489 [hep-ph]}}.
%%CITATION = ARXIV:1004.2489;%%.

\bibitem{Thaler:2010tr}
J.~Thaler and K.~Van~Tilburg, ``{Identifying Boosted Objects with
  N-subjettiness},'' \href{http://dx.doi.org/10.1007/JHEP03(2011)015}{{\em
  JHEP} {\bfseries 03} (2011) 015},
\href{http://arxiv.org/abs/1011.2268}{{\ttfamily arXiv:1011.2268 [hep-ph]}}.
%%CITATION = ARXIV:1011.2268;%%.

\bibitem{Khachatryan:2014vla}
{\bfseries CMS} Collaboration, V.~Khachatryan {\em et~al.}, ``{Identification
  techniques for highly boosted W bosons that decay into hadrons},''
  \href{http://dx.doi.org/10.1007/JHEP12(2014)017}{{\em JHEP} {\bfseries 12}
  (2014) 017},
\href{http://arxiv.org/abs/1410.4227}{{\ttfamily arXiv:1410.4227 [hep-ex]}}.
%%CITATION = ARXIV:1410.4227;%%.

\bibitem{Krohn:2013lba}
D.~Krohn, M.~D. Schwartz, M.~Low, and L.-T. Wang, ``{Jet Cleansing: Pileup
  Removal at High Luminosity},''
  \href{http://dx.doi.org/10.1103/PhysRevD.90.065020}{{\em Phys. Rev.}
  {\bfseries D90} no.~6, (2014) 065020},
\href{http://arxiv.org/abs/1309.4777}{{\ttfamily arXiv:1309.4777 [hep-ph]}}.
%%CITATION = ARXIV:1309.4777;%%.

\bibitem{Cacciari:2014gra}
M.~Cacciari, G.~P. Salam, and G.~Soyez, ``{SoftKiller, a particle-level pileup
  removal method},''
  \href{http://dx.doi.org/10.1140/epjc/s10052-015-3267-2}{{\em Eur. Phys. J.}
  {\bfseries C75} no.~2, (2015) 59},
\href{http://arxiv.org/abs/1407.0408}{{\ttfamily arXiv:1407.0408 [hep-ph]}}.
%%CITATION = ARXIV:1407.0408;%%.

\bibitem{Bertolini:2014bba}
D.~Bertolini, P.~Harris, M.~Low, and N.~Tran, ``{Pileup Per Particle
  Identification},'' \href{http://dx.doi.org/10.1007/JHEP10(2014)059}{{\em
  JHEP} {\bfseries 10} (2014) 059},
\href{http://arxiv.org/abs/1407.6013}{{\ttfamily arXiv:1407.6013 [hep-ph]}}.
%%CITATION = ARXIV:1407.6013;%%.

\bibitem{Komiske:2017ubm}
P.~T. Komiske, E.~M. Metodiev, B.~Nachman, and M.~D. Schwartz, ``{Pileup
  Mitigation with Machine Learning (PUMML)},''
  \href{http://dx.doi.org/10.1007/JHEP12(2017)051}{{\em JHEP} {\bfseries 12}
  (2017) 051},
\href{http://arxiv.org/abs/1707.08600}{{\ttfamily arXiv:1707.08600 [hep-ph]}}.
%%CITATION = ARXIV:1707.08600;%%.

\end{thebibliography}\endgroup
\bibliographystyle{utphys}

\end{document}